\documentclass[a4paper,11pt]{article}
\usepackage{jheppub}
\usepackage{lineno}
\usepackage{siunitx}
\usepackage{physics}
\usepackage{booktabs}
\usepackage{adjustbox}

\newcommand{\QRanges}{[q^2_{\text{min}},q^2_{\vphantom{\text{min}}\text{max}}]}

\usepackage[table]{xcolor}
\usepackage{cancel}
\usepackage{subcaption}
\usepackage{enumitem}
\xdefinecolor{tugreen}{RGB}{132, 184, 25}
\xdefinecolor{cyanLike}{RGB}{0,197,205}
\xdefinecolor{lightygreeny}{RGB}{144,238,144}

\def \azeL{{H_0^L}}

\def \apaL{{H_\parallel^L}}

\def \apeL{{H_\perp^L}}

\def \re{\textrm{Re}}
\def \im{\textrm{Im}}

\definecolor{LightGray}{gray}{0.91}
\definecolor{LightBlue}{rgb}{0.87, 0.94, 1}

\title{\boldmath Effective field theory analysis of rare $|\Delta c|=|\Delta u|=1$ charm decays}

\author[a,b,c]{Hector Gisbert,}
\emailAdd{hector.gisbert@universidadeuropea.es}
\author[d]{Gudrun Hiller}
\emailAdd{ghiller@physik.uni-dortmund.de}
\author[d]{and Dominik Suelmann}
\emailAdd{dominik.suelmann@tu-dortmund.de}
\affiliation[a]{Escuela de Ciencias, Ingenier\'ia y Dise\~{n}o, Universidad Europea de Valencia, \\
Passeig de la Petxina 2, 46008 Valencia, Spain}
\affiliation[b]{Istituto Nazionale di Fisica Nucleare (INFN), Sezione di Padova, \\
Via F. Marzolo 8, 35131 Padova, Italy}
\affiliation[c]{Dipartimento di Fisica e Astronomia `G.~Galilei', Universit\`a di Padova,
 \\ Via F. Marzolo 8, 35131 Padova, Italy}
\affiliation[d]{TU Dortmund University, Department of Physics,\\
Otto-Hahn-Str.4, D-44221 Dortmund, Germany}

\abstract{We perform a global analysis of $|\Delta c| = |\Delta u| = 1$  transitions using  recent data on 
$D^0 \to \mu^+\mu^-$, $D^+ \to \pi ^+\,\mu^+\mu^-$, $\Lambda_c \to p\,\mu^+\mu^-$, and $D^0 \to \pi^+\pi^-\,\mu^+\mu^-$ decays,
and work out constraints on new physics Wilson coefficients  $\mathcal{C}_{7,9,10}^{(\prime)}$. While  results are  consistent with the standard model,
we find sizeable room for new physics that can be cleanly signaled with null test observables, not probed with  searches in other sectors such as
kaon and $b$-decays.
The decay $D^0 \to \pi^+\pi^-\,\mu^+\mu^-$  requires better understanding of hadronic contributions to be competitive in the current  fit. 
Progress  can be achieved  by  precision study of  the double differential decay rate in the dipion and dimuon masses,  together with 
improved theory modelling and $D \to \pi \pi$  transition form factors.
On the other hand, the 4-body decay is an important contributor to the future global analysis due to
 its angular distributions that probe complementary combinations of Wilson coefficients, and as a QCD laboratory. The decay
 $\Lambda_c \to p\,\ell^+\ell^-$  is the rising star due to the simplicity of a 3-body decay with available  form factors from lattice QCD, sensitivity to both 4-fermion and electromagnetic dipole couplings and its null test forward-backward asymmetry.
 
}

\begin{document}
\maketitle


\section{Introduction}
\label{sec:intro}

Flavor-changing neutral currents (FCNCs)  are  suppressed in the Standard Model (SM) by  the Glashow-Iliopoulos-Maiani (GIM) mechanism,
an intricate  interplay of the  fermion masses and mixings. The suppression is parametrically enhanced for up-type quarks. Together with the
rich set of  couplings at work in radiative, semileptonic and dineutrino  decay modes  and  the experimental prospects at high luminosity flavor facilities, such as LHCb~\cite{Cerri:2018ypt}, Belle II~\cite{Belle-II:2018jsg}, BES III~\cite{BESIII:2020nme}, and possible future machines~\cite{Lyu:2021tlb,FCC:2018byv}
 this singles out  SM tests with  FCNC decays of charmed hadrons.

Charm physics is notoriously  challenging due to resonance pollution and the unfortunate value of the charm quark which  sits between the nominal regions of validity of effective field theories for light quarks  and
heavy quarks. Instead, sensitivity to new physics (NP)  comes from approximate symmetries of the SM. Notably, the  efficient GIM-suppression
\cite{Burdman:2001tf,Fajfer:2002gp,deBoer:2015boa}  gives directions to systematically identify clean null test observables from angular distributions of $D$-mesons 
 \cite{DeBoer:2018pdx} and  charm baryons \cite{Golz:2022alh}.

In this study we present the first global analysis of rare $|\Delta c|=|\Delta u|=1$ charm decays, based on  $D^0 \to \mu^+\mu^-$, $D^+ \to \pi ^+\,\mu^+\mu^-$, $\Lambda_c \to p\,\mu^+\mu^-$, and $D^0 \to \pi^+\pi^-\,\mu^+\mu^-$.
We  take advantage of the novel theory proposals  to test the SM, hadronic input from lattice QCD \cite{Meinel:2017ggx,FermilabLattice:2022gku} including recent data from the LHC~\cite{LHCb:2024hju,CMS-PAS-BPH-23-008}  to  extract limits on the Wilson coefficients governing $c\to u\,\mu^+\mu^-$ transitions.

While the null tests are able to  signal NP, the interpretation of the data in terms of NP, specifically when one aims to obtain limits,  is affected  by hadronic backgrounds. 
We discuss this for  angular null test observables in 
$\Lambda_c \to p\,\mu^+\mu^-$, and $D^0 \to \pi^+\pi^-\,\mu^+\mu^-$ decays.
We furthermore  analyze the impact of  binning in the dilepton invariant mass on the NP reach.

The paper is organized as follows:  We give the weak effective theory framework in Sec.~\ref{sec:EFT}. We present decay modes and their angular  observables in Sec.~\ref{sec:AngularObs}.
We also work out constraints from recent data on the branching ratio of $\Lambda_c \to p\,\mu^+\mu^-$ decays.
The phenomenological analysis and extraction of limits from data on  the 4-body $D^0 \to \pi^+\pi^-\,\mu^+\mu^-$ decays is given  in Sec.~\ref{sec:DPPfit}.
The results of the global $c \to  u \mu^+ \mu^-$ fit are presented in Sec.~\ref{sec:fit}.
In Sec.~\ref{sec:future} we study the future prospects for $\Lambda_c \to p\,\mu^+\mu^-$ decays and the physics potential of the forward-backward asymmetry for SM tests.
We investigate the impact of binning in the dilepton mass and discuss the complementarity with other rare charm null test observables.
In Sec.~\ref{sec:conclusions} we conclude.  Auxiliary information on form factors and the $D^0 \to \pi^+\pi^-\,\mu^+\mu^-$ angular observables is provided in the appendix.

\section{Effective Theory Framework}
\label{sec:EFT}

We give the weak effective theory framework for  semileptonic $|\Delta c| = |\Delta u| = 1$ transitions. They  are described  by the  effective  low energy Hamiltonian~\cite{Greub:1996wn, Fajfer:2002gp,deBoer:2015boa,deBoer:2016dcg}
\begin{align}\label{eq:Heff}
\mathcal{H}_{\rm eff} = -\frac{4\,G_F}{\sqrt{2}} \frac{\alpha_e}{4\pi} &\left[\sum_{q=d,s}\lambda_q\sum_{i=1,2}c_i^{(q)}(\mu)\,\mathcal{O}_i^{(q)}(\mu) + \sum_{i\neq T,\,T_5} \biggl( c_i(\mu)\, \mathcal{O}_i(\mu) + c_i^{\prime}(\mu)\, \mathcal{O}_i^{\prime}(\mu) \biggr) \right. \nonumber\\
&\left. + \sum_{i=T,\,T_5} c_i(\mu)\, \mathcal{O}_i(\mu) \right],
\end{align}
with  dimension-six operators~\cite{Chetyrkin:1996vx, Bobeth:1999mk, Gambino:2003zm}
\begin{align}
\begin{aligned}
\mathcal{O}_2^{(q)} &= (\bar{u}_L \gamma_\mu q_L)(\bar{q}_L \gamma^\mu c_L)~, &
\mathcal{O}_1^{(q)} &= (\bar{u}_L \gamma_\mu T^a q_L)(\bar{q}_L \gamma^\mu T^a c_L)~, \\
\mathcal{O}_7 &= \frac{m_c}{e} (\bar{u}_L \sigma_{\mu\nu} c_R)F^{\mu\nu}~, &
\mathcal{O}_8 &= \frac{m_c\, g_s}{e^2} (\bar{u}_L \sigma_{\mu\nu} T^a c_R) G^{\mu\nu}_a~, \\
\mathcal{O}_9 &= (\bar{u}_L \gamma_\mu c_L)(\bar{\ell} \gamma^\mu \ell)~, &
\mathcal{O}_{10} &= (\bar{u}_L \gamma_\mu c_L)(\bar{\ell} \gamma^\mu \gamma_5 \ell)~, \\
\mathcal{O}_{S\,(P)} &= (\bar{u}_L c_R)(\bar{\ell}(\gamma_5) \ell)~, &
\mathcal{O}_{T\,(T_5)} &= \frac{1}{2} (\bar{u} \sigma_{\mu\nu} c)(\bar{\ell} \sigma^{\mu\nu} (\gamma_5) \ell)~.
\end{aligned}
\label{eq:operators}
\end{align}
Here, $q_{L,R} = \frac{1}{2}(1 \mp \gamma_5)q$ represent the chiral quark fields, $\sigma^{\mu\nu} = \frac{i}{2}[\gamma^\mu, \gamma^\nu]$~, $T^a$ denote the generators of $SU(3)_C$ and $g_s$ is the strong coupling constant. $G_F$ denotes  Fermi's constant and  $\lambda_q = V_{cq}^* V_{uq}$, where $V_{cq}$ and $V_{uq}$ are 
elements of the Cabibbo-Kobayashi-Maskawa (CKM) matrix.
The fine-structure constant, with the electromagnetic coupling $e$, reads $\alpha_e = e^2/(4\pi)$. The field strength tensors of  the electromagnetic and gluonic fields are 
denoted by $F^{\mu\nu}$ and $G_a^{\mu\nu}$, respectively, where $a=1,\ldots,8$. 

Primed  operators $\mathcal{O}_i^\prime$  are obtained  from the $\mathcal{O}_i$ by swapping  the chiral projectors of  the quark fields  $L(R) \to R(L)$.
We distinguish  the purely NP contribution  to the Wilson coefficients  $ \mathcal{C}_{i}$ (capital letter)   from the total one $c_i$ (small letter) as
$c_{i}^{(\prime)} = \mathcal{C}_{i}^{(\prime) \,\text{SM}} + \mathcal{C}_{i}^{(\prime)}$, 
with the SM coefficients denoted by $ \mathcal{C}_{i}^{(\prime) \,\text{SM}} $.
In the SM and, more general in models with minimal flavor violation,  the primed coefficients are suppressed  by $m_u/m_c$, the ratio of up to charm mass, which is negligible. Scalar, pseudoscalar and tensor operators also vanish in the SM, 
$\mathcal{C}_{S,P}^{(\prime)\text{SM}}= \mathcal{C}_{T,T5}^\text{SM}=0$.

In the SM   $c \to u \gamma, \ell^+ \ell^-$ induced decays are dominated by contributions from  the four-quark operators $\mathcal{O}^{(q)}_{1,2}$.
They enter with single Cabibbo-suppression, see the corresponding CKM-factors $\lambda_q$. The other Wilson coefficients are protected by the GIM-mechanism,
and vanish at the weak scale $\mu_W$.
$\mathcal{O}^{(q)}_{1,2}$ are  induced by tree-level $W$-boson exchange at $\mu_W$, and  mix  under QCD  and running to the $b$-quark scale.
At the $b$-quark mass scale  the $b$-quark is integrated out, and contributions to  ${\mathcal{O}_{7,8, 9}} $ and additional  four-quark penguin operators, not spelled out explicitly in (\ref{eq:operators}) and negligible for $c \to u \ell^+ \ell^-$ transitions, are induced. 
The coefficients are then evolved to the  charm mass scale, $\mu = m_c$ \cite{deBoer:2015boa,deBoer:2017dgn,deBoer:2016dcg}.
While the coefficients of $\mathcal{O}^{(q)}_{1,2}$ are order one, and all others  assume sub-few percent yet finite values, the coefficient of $\mathcal{O}_{10}$
remains protected, $\mathcal{C}_{10}^\text{SM}=0$. Finite contributions to $\mathcal{O}_{10}$ are expected at higher order in the electromagnetic interaction  \cite{DeBoer:2018pdx}. Effects are estimated at about per mille level, which is far beyond experimental sensitivity.

The  operators $\mathcal{O}_{7,9}$ also receive  perturbative contributions from matrix elements of four-quark operators  involving electromagnetic and gluonic corrections,
included in  effective coefficients. 
At $\mu = m_c$, and at next-to-next-to-leading order (NNLO)
they assume the following values~\cite{deBoer:2015boa,  deBoer:2017dgn}, 
\begin{align}
|\mathcal{C}_9^{\,\text{eff}}(q^2)|  \lesssim \mathcal{O}(0.1) \quad  (\lesssim \,  0.01)~, \quad
|\mathcal{C}_7^{\,\text{eff}}(q^2)| \lesssim \, 0.01  \quad  ( \simeq \mathcal{O}(0.001)) ~,
\end{align}
with  entries   in parentheses  specific to the region of high dilepton invariant mass squared,  $q^2 \gtrsim 1 \, \text{GeV}^2$.
These  are all small and subdominant  for SM phenomenology. 

We also  mention additional contributions worked out in the framework of QCD factorization, notably
weak annihilation  for $D \to \rho \ell^+  \ell^-$ \cite{Feldmann:2017izn}, $D^+ \to \pi^+ \ell^+ \ell^-$  \cite{Bharucha:2020eup},  and multibody  radiative  decays $D \to  P _1 P_2  \gamma, P_{1,2}=\pi,K$ \cite{Adolph:2020ema}.  The annihilation of a charged  (neutral) $D$ meson into a $W$-boson  is color favored (suppressed), hence larger  for $D^\pm$ and $D_s$
than $D^0$ decays.  Quantitatively, it  has  only  a subleading effect in the  $D^+ \to \pi^+ \mu^+ \mu^-$ analysis  \cite{Bharucha:2020eup} with current experimental binning and accuracy~\cite{LHCb:2020car}. Significant corrections are also expected on general grounds due to the small separation between the charm quark mass and the scale of QCD.

The  phenomenology of $c \to u \ell^+ \ell^-$ transitions in the SM is governed by resonance contributions.
These are predominantly sourced by  charged-current operators  $\mathcal{O}^{(q)}_{1,2}$, which have order one Wilson coefficients and  give  large rates of  charm  to light hadrons.
The latter can decay electromagnetically to $\ell^+ \ell^-$, such as the $\rho, \omega$, or $\phi$-mesons, and therefore effectively  contribute to $\mathcal{O}_9$ with coefficient denoted by $\mathcal{C}_9^R$.
Due to their non-perturbative nature, these long-distance, resonant effects pose a challenge in controlling decay amplitudes and require improved data-driven modelling.

We model resonances with a phenomenological Breit-Wigner ansatz, fitting  parameters to data.
An alternative way would be to use  data on the cross section of $e^+ e^-$ to hadrons  plus dispersion relations, and  assuming factorization, a method  put forward by   \cite{Kruger:1996cv}. Historically, it has been employed in rare $B$-decays to avoid double-counting of short-and long-distance contributions to
$\mathcal{O}_9$. However, in rare charm decays, the short-distance contribution is essentially negligible, so this not an issue here.
Recent analysis also reveals  significant uncertainties from the dispersive analysis \cite{Bharucha:2020eup}. In addition, pseudoscalar resonances such as $\eta, \eta^\prime$ 
decaying to $\ell^+ \ell^-$ are not
included. 
We therefore employ the simpler, yet flexible ansatz to explore
the phenomenology of  $\Lambda_c \to p \mu^+ \mu^-$ decays, and in particular,  the  4-body $D^0 \to \pi^+ \pi^- \mu^+ \mu^-$ decays . The latter has resonance structure not just in the dilepton mass, hence requires further modelling beyond the approach \cite{Kruger:1996cv} anyway.

We study observables that are especially sensitive to physics beyond the SM.
The  key feature to do so,
which  sets charm FCNCs apart  from those in the kaon and $B$-sector  is the GIM-protection of  $\mathcal{C}_{10}$,
and  that the operators $\mathcal{O}_{10}, \mathcal{O}_{10}^\prime$, characterized by an axial-vector coupling in the leptonic sector, are not switched on by QCD.
In essence, this  nullifies  $V - A$ structure at low energies within the SM and  observables  proportional to this operator  vanish  \cite{DeBoer:2018pdx}.
This opportunity invites dedicated searches with concrete null test observables in $D^0 \to \pi^+ \pi^- \mu^+ \mu^-$  \cite{DeBoer:2018pdx} and $\Lambda_c \to p \mu^+ \mu^-$ \cite{Golz:2021imq} decays.

\section{Rare charm decays}
\label{sec:AngularObs}

We discuss the $c \to u\,\ell^+\ell^-$ decay modes considered in the global analysis.
The present best bounds from semileptonic decays stem from muons. However, we give the distributions for $\ell=e, \mu$. Tau-pairs are too heavy to be produced in the decays
of charmed hadrons with the sole exception of the lepton-flavor violating decay $D \to e \tau$.

In Sec.~\ref{sec:Dll} we discuss $D^0 \to \mu^+\mu^-$ and $D \to \pi \,\mu^+\mu^-$ decays and give the current limits on Wilson coefficients.
We discuss baryonic $\Lambda_c \to p\,\mu^+\mu^-$ decays and work out constraints on the Wilson coefficients using recent data from LHCb in
Sec.~\ref{sec:Lc}.
Distributions of  4-body $D^0 \to \pi^+\pi^-\,\mu^+\mu^-$ decays, which have the richest angular structure and offer the most observables from full angular analysis are given in Sec.~\ref{sec:4body}.
Due to its complicated nature  we perform the phenomenological analysis and extractions of limits  in Sec.~\ref{sec:DPPfit}.

We focus  on NP effects  from semileptonic and radiative  operators ${\mathcal{O}_{7,9,10}}^{(\prime)}$,
but give  here also limits  on scalar, pseudoscalar and tensor operators.

\subsection[\texorpdfstring{$D \to \ell^+ \ell^-$ and $D \to \pi \ell^+ \ell^-$}{D -> l+l- and D -> pi l+ l-}]{$D \to \ell^+ \ell^-$ and $D \to \pi \ell^+ \ell^-$ \label{sec:Dll}}

The double differential distribution of $D\to P\,\ell^+\ell^-$, $P=\pi,K$ decays can be written  as~\cite{Bobeth:2007dw}
\begin{align}\label{eq:ang_DPll}
    \frac{\text{d}^2\Gamma}{\text{d}q^2\text{d}\cos\theta_\ell} = a(q^2) + b(q^2)\cos\theta_\ell + c(q^2)\cos^2\theta_\ell~,
\end{align}
where $\theta_\ell$ is the angle between the momentum of the  $\ell^+$ and the  negative direction of flight of the parent $D$-meson  in the dilepton $\ell^+\ell^-$ rest frame. 
The distribution \eqref{eq:ang_DPll} leads to null test observables of the SM,  the forward-backward asymmetry of the lepton pair
\begin{align}\label{eq:AFBDPll}
    A^{D \to P \ell \ell}_\text{FB}(q^2) = \frac{1}{\Gamma}\left[\int^1_0 - \int_{-1}^0\right]\frac{\text{d}^2\Gamma}{\text{d}q^2\text{d}\cos\theta_\ell}\,\text{d}\cos\theta_\ell = \frac{b(q^2)}{\Gamma}~,
\end{align}
and the ``flat'' term,
\begin{align}\label{eq:flatDPll}
    F_H(q^2) = \frac{2}{\Gamma}\left(a(q^2) + c(q^2)\right)~.
\end{align}
$A^{D \to P \ell \ell}_\text{FB}$ and $F_H$ are normalized to the decay rate $\Gamma$,
which in general depends on the $q^2$-bin, $q^2 \in \QRanges$,
\begin{align}\label{eq:normDPll}
    \Gamma = \int_{q^2_\text{min}}^{q^2_\text{max}}\text{d}q^2  \int_{-1}^1 \frac{\text{d}^2\Gamma}{\text{d}q^2\text{d}\cos\theta_\ell}\,\text{d}\cos\theta_\ell 
    =2\int_{q^2_\text{min}}^{q^2_\text{max}}\text{d}q^2\left(a(q^2) + \frac{c(q^2)}{3}\right)~.
\end{align}
Explicit expressions for  the angular functions $a, b, c$ in terms of the NP Wilson coefficients are given in Ref.~\cite{Bause:2019vpr}.
$F_H(q^2)$ and $A^{D \to P \ell \ell}_\text{FB}(q^2)$ are sensitive to (pseudo-)scalar and tensor operators. None of them is measured.

The most stringent constraint on $D\to P\,\ell^+\ell^-$ branching ratios is  $\mathcal{B}(D^+\to\pi^+\,\mu^+\mu^-)<6.7\cdot 10^{-8}$  @ 90\% C.L. from LHCb~\cite{LHCb:2020car} in the full-$q^2$ region, $4 m_\mu^2 \leq q^2 \leq (m_{D}-m_\pi)^2$, where $m_{D}$, $m_\pi$ and $m_\ell$  denote mass of  the $D$-meson, pion and lepton $\ell=e,\mu$, respectively.\footnote{Ref.~\cite{LHCb:2020car}  excluded the resonant region $\sqrt{q^2}\in[0.525,1.250]\,\mathrm{GeV}$ and extrapolated to obtain the limits for the full-$q^2$ region.} We obtain
\begin{align}\label{eq:DPllcons}
\begin{split}
1.2&\,\left|\mathcal{C}_7\right|^2 + 1.4\,\left|\mathcal{C}_9\right|^2 + 1.4\,\left|\mathcal{C}_{10}\right|^2 + 2.8\,\left|\mathcal{C}_S\right|^2 \\ 
+\, 2.8&\,\left|\mathcal{C}_P\right|^2 + 0.4\left|\mathcal{C}_T\right|^2 + 0.3\,\left|\mathcal{C}_{T_5}\right|^2  + 0.6\,\mathrm{Re}\left[\mathcal{C}_9 \,{\mathcal{C}_T}^*\right]\\
+\, 1.1&\,\mathrm{Re}\left[\mathcal{C}_{10}\, {\mathcal{C}_P}^*\right] + 2.6\,\mathrm{Re}\left[\mathcal{C}_7 \,{\mathcal{C}_9}^*\right] + 0.6\,\mathrm{Re}\left[\mathcal{C}_7\, {\mathcal{C}_T}^*\right] \lesssim 1 \,,
\end{split}
\end{align}
using  $D \to \pi$ lattice  form factors~\cite{FermilabLattice:2022gku,Lubicz:2018rfs}. Contributions from primed Wilson coefficients have not been spelled out to avoid clutter but they can be included by replacing $\mathcal{C}_i \to \mathcal{C}_i + \mathcal{C}^{\prime}_i$. Results are consistent with 
previous analyses~\cite{Gisbert:2020vjx, Bause:2019vpr,Bharucha:2020eup}.

The purely leptonic decay $D^0\to\ell^+\ell^-$ offers complementary information to $D\to P\,\ell^+\ell^-$ as it is sensitive to the difference $\mathcal{C}_{10,S,P} - \mathcal{C}^\prime_{10,S,P}$ of Wilson coefficients,
\begin{align}
\mathcal{B}(D^0\to\ell^+\ell^-) &= \frac{G_F^2\,\alpha_e^2\, m_D^5\, f_D^2}{64\,\pi^3\, m_c^2\, \Gamma_D} \sqrt{1 - \frac{4\,m_\ell^2}{m_D^2}} \left[\left(1 - \frac{4 \,m_\ell^2}{m_D^2}\right)\left|\mathcal{C}_S - \mathcal{C}_S^\prime\right|^2\right.\nonumber\\
&\left.+ \left|\mathcal{C}_P - \mathcal{C}_P^\prime + \frac{2\,m_\ell\, m_c}{m_D^2}(\mathcal{C}_{10} - \mathcal{C}_{10}^\prime)\right|^2\right] + \mathcal{B}_{\text{LD}}~.
\end{align}
Here,  the leading long-distance contribution $\mathcal{B}_{\text{LD}}$ comes from two-photon intermediate states and can be related to the decay $D^0\to \gamma \gamma$ \cite{Burdman:2001tf}. 
Using $\mathcal{B}(D^0\to \gamma \gamma) < 8.5\cdot 10^{-7}$ @ 90\% C.L. by Belle \cite{Belle:2015pzk} one obtains  $\mathcal{B}_{\text{LD}}(D^0\to\mu^+\mu^-)<4\cdot 10^{-11}$.
This is  about two orders of magnitude below 
the upper limit  @ 90\% C.L. by CMS~\cite{CMS-PAS-BPH-23-008}
\begin{align} \label{eq:CMS}
\mathcal{B}(D^0\to\mu^+\mu^-)< 2.2\cdot 10^{-9} \, , 
\end{align}
which  yields 
\begin{align}\label{eq:Dllcons}
    |\mathcal{C}_{10} - \mathcal{C}_{10}^\prime| < 0.52~,
\end{align}
where we used the value of the  $D$-meson decay constant  $f_D= 209.0 \pm 2.4 \, \text{MeV}$  \cite{FlavourLatticeAveragingGroupFLAG:2021npn}. 
The SM estimate $\mathcal{B}_{\text{SM}}(D^0\to \gamma \gamma) \sim 10^{-8}$ \cite{Burdman:2001tf,Fajfer:2001ad} corresponding to $\mathcal{B}_{\text{SM}}(D^0\to \mu^+\mu^-) \sim 10^{-12}$ is  even lower albeit with large theoretical uncertainties.
Therefore, $\mathcal{B}_{\text{LD}}$  can be safely neglected   but it ultimately limits the sensitivity  to test $\mathcal{C}_{10}-\mathcal{C}_{10}^\prime$  below  $\sim 0.01$.
Assuming that only $\mathcal{C}_{10}$, $\mathcal{C}_{10}^\prime$ contributions are present in Eq.~\eqref{eq:DPllcons} it follows
   $ |\mathcal{C}_{10} + \mathcal{C}_{10}^\prime| \lesssim 0.85$.
Combination  with Eq.~\eqref{eq:Dllcons} allows one to put  limits on the individual coefficients as 
\begin{equation} \label{eq: C10max}
    |\mathcal{C}_{10}|,|\mathcal{C}_{10}^{\prime}| \lesssim 0.7~.
\end{equation}
Similarly, combining $(|\mathcal{C}_{S} + \mathcal{C}_{S}^\prime|^2+|\mathcal{C}_{P} + \mathcal{C}_{P}^\prime|^2)^{1/2}  < 0.60$ from Eq.~\eqref{eq:DPllcons} 
 with 
\begin{align}\label{eq:SPwcs}
 (   |\mathcal{C}_{S} - \mathcal{C}_{S}^\prime|^2+  |\mathcal{C}_{P} - \mathcal{C}_{P}^\prime|^2 )^{1/2} < 0.04~,
\end{align}
from  (\ref{eq:CMS})  and the charm MS-bar mass at the charm mass scale $m_c(\mu_c) \simeq 1.27$ GeV
 we obtain  for the pseudo-scalar and  scalar Wilson coefficients 
\begin{equation}
    |\mathcal{C}_{S}|,|\mathcal{C}_{S}^{\prime}|, |\mathcal{C}_{P}|,|\mathcal{C}_{P}^{\prime}|  \lesssim 0.4~.
\end{equation}
The bound (\ref{eq:SPwcs})  is an order of magnitude stronger than the one on the vector operators  Eq.~\eqref{eq:Dllcons}, as it is not affected by  chiral suppression $\sim m_\mu/m_D$.
Tensor operators are bounded by $D^+ \to \pi^+ \mu^+ \mu^-$ decays, see Eq.~\eqref{eq:DPllcons}, as follows
\begin{equation} \label{eq: tensorsmax}
    |\mathcal{C}_{T}|  \lesssim 1.6~,   |\mathcal{C}_{T5}|  \lesssim 1.8 \, . 
\end{equation}

\subsection[\texorpdfstring{$\Lambda_c\to p\,\ell^+\ell^-$}{Lambda c -> p l+ l-}]{$\boldsymbol{\Lambda_c\to p\,\ell^+\ell^-}$ \label{sec:Lc}}

The double differential distribution for $\Lambda_c\to p\,\ell^+\ell^-$ can be written as~\cite{Golz:2021imq}
\begin{align}\label{eq:ang_Lcpll}
    \frac{\text{d}^2\Gamma}{\text{d}q^2\text{d}\cos\theta_\ell} = \frac{3}{2} \left(K_{1ss}(q^2)\sin^2\theta_\ell + K_{1cc}(q^2)\cos^2\theta_\ell + K_{1c}(q^2)\cos\theta_\ell\right)~.
\end{align}
Here,  $\theta_\ell$ is defined as the angle between the negative direction of flight of the parent charm baryon and the $\ell^+$ in the dilepton center of mass frame. 
The differential decay rate is obtained after integrating over the lepton angle as
\begin{align}
\frac{\text{d}\Gamma}{\text{d}q^2}=\int_{-1}^{1}\frac{\text{d}^2\Gamma}{\text{d}q^2\text{d} \! \cos\theta_\ell}\,\text{d} \! \cos\theta_\ell=2\,K_{1ss}(q^2) + K_{1cc}(q^2) \, . 
\label{eq:single-Lc}
\end{align}
In addition, baryonic  $\Lambda_c\to p\,\ell^+\ell^-$ decays give rise to angular observables $A_\text{FB}$ and $F_L$
\begin{align}
A_\text{FB}(q^2)  = \frac{3}{2}\,\frac{K_{1c}(q^2)}{2K_{1ss} (q^2)  +K_{1cc}(q^2)}  ~,\quad F_L (q^2)  = \frac{2K_{1ss} (q^2)  -K_{1cc}(q^2)}{2K_{1ss}(q^2)  +K_{1cc}(q^2)  }~,
\end{align}
Additional observables exist if the $\Lambda_c$'s are polarized  \cite{Golz:2022alh}. In case they are polarized, for instance if originating from $\Lambda_b$- or $Z$-decays but the
polarization is  not measured, the differential distribution (\ref{eq:ang_Lcpll}) remains the same.
The dilepton mass squared is kinematically limited to be within $4 m_\ell^2 \leq q^2 \leq (m_{\Lambda_c}-m_p)^2$, where $m_{\Lambda_c}$ $(m_p)$ denotes the $\Lambda_c$ (proton) mass.

The longitudinal polarization fraction obeys $0 \leq F_L (q^2) \leq 1$. Together with  (\ref{eq:single-Lc}) one finds that
\begin{align}
0 \leq K_{1cc}  (q^2)  \leq \frac{1}{2} \frac{\text{d}\Gamma}{\text{d}q^2}  \, , \quad \frac{1}{4} \frac{\text{d}\Gamma}{\text{d}q^2}  \leq K_{1ss}  (q^2)  \leq \frac{1}{2} \frac{\text{d}\Gamma}{\text{d}q^2} \, . 
\end{align}
The NP coefficients are encoded in the angular functions $K_{1ss}$, $K_{1cc}$, and $K_{1c}$, see Ref.~\cite{Golz:2021imq} for details. 
The baryonic decay mode is phenomenologically richer than $D\to P\,\ell^+\ell^-$ due to 
 the higher spin. In particular, the angular functions receive contributions from both $\mathcal{C}_i + \mathcal{C}_i^\prime$ and $\mathcal{C}_i - \mathcal{C}_i^\prime$, and
 are sensitive to dipole operators.
 Importantly,  in $\Lambda_c$-decays the forward-backward asymmetry $A_{\text{FB}} \propto K_{1c} \propto \mathcal{C}_{10}$ is a  clean null test of the SM.

\begingroup
\renewcommand*{\arraystretch}{1.6}
\begin{table}[h!]
    \centering
     \resizebox{\textwidth}{!}{
    \begin{tabular}{c|c|c|c|c}
        \hline\hline
        \rowcolor{LightBlue} bin & $q^2\,/\,\mathrm{GeV}^2$ region & $\mathcal{B}_{\text{exp},\text{bin}}$ & $\mathcal{B}_{\text{SM},\text{bin}}$ & $\langle F_{L} \rangle_{\text{SM},\text{bin}}$\\
        \hline\hline
        low-$q^2$ & $[0.045,0.258]$ & $< 0.93 \times 10^{-8}$ & $[0.26,0.83]\times 10^{-8}$ & $[0.62,0.71]$\\
        \hline
        high-$q^2$ & $[1.122,1.817]$ & $< 3.0 \times 10^{-8}$ & $ [0.02,2.3]\times 10^{-8}$ & $[0.00,0.91]$\\
        \hline
        low \& high-$q^2$ & $[0.045,0.258]$ & $< 2.9 \times 10^{-8}$ & $[0.3,3.1]\times 10^{-8}$ & $[0.48,0.71]$\\
         & $\cup [1.122,1.817]$ & & & \\
        \hline
        $\omega$ & $[0.552,0.677]$ & $(7.3\pm2.9)\times 10^{-8}\phantom{1}^\dagger$ & $[5.1,9.6]\times 10^{-8}$ & $[0.57,0.67]$\\
        \hline
        $\rho$ & $[0.346,0.552]$ & $(6.9\pm2.0)\times 10^{-8}\phantom{1}^\dagger$ & $[4.3,11.9]\times 10^{-8}$ & $[0.59,0.66]$\\
         & $\cup [0.677,0.932]$ & & & \\
        \hline
        $\phi$ & $[0.959,1.122]$ & $(3.02\pm0.45)\times 10^{-7}\phantom{1}^\dagger$ & $[2.3,3.7]\times 10^{-7}$ & $[0.48,0.53]$\\
        \hline
        full-$q^2$ & $[4m_\mu^2,(m_{\Lambda_c}-m_p)^2]$ & - & $[3.6,6.0]\times 10^{-7}$ & $[0.51,0.57]$\\
        \hline
    \end{tabular}
    }
    \caption{Experimental $90\%\, \text{C.L.}$ upper  limits  on  the  $\Lambda_c \to p\,\mu^+\mu^-$  branching fraction  for  different $q^2$ bins~\cite{LHCb:2024hju}.
    We further provide SM predictions for branching ratios and the longitudinal polarization fraction $\langle F_L \rangle_{\text{bin}}$. 
    $^\dagger$ Values extracted from Eq.~(\ref{eq:R-region}).}
    \label{tab:FL_SM}
\end{table}
\endgroup

 For the phenomenological analysis it is useful to define  binned observables 
\begin{align}
\begin{split}
    \mathcal{B}_{\QRanges} &= \tau_{\Lambda_c}\int_{q^2_{\text{min}}}^{q^2_{\text{max}}}(2K_{1ss}(q^2)+K_{1cc}(q^2))\,\mathrm{d}q^2~, \\
    \langle A_\text{FB} \rangle_{\QRanges} &= \frac{3}{2}\,\frac{\int_{q^2_{\text{min}}}^{q^2_{\text{max}}}K_{1c}(q^2)\,\mathrm{d}q^2}{\int_{q^2_{\text{min}}}^{q^2_{\text{max}}}(2K_{1ss}(q^2)+K_{1cc}(q^2)) \,\mathrm{d}q^2}~,\\
    \langle F_L \rangle_{\QRanges} &= \frac{\int_{q^2_{\text{min}}}^{q^2_{\text{max}}}(2K_{1ss}(q^2)-K_{1cc}(q^2))\,\mathrm{d}q^2}{\int_{q^2_{\text{min}}}^{q^2_{\text{max}}}(2K_{1ss}(q^2)+K_{1cc}(q^2))\,\mathrm{d}q^2}~,
\end{split}
\label{eq:AFB}
\end{align}
where $\tau_{\Lambda_c}$ denotes the lifetime of the $\Lambda_c$-baryon.  Note,  $0 \leq \langle F_L \rangle_{\QRanges}  \leq 1$.

In the presence of CP-violation the decays of  $\Lambda_c$ and $\bar  \Lambda_c$ differ.
Observables for the conjugated mode  $\bar  \Lambda_c \to \bar  p\,\mu^+\mu^-$  are denoted by a bar. They can be constructed  from the CP-conjugate angular distribution
$ \text{d}^2 \bar \Gamma /\text{d}q^2\text{d}\cos\theta_\ell = \frac{3}{2} \left(\bar K_{1ss} \sin^2\theta_\ell + \bar  K_{1cc}\cos^2\theta_\ell - \bar K_{1c}\cos\theta_\ell\right)$, where the 
$\bar K_{1ss}, \bar K_{1cc}, \bar K_{1c}$  are  obtained from the  $K_{1ss}, K_{1cc},K_{1c}$
  by conjugating the weak phases, respectively. As we keep the definition of $\theta_\ell$ with its reference to the $\ell^+$ for the decay and its conjugate identical,
  there is a minus sign in front of the $\cos\theta_\ell$-term.
  This gives, for instance,  the  forward-backward asymmetry of  $\bar  \Lambda_c \to \bar  p\,\mu^+\mu^-$  decays as
    \begin{align}
\begin{split}
    \langle \bar A_\text{FB} \rangle_{\QRanges} &= - \frac{3}{2}\,\frac{\int_{q^2_{\text{min}}}^{q^2_{\text{max}}} \bar  K_{1c}(q^2)\,\mathrm{d}q^2}{\int_{q^2_{\text{min}}}^{q^2_{\text{max}}}(2\bar K_{1ss}(q^2)+\bar K_{1cc}(q^2)) \,\mathrm{d}q^2}~.\\
\end{split}
\label{eq:AFBbar}
\end{align}

\subsubsection{Modelling resonances} 

To model the resonance contributions $\Lambda_c \to p R ( \to \mu^+ \mu^-)$, $R=\rho, \omega,\phi$,  we employ a phenomenological ansatz 
\begin{align}\label{eq:BWwcLambdaC}
\mathcal{C}_9^{R}(q^2)\,=\,\frac{a_{\rho} \,\text{e}^{i\,\delta_{\rho}} }{q^2-m_\rho^2+i\,m_\rho\,\Gamma_\rho}+\frac{a_{\omega} \text{e}^{i\,\delta_{\omega}}}{q^2-m_\omega^2+i\,m_\omega\,\Gamma_\omega}+\frac{a_{\phi} \,\text{e}^{i\,\delta_{\phi}}}{q^2-m_\phi^2+i\,m_\phi\,\Gamma_\phi}  \, , 
\end{align}
with  the resonance  masses $m_R$ and  decay widths $\Gamma_R$, which we take from Ref.~\cite{Workman:2022ynf},  and the a priori unknown strong phases $\delta_R$.
The modulus of the resonance parameters $a_R$ are determined  using data from LHCb~\cite{LHCb:2024hju}
\begin{align} \nonumber
    \frac{\mathcal{B}(\Lambda_c^+ \to p\, \mu^+\mu^-)_{\text{$\omega$-region}}}{\mathcal{B}(\Lambda_c^+ \to p\, \mu^+\mu^-)_{\text{$\phi$-region}}} = 0.240\pm 0.035 \,,\\
    \frac{\mathcal{B}(\Lambda_c^+ \to p\, \mu^+\mu^-)_{\text{$\rho$-region}}}{\mathcal{B}(\Lambda_c^+ \to p\, \mu^+\mu^-)_{\text{$\phi$-region}}} = 0.229\pm 0.056 \,,
    \label{eq:R-region}
\end{align}
with the  $q^2$-regions specified  in  Table~\ref{tab:FL_SM} and assuming uncorrelated statistical and systematic uncertainties.
We obtain 
\begin{align}\label{eq:fitLambdac_resonance}
    a_\phi\,=\,0.108^{+0.008}_{-0.008} \,  \text{GeV}^2,\,
    a_\omega\,=\,0.074^{+0.012}_{-0.015} \,  \text{GeV}^2,\,
    a_\rho\,=\,0.50^{+0.06}_{-0.06} \,  \text{GeV}^2 \, ,
\end{align}
where we used the narrow-width approximation for $\mathcal{B}(\Lambda_c^+\to p\,\mu^+\mu^-)_{\text{$\phi$-region}} \approx \mathcal{B}(\Lambda_c^+\to p\,\phi)\times \mathcal{B}(\phi\to \mu^+\mu^-)= (3.02\pm0.45)\times 10^{-7}$~\cite{Workman:2022ynf}. 
In the isospin limit $a_\rho/a_\omega = 3$  and $\delta_\rho - \delta_\omega = \pi$, as used in \cite{Golz:2021imq}.
The  ratio $a_\rho/a_\omega = 6.8^{+2.7}_{-1.6}$  obtained from the  fit therefore  mildly   hints at  isospin violation in this decay.  Here, we used  $\delta_\omega - \delta_\rho = \pi$ and $\delta_\phi - \delta_\rho = 0$, but note that other choices for the strong phases give values that agree within uncertainties. 

The $\eta$ and $\eta^\prime$ are very narrow and less pronounced than the vector meson resonances  in the dilepton mass distributions,
see \cite{deBoer:2015boa,Fajfer:2015mia} for study in  $D \to \pi \ell^+ \ell^-$ decays.  This is consistent with the observed hierarchy of the $\Lambda_c$-branching ratios
$ \frac{\mathcal{B}(\Lambda_c^+ \to p\, \mu^+\mu^-)_{\text{$\eta$-region}}}{\mathcal{B}(\Lambda_c^+ \to p\, \mu^+\mu^-)_{\text{$\phi$-region}}}=0.032 \pm 0.014$~\cite{LHCb:2024hju}.
We therefore refrain in this study from including the pseudoscalar resonances in our model, noting that also in the future their use for resonance enhanced new physics signals would  require very precise $q^2$-binning.  

\begin{figure}
    \centering
    \includegraphics[width=0.6\textwidth]{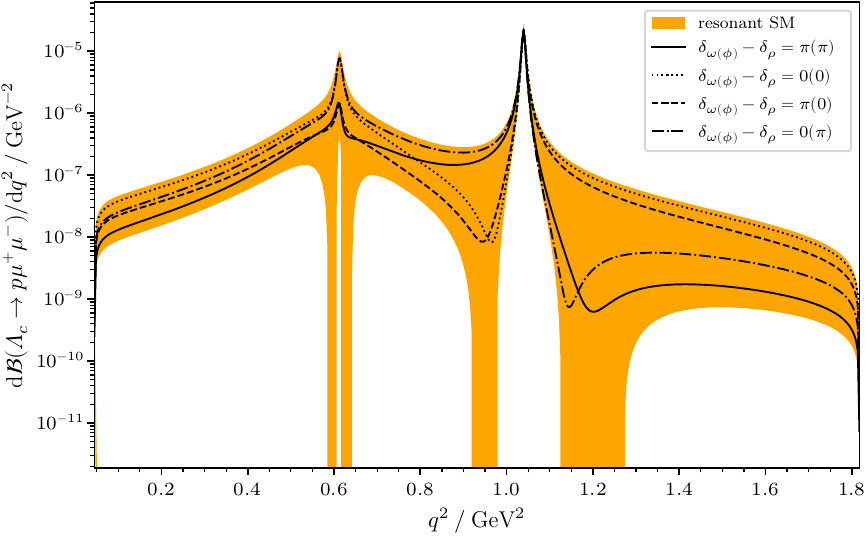}
    \caption[Caption]{
 The differential branching ratio of $\Lambda_c \to p\,\mu^+\mu^-$ decays including uncertainties predominantly caused by strong phases (orange band). The black curves correspond to different relative strong phases
 $\delta_\omega-\delta_\rho$ and $\delta_\phi - \delta_\rho$  and illustrate the sensitivity to them in the different $q^2$-regions.}
    \label{fig:Lc-br}
\end{figure}

The relative strong phases $\delta_\omega-\delta_\rho$ and $\delta_\phi - \delta_\rho$ cannot be determined from the fit, because the given $q^2$ regions depend only mildly  on the strong phases and  hence a fit is not possible with the current precision and binning. Additional bins in regions, where resonances overlap, could  pin down these phases in the future. 
 In Fig.~\ref{fig:Lc-br} we show the $\Lambda_c \to p\,\mu^+\mu^-$ differential branching ratio, including curves (black) with fixed relative strong phases
$\delta_{\omega(\phi)}-\delta_\rho$.
We learn that the sensitivity is large  in  the high-$q^2$  region, however, rates are small.
At low $q^2$, on the other hand,  rates are large but the variation with phases is smaller.
In between the  $\rho/\omega$ and $\phi$ resonances  the impact of the relative strong phases is also significant, and invites further study.

In Tab.~\ref{tab:FL_SM}  we present the SM predictions and data \cite{LHCb:2024hju} for the branching ratio  $\mathcal{B}_\text{SM}$, as well as  $\langle F_L \rangle_{\text{SM}}$, for different $q^2$ bins for $\Lambda_c \to p\,\mu^+\mu^-$ decays. 
The  SM uncertainties are predominantly  driven by the  strong phases and cancel partially in $\langle F_L \rangle$.

The forward-backward asymmetry vanishes in the SM because it is proportional to the product  $\mathcal{C}_9^{R}\, \mathcal{C}_{10}$. Consequently, even a small NP contribution from $\mathcal{C}_{10}$ can be significantly enhanced through interference with a resonance contribution $\mathcal{C}_9^R$. 
We analyze the NP reach  and prospects of $A_\text{FB}$   in Sec.~\ref{sec:Lcfuture}.
For further details, see Ref.~\cite{Golz:2021imq}.

\subsubsection{Limits on  NP}

Using the upper $90\%$ C.L. limits on the branching fractions of $\Lambda_c\to p \mu^+\mu^-$ in the low-$q^2$ region, the high-$q^2$ region, and the combined region~\cite{LHCb:2024hju} given in Tab.~\ref{tab:FL_SM},
and the  resonance parameters given in Eq.~\eqref{eq:fitLambdac_resonance}, we obtain  the following constraints on the Wilson coefficients in the low $q^2$ region
\begin{align}\label{eq:Lambdacons1}
&0.28 + 0.27 \left|\mathcal{C}_{10}-\mathcal{C}_{10}^\prime\right|^2 + 0.31 \left|\mathcal{C}_{10}+\mathcal{C}_{10}^\prime\right|^2 + 18 \left|\mathcal{C}_{7}-\mathcal{C}_{7}^\prime\right|^2 \nonumber\\
&+ 17 \left|\mathcal{C}_{7}+\mathcal{C}_{7}^\prime\right|^2 + 0.29 \left|\mathcal{C}_{9}-\mathcal{C}_{9}^\prime\right|^2 + 0.31 \left|\mathcal{C}_{9}+\mathcal{C}_{9}^\prime\right|^2\nonumber \\
&+ 1.8 \,\mathrm{Re}\left[ (\mathcal{C}_7 + \mathcal{C}_7^\prime) (\mathcal{C}_{9} + \mathcal{C}_{9}^\prime)^\ast \right] 
+ 2.4 \,\mathrm{Re}\left[ (\mathcal{C}_7 - \mathcal{C}_7^\prime) (\mathcal{C}_{9} - \mathcal{C}_{9}^\prime)^\ast \right] \\
&- 0.23 \,\mathrm{Re}\left[\mathcal{C}_9 - \mathcal{C}_9^\prime \right] - 0.06 \,\mathrm{Im}\left[\mathcal{C}_9 - \mathcal{C}_9^\prime \right] - 0.25 \,\mathrm{Re}\left[\mathcal{C}_9 + \mathcal{C}_9^\prime \right] - 0.07 \,\mathrm{Im}\left[\mathcal{C}_9 + \mathcal{C}_9^\prime \right] \nonumber\\
&- 0.7 \,\mathrm{Re}\left[\mathcal{C}_7 + \mathcal{C}_7^\prime \right] - 0.20 \,\mathrm{Im}\left[\mathcal{C}_7 + \mathcal{C}_7^\prime \right] - 0.9 \,\mathrm{Re}\left[\mathcal{C}_7 - \mathcal{C}_7^\prime \right] - 0.26 \,\mathrm{Im}\left[\mathcal{C}_7 - \mathcal{C}_7^\prime \right] \lesssim 1 \,,\nonumber
\end{align}
in the high $q^2$ region
\begin{align} \label{eq:Lambdacons-hi}
&0.5 \left|\mathcal{C}_{10} - \mathcal{C}_{10}^\prime\right|^2 + 0.15 \left|\mathcal{C}_{10} + \mathcal{C}_{10}^\prime\right|^2 + 1.9 \left|\mathcal{C}_{7} - \mathcal{C}_{7}^\prime\right|^2 \nonumber\\
&+ 0.6 \left|\mathcal{C}_{7} + \mathcal{C}_{7}^\prime\right|^2 + 0.5 \left|\mathcal{C}_{9} - \mathcal{C}_{9}^\prime\right|^2 + 0.14 \left|\mathcal{C}_{9} + \mathcal{C}_{9}^\prime\right|^2 \nonumber\\
&+ 0.5 \,\mathrm{Re}\left[ (\mathcal{C}_7 + \mathcal{C}_7^\prime) (\mathcal{C}_{9} + \mathcal{C}_{9}^\prime)^\ast \right] + 1.9 \,\mathrm{Re}\left[ (\mathcal{C}_7 - \mathcal{C}_7^\prime) (\mathcal{C}_{9} - \mathcal{C}_{9}^\prime)^\ast \right] \\
&+ 0.5 \,\mathrm{Re}\left[\mathcal{C}_9 - \mathcal{C}_9^\prime \right] - 0.04 \,\mathrm{Im}\left[\mathcal{C}_9 - \mathcal{C}_9^\prime \right] + 0.19 \,\mathrm{Re}\left[\mathcal{C}_9 + \mathcal{C}_9^\prime \right] - 0.016 \,\mathrm{Im}\left[\mathcal{C}_9 + \mathcal{C}_9^\prime \right] \nonumber\\
&+ 0.4 \,\mathrm{Re}\left[\mathcal{C}_7 + \mathcal{C}_7^\prime \right] - 0.030 \,\mathrm{Im}\left[\mathcal{C}_7 + \mathcal{C}_7^\prime \right] + 1.0 \,\mathrm{Re}\left[\mathcal{C}_7 - \mathcal{C}_7^\prime \right] - 0.09 \,\mathrm{Im}\left[\mathcal{C}_7 - \mathcal{C}_7^\prime \right] \lesssim 1 \,,\nonumber
\end{align}
and for the combined region:
\begin{align}  \label{eq:Lambdacons2}
&0.10 + 0.6 \left|\mathcal{C}_{10} - \mathcal{C}_{10}^\prime\right|^2 + 0.25 \left|\mathcal{C}_{10} + \mathcal{C}_{10}^\prime\right|^2 + 8 \left|\mathcal{C}_{7} - \mathcal{C}_{7}^\prime\right|^2 \nonumber\\
&+ 6 \left|\mathcal{C}_{7} + \mathcal{C}_{7}^\prime\right|^2 + 0.6 \left|\mathcal{C}_{9} - \mathcal{C}_{9}^\prime\right|^2 + 0.25 \left|\mathcal{C}_{9} + \mathcal{C}_{9}^\prime\right|^2 \nonumber\\
&+ 1.1 \,\mathrm{Re}\left[ (\mathcal{C}_7 + \mathcal{C}_7^\prime) (\mathcal{C}_{9} + \mathcal{C}_{9}^\prime)^\ast \right] + 2.7 \,\mathrm{Re}\left[ (\mathcal{C}_7 - \mathcal{C}_7^\prime) (\mathcal{C}_{9} - \mathcal{C}_{9}^\prime)^\ast \right] \\
&+ 0.5 \,\mathrm{Re}\left[\mathcal{C}_9 - \mathcal{C}_9^\prime \right] - 0.06 \,\mathrm{Im}\left[\mathcal{C}_9 - \mathcal{C}_9^\prime \right] + 0.11 \,\mathrm{Re}\left[\mathcal{C}_9 + \mathcal{C}_9^\prime \right] - 0.04 \,\mathrm{Im}\left[\mathcal{C}_9 + \mathcal{C}_9^\prime \right] \nonumber\\
&+ 0.13 \,\mathrm{Re}\left[\mathcal{C}_7 + \mathcal{C}_7^\prime \right] - 0.10 \,\mathrm{Im}\left[\mathcal{C}_7 + \mathcal{C}_7^\prime \right] + 0.8 \,\mathrm{Re}\left[\mathcal{C}_7 - \mathcal{C}_7^\prime \right] - 0.17 \,\mathrm{Im}\left[\mathcal{C}_7 - \mathcal{C}_7^\prime \right] \lesssim 1 \,.\nonumber
\end{align}
The strongest constraints on $\mathcal{C}_{9,10}$ are from the joint region, whereas those on the dipole operator are from the low $q^2$ region.
Comparison with the  constraints from $D \to \pi \mu^+ \mu^-$ and $D \to \mu^+ \mu^-$, given in Eqs.~\eqref{eq:DPllcons} and \eqref{eq:Dllcons}, shows that the limits from $\Lambda_c \to p\, \mu^+ \mu^-$ decays are generally weaker. On the other hand,  the baryonic decay is sensitive at low $q^2$ to the photon pole via $\Lambda_c \to p \gamma^*$, yielding
constraints  from (\ref{eq:Lambdacons1})  on the electromagnetic  dipole operators $\mathcal{O}_7^{(\prime)}$ as
\begin{align} |\mathcal{C}_{7}|,  |\mathcal{C}_{7}^{\prime}|\lesssim 0.2 \, ,  \label{eq:c7bound}
\end{align} 
which are  stronger than the ones  from radiative decay $D^0 \to \rho^0 \gamma$, $|\mathcal{C}_{7}^{(\prime)}|\lesssim 0.3$~\cite{DeBoer:2018pdx}.

Combining data on the branching ratios of  $D \to \pi \mu^+ \mu^-$ (\ref{eq:DPllcons}) and $\Lambda_c$ decays (\ref{eq:Lambdacons2}) allows one  to constrain
$\mathcal{C}_9$ and $\mathcal{C}_9^\prime$ individually as
\begin{align} \label{eq:c9bound}
|\mathcal{C}_9|, |\mathcal{C}_9^\prime| \lesssim 1.1 \, ,
\end{align}
assuming  that no other NP coefficients are present.

 At low $q^2$ and therefore also in the combined region the SM contributions are  not entirely negligible compared to the  experimental limits.
We therefore indicate them by the constant terms on the left hand sides of  (\ref{eq:Lambdacons1}) and    (\ref{eq:Lambdacons2}), noting that they have larger uncertainties than the NP terms. The finite SM branching ratios  tighten the constraints on NP, however, this is within the accuracy of the limits  (\ref{eq:c7bound}), (\ref{eq:c9bound}).

The difference in constraints from  (\ref{eq:Lambdacons1}) and  (\ref{eq:Lambdacons-hi})  highlight the importance of $q^2$-binning. We discuss this in detail for the
null test observable $A_{\text{FB}}$  in Sec.~\ref{sec:Lcfuture}, where we study  the future NP potential of   $\Lambda_c \to p\, \ell^+ \ell^-$ decays.

\subsection[\texorpdfstring{$D\to 
P_1 P_2\,\ell^+\ell^-$}{D -> 
P1 P2 l+ l-}]{$\boldsymbol{D\to 
P_1 P_2\,\ell^+\ell^-}$     \label{sec:4body}}

We discuss observables   of the 4-body decay $D \,\to\, P_1\, P_2\, \ell^+\, \ell^-$. In  Sec.~\ref{subsec:fullangular}  we give  the full angular  distribution and detail its SM predictions.
CP-symmetries and asymmetries of the angular coefficients are discussed in  Sec.~\ref{subsec:observables}.

\subsubsection{Full angular distribution}\label{subsec:fullangular}

The 5-differential  decay  distribution of $D\,\to\, P_1\, P_2\, \ell^+\,\ell^-$ can be written as 
\begin{align}   \label{eq:full}
    \frac{\text{d}^5\Gamma}{\text{d}q^2\,\text{d}p^2\,\text{d}\cos\theta_{P_1}\,\text{d}\cos\theta_\ell\, \text{d}\phi} 
    &=\frac{1}{ 2 \, \pi}  \sum_{i=1}^9 c_i(\theta_\ell,\phi)\, I_i (q^2,p^2,\cos \theta_{P_1})\,,
\end{align}
where  $\pi^+$  equals $P_1$ in $D \to \pi^+ \pi^- \mu^+ \mu^-$ decays, and $q^2$ and $p^2$ represent the invariant mass-squared of the dilepton and the ($P_1 P_2$)-subsystem, respectively. $\theta_\ell$, $\theta_{P_1}$, and $\phi$ are the angular variables defined in Ref.~\cite{DeBoer:2018pdx}. $\theta_\ell$ denotes the angle between the $\ell^-$-momentum and the $D$-momentum in the dilepton center-of-mass system (cms), while $\theta_{P_1}$ represents the angle between the $P_1$-momentum and the negative direction of flight of the $D$-meson in the ($P_1 P_2$)-cms.\footnote{The relation between  $\phi$ and the angle $\phi_\text{LHCb}$ used in the experimental analysis~\cite{LHCb:2021yxk} by the LHCb collaboration is $\phi=\pi+\phi_\text{LHCb}$. 
Therefore, $I_{4,5,7,8}=- I_{4,5,7,8}^\text{LHCb}$. \label{foot:angles} } The definition of the angles is identical for $D$ and CP-conjugated $\bar D$ decays.  Specifically,
also in $\bar D \to \pi^+ \pi^- \mu^+ \mu^-$ decays  $\pi^+=P_1$.
The kinematically accessible phase space  is given by
\begin{align}\label{eq:phasespace}
    &4\,m_\ell^2\,< q^2\leq \left(m_D - m_{P_1}-m_{P_2}\right)^2~,\quad \left(m_{P_1}+m_{P_2}\right)^2\,< p^2\leq \left(m_D - \sqrt{q^2}\right)^2~,\nonumber\\ &-1<\cos \theta_{P_1}\leq 1~, \quad -1<\cos \theta_\ell\leq 1~,\quad 0<\phi\leq2 \pi~.
\end{align}
The functions $c_i(\theta_\ell,\phi)$ encode the dependence on the angles $\theta_\ell$ and $\phi$
\begin{align}
    c_1 & =1\,, \quad c_2=\cos 2\,\theta_\ell\,, \quad c_3=\sin^2\theta_\ell\,\cos 2\phi\,, \quad c_4=\sin 2\,\theta_\ell\, \cos \phi\,, \quad c_5=\sin\theta_\ell\,\cos\phi\,, \nonumber \\ c_6& =\cos\theta_\ell\,, \quad c_7=\sin\theta_\ell\,\sin\phi\,, \quad c_8=\sin 2\,\theta_\ell\,\sin\phi\,, \quad c_9=\sin^2\theta_\ell\,\sin2\,\phi \, ,
    \label{eq:ci}
\end{align}
while the angular coefficients $I_i \equiv I_i(q^2,p^2,\cos \theta_{P_1})$ are functions of $q^2$, $p^2$, and $\theta_{P_1}$. The angular coefficients can be expressed in terms of transversity amplitudes, $H^{L,R}_x, x=\perp, \parallel,0$ \cite{DeBoer:2018pdx} as
\begin{align}  \nonumber
I_1 & = \phantom{-}\frac{1}{16} \bigg[ |\azeL|^2 +(L\to R) + \frac{3}{2}\sin^2 \theta_{P_1} \left\lbrace |\apeL|^2 + |\apaL|^2 + (L\to R) \right\rbrace \bigg]\,,
\\  \nonumber
  I_2 & = -\frac{1}{16} \bigg[ |\azeL|^2 + (L\to R)  -\frac{1}{2} \sin^2 \theta_{P_1} \left\lbrace |\apeL|^2+ |\apaL|^2 + (L\to R) \right\rbrace\bigg]\,,
\\  \nonumber
  I_3 &=  \phantom{-}\frac{1}{16}   \bigg[ |\apeL|^2 - |\apaL|^2  + (L\to R)\bigg]  \sin^2\theta_{P_1}\,,
\\  \nonumber
  I_4 & = -\frac{1}{8}  \bigg[\re\left[\azeL^{}\apaL^*\right] + (L\to R)\bigg] \sin\theta_{P_1}\,,
\\  \label{eq:IH}
  I_5 & = -\frac{1}{4}  \bigg[\re\left[\azeL^{}\apeL^*\right] - (L\to R)\bigg] \sin\theta_{P_1}\,,
\\  \nonumber
  I_6 & = \phantom{-}\frac{1}{4}   \bigg[\re \left[\apaL^{}\apeL^*\right] - (L\to R)\bigg] \sin^2\theta_{P_1}\,,
\\  \nonumber
  I_7 &=- \frac{1}{4} \bigg[\im \left[\azeL^{}\apaL^*\right] - (L\to R)\bigg] \sin\theta_{P_1}\,,
\\  \nonumber
  I_8 & =-\frac{1}{8}  \bigg[\im\left[\azeL^{}\apeL^*\right] + (L\to R)\bigg] \sin\theta_{P_1}\,,
\\  \nonumber
  I_9 & =\phantom{-}\frac{1}{8}   \bigg[\im \left[\apaL^{*} \apeL\right] + (L\to R)\bigg] \sin^2\theta_{P_1}\,,
\end{align}
where tensor and (pseudo)-scalar operators are not included, and the lepton mass is neglected.
The subscripts $0$, $\parallel$, and $\perp$ represent longitudinal, parallel, and perpendicular polarizations, respectively. Additionally, $L$ and $R$ denote the chirality of the lepton current. By examining the relative signs between the left-handed and right-handed contributions in Eq.~\eqref{eq:IH}, it follows  that finite $I_{5,6,7}$  require non-vanishing axial-vector contributions. These are induced by $\mathcal{O}_{10}^{(\prime)}$, or at higher order from electromagnetic loops, and are very small, hence
$I_{5,6,7}$  serve as clean null tests of the SM~\cite{DeBoer:2018pdx}. The transversity amplitudes can be written, for the non-resonant part to
lowest order in $1/m_c$~\cite{Das:2014sra}, as 
\begin{align}\label{eq:Hamp}
\begin{split}
    H_i^{L/R} &\,=\, \mathcal{C}_-^{L/R}(q^2)\, \mathcal{F}_i (q^2,p^2,\theta_{P_1})\,+\, \sum_{J=S,\,P,\,D,\,...}\mathcal{C}_9^{\,\mathcal{R},\,J}(q^2)\, \mathcal{F}_{i,\,J}^{\,\text{res}}(q^2,p^2,\theta_{P_1})~,\quad i=0,\,\parallel\,, \\ 
    H_\perp^{L/R} &\,=\, \mathcal{C}_+^{L/R}(q^2)\, \mathcal{F}_\perp (q^2,p^2,\theta_{P_1})\,+\, \sum_{J=S,\,P,\,D,\,...}\mathcal{C}_9^{\,\mathcal{R},\,J}(q^2)\, \mathcal{F}_{\perp,\,J}^{\,\text{res}}(q^2,p^2,\theta_{P_1})~,
\end{split}
\end{align}
where 
\begin{align}
\begin{split}
    \mathcal{C}_\pm^{L}(q^2)\,&=\,\mathcal{C}_9^{\,\text{eff}}(q^2)\,+\,\mathcal{C}_9\,\pm\,\mathcal{C}_9^{\,\prime}\,-\,\left(\mathcal{C}_{10}\,\pm\,\mathcal{C}_{10}^{\,\prime}\right)\,+\,\kappa\, \frac{2 m_c m_D}{q^2}\left(\mathcal{C}_7 \pm \mathcal{C}_7^\prime\right)~,\\
    \mathcal{C}_\pm^{R}(q^2)\,&=\,\mathcal{C}_9^{\,\text{eff}}(q^2)\,+\,\mathcal{C}_9\,\pm\,\mathcal{C}_9^{\,\prime}\,+\,\mathcal{C}_{10}\,\pm\,\mathcal{C}_{10}^{\,\prime}\,+\,\kappa\, \frac{2 m_c m_D}{q^2}\left(\mathcal{C}_7 \pm \mathcal{C}_7^\prime\right)~,    
\end{split}
\end{align}
with $\kappa = 1 - 2 \alpha_s /(3\pi) \ln (\mu/m_c)$~\cite{DeBoer:2018pdx},  and 
\begin{align}\label{eq:BWwc}
\mathcal{C}_9^{\,\mathcal{R},\,J }(q^2)\,=\,\frac{a_{J,\,\rho} \,\text{e}^{i\,\delta_{J,\,\rho}} }{q^2-m_\rho^2+i\,m_\rho\,\Gamma_\rho}+\frac{a_{J,\,\omega} \,\text{e}^{i\,\delta_{J,\,\omega}}}{q^2-m_\omega^2+i\,m_\omega\,\Gamma_\omega}+\frac{a_{J,\,\phi} \,\text{e}^{i\,\delta_{J,\phi}}}{q^2-m_\phi^2+i\,m_\phi\,\Gamma_\phi}  \, .
\end{align}
This  phenomenological ansatz  is identical to   \eqref{eq:BWwcLambdaC} with resonance  parameters  $a_{J,\mathcal{R}}$, $\delta_{J,\mathcal{R}}$   taking on different values for different decay modes.
The additional index $J$ for $D^0 \to \pi^+ \pi^- \mu^+ \mu^-$ decays is needed to take into account resonances in $p^2$ with spin $J=0,1,..$ that decay to  $\pi^+ \pi^-$.
The parameters $a_{J,\mathcal{R}}$ and phases $\delta_{J,\mathcal{R}}$ are extracted in Sec.~\ref{subsec:resonancefit} from a fit to  data on the $q^2$ and $p^2$ differential distributions of $D^0 \to \pi^+ \pi^- \mu^+ \mu^-$ from the LHCb collaboration~\cite{LHCb:2021yxk}.
The resonance coefficient $\mathcal{C}_9^{\,\mathcal{R} ,\, J}$ does not depend on the  transversity,
an assumption which can be investigated further  using data on hadronic $D$-decays with polarization information. 
The functions $\mathcal{F}_i$ and $\mathcal{F}_{i,J}^{\,\text{res}}$ with $i=0, \,\parallel,\, \perp$  represent the transversity form factors  for non-resonant and resonant contributions, respectively, with details provided  in Appendix~\ref{app:FFs}.
Requisite $D \to \pi \pi$  transition form factors  are  available only from heavy-hadron chiral perturbation theory \cite{Lee:1992ih}, subject to large uncertainties.
Improved computations from other non-perturbative means would be desirable.
Explicit expressions of  the $I_{1-9}$ using   the transversity amplitudes  in Eq.~\eqref{eq:Hamp}   are given in App.~\ref{app:full}.

Due to the smallness of the short-distance contributions in $c \to u \ell^+ \ell^-$ in the SM,  as reviewed in  Sec.~\ref{sec:EFT}, the decay rate and angular observables of  $D\to P_1\,P_2\,\ell^+\,\ell^-$ decays are  primarily induced by the resonances, that is $\mathcal{C}_9^{\,\mathcal{R}\,,J}(q^2)$,  as
\begin{align}
    I_1^{\,\text{SM}}&\simeq
+\frac{1}{8} \left|\mathcal{C}_9^{\,\mathcal{R},\,S}\right|^2 \left|\mathcal{F}_{0,\,S}^{\,\text{res}}\right|^2
+\left| \mathcal{C}_9^{\,\mathcal{R},\,P}\right|^2 \left(\frac{1}{8}\left|\mathcal{F}_{0,\,P}^{\,\text{res}}\right|^2
+\frac{3}{16} \left(\left|\mathcal{F}_{\parallel,\,P}^{\,\text{res}}\right|^2+\left|\mathcal{F}_{\perp,\,P}^{\,\text{res}}\right|^2\right)
\sin^2\theta_{P_1}\right)\nonumber
\\  &\quad 
+\frac{1}{4}{\rm Re} \left\{\mathcal{C}_9^{\mathcal{R},P} \left(\mathcal{C}_9^{\mathcal{R},S}\right)^\ast\right\}\, {\rm Re}[\mathcal{F}_{0,\,S}^{\,\text{res}} \mathcal{F}_{0,\,P}^{\,\text{res},\ast}]
+\frac{1}{4}{\rm Im} \left\{\mathcal{C}_9^{\mathcal{R},P} \left(\mathcal{C}_9^{\mathcal{R},S}\right)^\ast\right\}\, {\rm Im}[\mathcal{F}_{0,\,S}^{\,\text{res}} \mathcal{F}_{0,\,P}^{\,\text{res},\ast}]\,,\nonumber\\    
    I_2^{\,\text{SM}}&\simeq
-\frac{1}{8} \left|\mathcal{C}_9^{\,\mathcal{R},\,S}\right|^2 \left|\mathcal{F}_{0,\,S}^{\,\text{res}}\right|^2
+\left| \mathcal{C}_9^{\,\mathcal{R},\,P}\right|^2 \left(
    -\frac{1}{8} \left|\mathcal{F}_{0,\,P}^{\,\text{res}}\right|^2
+\frac{1}{16}
\left(\left|\mathcal{F}_{\parallel,\,P}^{\,\text{res}}\right|^2+\left|\mathcal{F}_{\perp,\,P}^{\,\text{res}}\right|^2\right) \sin^2\theta_{P_1}\right)
\nonumber\\  &\quad \nonumber
-\frac{1}{4} {\rm Re} \left\{\mathcal{C}_9^{\mathcal{R},P} \left(\mathcal{C}_9^{\mathcal{R},S}\right)^\ast\right\}\, {\rm Re}[\mathcal{F}_{0,\,S}^{\,\text{res}} \mathcal{F}_{0,\,P}^{\,\text{res},\ast}]
-\frac{1}{4} {\rm Im} \left\{\mathcal{C}_9^{\mathcal{R},P} \left(\mathcal{C}_9^{\mathcal{R},S}\right)^\ast\right\}\, {\rm Im}[\mathcal{F}_{0,\,S}^{\,\text{res}} \mathcal{F}_{0,\,P}^{\,\text{res},\ast}]\,,\\
    I_3^{\,\text{SM}}&\simeq  
-\frac{1}{8} \left| \mathcal{C}_9^{\,\mathcal{R},\,P}\right|^2 \left(\left|\mathcal{F}_{\parallel,\,P}^{\,\text{res}}\right|^2-\left|\mathcal{F}_{\perp,\,P}^{\,\text{res}}\right|^2\right) \, \sin^2\theta_{P_1} 
    \,,\nonumber 
    \\
        I_4^{\,\text{SM}}&\simeq
\biggr[-\frac{1}{4} \left| \mathcal{C}_9^{\,\mathcal{R},\,P}\right|^2 {\rm Re}[\mathcal{F}_{0,\,P}^{\,\text{res}} \mathcal{F}_{\parallel,\,P}^{\,\text{res},\ast}]
\nonumber\\ &\quad \nonumber
+\frac{1}{4} {\rm Im} \left\{\mathcal{C}_9^{\mathcal{R},P} \left(\mathcal{C}_9^{\mathcal{R},S}\right)^\ast\right\} \,
{\rm Im}[\mathcal{F}_{\parallel,\,P}^{\,\text{res}} \mathcal{F}_{0,\,S}^{\,\text{res},\ast}] 
-\frac{1}{4} {\rm Re} \left\{\mathcal{C}_9^{\mathcal{R},P} \left(\mathcal{C}_9^{\mathcal{R},S}\right)^\ast\right\} \, {\rm Re}[\mathcal{F}_{0,\,S}^{\,\text{res}} \mathcal{F}_{\parallel,\,P}^{\,\text{res},\ast}] 
\nonumber 
\biggr]\sin\theta_{P_1} 
    \,,\\
I_8^{\,\text{SM}}&\simeq
\biggl[+\frac{1}{4} \left| \mathcal{C}_9^{\,\mathcal{R},\,P}\right|^2 \cancelto{\approx 0}{{\rm Im}[\mathcal{F}_{\perp,\,P}^{\,\text{res}} \mathcal{F}_{0,\,P}^{\,\text{res},\ast}]}
\nonumber\\ &\quad \nonumber
+\frac{1}{4} {\rm Re} \left\{\mathcal{C}_9^{\mathcal{R},P} \left(\mathcal{C}_9^{\mathcal{R},S}\right)^\ast\right\} {\rm Im}[\mathcal{F}_{\perp,\,P}^{\,\text{res}} \mathcal{F}_{0,\,S}^{\,\text{res},\ast}]
+\frac{1}{4} {\rm Im} \left\{\mathcal{C}_9^{\mathcal{R},P} \left(\mathcal{C}_9^{\mathcal{R},S}\right)^\ast\right\} {\rm Re}[\mathcal{F}_{0,\,S}^{\,\text{res}}
\mathcal{F}_{\perp,\,P}^{\,\text{res},\ast}]
\biggr] \sin\theta_{P_1} 
    \,,\\ 
I_9^{\,\text{SM}}&\simeq 
-\frac{1}{4} \left| \mathcal{C}_9^{\,\mathcal{R},\,P}\right|^2 \cancelto{\approx 0}{{\rm Im}[\mathcal{F}_{\parallel,\,P}^{\,\text{res}} \mathcal{F}_{\perp,\,P}^{\,\text{res},\ast}]}
 \sin^2\theta_{P_1} 
    \,.   \label{eq:IiSM}
\end{align}
Therefore,  while  $I_{5,6,7}^{\text{SM}}=0$ due to the GIM-mechanism,  $I_{1,2,3,4,8,9}^{\text{SM}}$ are finite. 
As indicated by the crossed out terms above in Eq.~(\ref{eq:IiSM}),  the product of P-wave form factors 
$\mathcal{F}_{i,\,P}^{\,\text{res}} \,\mathcal{F}_{j,\,P}^{\,\text{res},\,\ast}$  is real assuming universality between different  polarizations $i,j$.
In this case $I_9^{\text{SM}}$ also vanishes.
$I_8^{\text{SM}}$ receives contributions from  S-P-interference. Its associated angular observable $\langle I_8^{\text{SM}} |_{\text{S-P}}\rangle$, defined  in Sec.~\ref{subsec:observables}, vanishes after integration over 
$\cos\theta_{P_1}$, schematically $\left(\int_0^1 -\int_{-1}^0\right) I_8^{\text{SM}} |_{\text{S-P}}  \,\mathrm{d}\!\cos\theta_{P_1}=0$ \cite{Fajfer:2023tkp}.
On the other hand, the pure P-wave contribution to $I_8^{\text{SM}}$ would be finite for non-universal form factors,
and does {\it not} vanish after $\cos\theta_{P_1}$-integration. 

The expressions (\ref{eq:IiSM})  allow  to probe the resonance structure  and long-distance dynamics  underlying $D\to P_1\,P_2\,\ell^+\,\ell^-$ decays. 
For instance, $I_3^{\,\text{SM}}$ exclusively captures P-wave contributions, while $I_1^{\,\text{SM}}+I_2^{\,\text{SM}}$ eliminates S-wave contributions. The difference $I_1^{\,\text{SM}}-I_2^{\,\text{SM}}$ is sensitive to these S-wave contributions, retaining a P-wave contribution and S-P-interference terms. The latter are also present in $I_4^{\,\text{SM}}$. 
Details on how the angular observables $I_i$ can be extracted from the 5-differential distribution~\eqref{eq:full} are discussed in Ref.~\cite{DeBoer:2018pdx}.
The distributions including  D-waves have been analyzed in \cite{Das:2014sra}.

\subsubsection{CP-symmetries and asymmetries}\label{subsec:observables}

We present the angular observables related to the angular coefficients $I_i$ given in  Eqs.~\eqref{eq:I1}-\eqref{eq:I9}. We use the same conventions as LHCb Ref.~\cite{LHCb:2021yxk}, that is
\begin{align}\label{eq:I_LHCb}
    \langle S_i\rangle_{\QRanges}\,&=\,\frac{1}{2}\left[\left\langle I_i\right\rangle_{\QRanges}+(-)\left\langle \bar{I}_i\right\rangle_{\QRanges}\right]~,\nonumber\\
    \langle A_i\rangle_{\QRanges}\,&=\,\frac{1}{2}\left[\left\langle I_i\right\rangle_{\QRanges}-(+)\left\langle \bar{I}_i\right\rangle_{\QRanges}\right]~,    
\end{align}
for the CP-even (CP-odd) coefficients $\left\langle I_{2,3,4,7}\right\rangle \left(\left\langle I_{5,6,8,9}\right\rangle\right)$, with
\begin{align} \label{eq:Ij_of_q2}
    \left\langle I_{2,3,6,9}\right\rangle(q^2)\,&=\,\frac{1}{\Gamma_{\QRanges}}\,\int_{(m_{P_1}+m_{P_2})^2}^{(m_D-\sqrt{q^2})^2}\,\text{d}p^2\,\int^1_{-1} \text{d}\cos\theta_{P_1} \, I_{2,3,6,9}\,,\\
        \left\langle
 I_{4,5,7,8}\right\rangle(q^2)\,&=\,\frac{1}{\Gamma_{\QRanges}}\,\int_{(m_{P_1}+m_{P_2})^2}^{(m_D-\sqrt{q^2})^2}\,\text{d}p^2\,\left[ \int^1_0 \text{d}\cos\theta_{P_1} - \int_{-1}^0 \text{d}\cos\theta_{P_1} \right]\, I_{4,5,7,8}\,, \nonumber
\end{align}
for  $q^2\in \QRanges$ 
and the binned observables 
\begin{align}\label{eq:Ij_binned}
\left\langle I_{2-9}\right\rangle_{\QRanges}\,&=\,\int^{q^2_{\text{max}}}_{q^2_{\text{min}}}\,\text{d}q^2\,\left\langle I_{2-9}\right\rangle(q^2)\,.
\end{align}
The $q^2$-binned decay rate  $\Gamma_{\QRanges}$ is obtained after  phase space integration, see Eqs.~\eqref{eq:phasespace}, as
\begin{align}\label{eq:Gamma_binned}
\Gamma_{\QRanges}\,=\,\int^{q^2_{\text{max}}}_{q^2_{\text{min}}}\,\text{d}q^2\,\int_{(m_{P_1}+m_{P_2})^2}^{(m_D-\sqrt{q^2})^2}\,\text{d}p^2\,\int^1_{-1} \text{d}\cos\theta_{P_1} \, 2\left(I_1 - \frac{I_2}{3}\right) 
~.
\end{align}

In our analysis we compute  the decay rate $\Gamma_{\QRanges}$, unless stated otherwise, including NP effects from the Wilson coefficients.  While viable NP-effects  in the decay rate are too small to be
noticeable this enables to take into account correlations in the fit  in ratio-type observables such as $S_i,A_i$ in a consistent manner.
The  $\left\langle \bar{I}_i\right\rangle$ in Eq.~\eqref{eq:I_LHCb} are obtained by conjugating the NP weak phases. In addition also  the SM phases in  the CKM factors $V_{cd}^\ast V_{ud}^{\phantom{\ast}}$ and $V_{cs}^\ast V_{us}^{\phantom{\ast}}$  in the resonance contributions 
should be conjugated, however, CP-violation  is here very small and this has a negligible effect.
Therefore, we   complex conjugate the NP weak phases, that is, conjugating $\mathcal{C}_{7, 9,10}^{\,(\prime)}$
\begin{align}\label{eq:I_CPconj}
    \left\langle \bar{I}_i\right\rangle\equiv + (-)\left\langle I_i(\mathcal{C}_{7, 9,10}^{\,(\prime) *} )\right\rangle~,
\end{align}
for CP-even (CP-odd) angular coefficients.
In the CP-limit holds $\langle S_i\rangle=\langle I_i\rangle$ and $\langle A_i\rangle=0$.

\begin{figure}
    \centering
    \includegraphics[width=\textwidth]{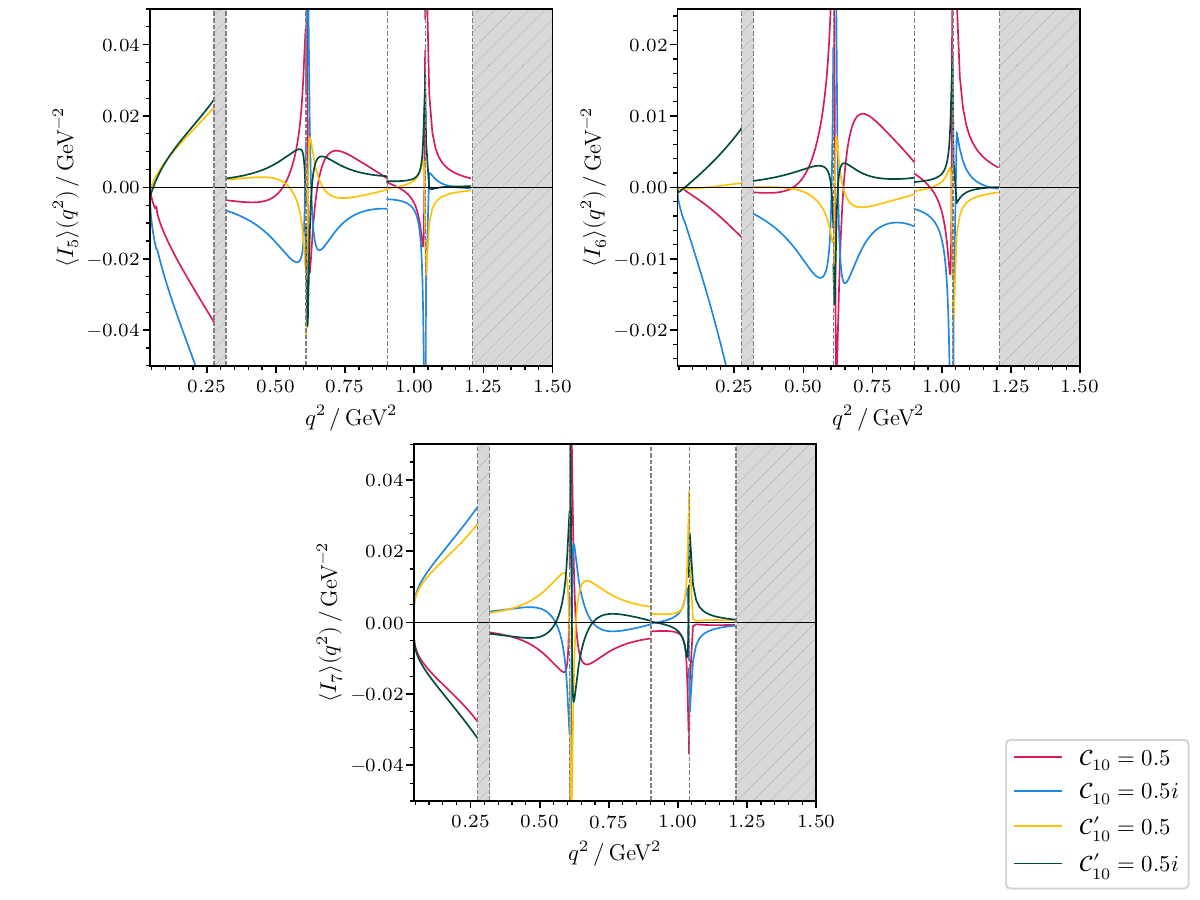}
    \caption[Caption]{$\langle I_{5,6,7}\rangle(q^2)$ with normalization $\Gamma_{\QRanges}$ depending on the $q^2$-region and corresponding to the bins, indicated by the gray dashed vertical lines, used by LHCb \cite{LHCb:2021yxk}. 
    NP signals are illustrated with 
   $ \mathcal{C}_{10}=0.5$ (red),  $\mathcal{C}_{10}^{\prime}=0.5$ (yellow), $\mathcal{C}_{10}=0.5i$ (blue)  and $\mathcal{C}_{10}^{\prime}=0.5i$ (green). 
       No data are reported by the experiment in the gray shaded areas.  } 
    \label{fig:nulltests_binned}
\end{figure}

In Fig.~\ref{fig:nulltests_binned}, we illustrate the impact of  NP in the null test distributions $\langle I_{5,6,7}\rangle(q^2)$, 
\footnote{Private communication revealed  an error in Fig.~3 of \cite{DeBoer:2018pdx}. The square of $m_{P_1}^2$ below Eq.~(44) was missing in the numerics of the plots of  $I_{5,6,7}$ in \cite{DeBoer:2018pdx}. 
Once this is corrected, the curves change, but remain at  the same order of magnitude. We are grateful to Stefan de Boer for correspondence and confirmation.}
as defined in Eqs.~\eqref{eq:Ij_of_q2},  using the  central values of resonance parameters of scenario 4 given in Tab.~\ref{tab:SMfit} . We employ  the same binning as in  the LHCb analysis \cite{LHCb:2021yxk}. We display different benchmarks 
$\mathcal{C}_{10}=0.5$ (red),  $\mathcal{C}_{10}^{\prime}=0.5$ (yellow), $\mathcal{C}_{10}=0.5i$ (blue)  and $\mathcal{C}_{10}^{\prime}=0.5i$ (green).  No data are reported in \cite{LHCb:2021yxk}  around the $\eta$ resonance with $q^2 = m_\eta^2 \sim 0.3\,\mathrm{GeV}^2$  and the  high-$q^2$ endpoint region.
The vertical dashed lines denote the boundaries of the bins. The interference between SM resonances and NP contributions gives the largest effects to the
   null tests  $\langle I_{5,6,7}\rangle$    near $q^2 \simeq m_\rho^2$ or $m_\phi^2$.

\section[\texorpdfstring{Phenomenological analysis of $\boldsymbol{D^0\to\pi^+\pi^-\mu^+\mu^-}$ observables}{Analysis of D0 -> pi+pi-mu+mu-} ]{Analysis of $\boldsymbol{D^0\to\pi^+\pi^-\mu^+\mu^-}$ decays}
\label{sec:DPPfit}

We present a detailed phenomenological analysis of  $D^0\to\pi^+\pi^-\mu^+\mu^-$ observables.
In the appendix in Tab.~\ref{tab:4bodydata} we give the available measurements in the categories ``SM measurements", with subset ``good" bins, which are dominated by SM contributions,
and NP measurements, with observables that  are negligible in the SM.
We use the former to determine resonance parameters  in the ansatz~\eqref{eq:BWwc}  in Sec.~\ref{subsec:resonancefit}, and the latter to work out constraints on  NP from $D^0\to\pi^+\pi^-\mu^+\mu^-$ observables  in Sec.~\ref{subsec:fitsNPDPPmumu}. 
In Sec.~\ref{sec:beyondSP} we discuss contributions beyond the ansatz~\eqref{eq:BWwc} with S-and P-wave $\pi \pi$-resonances.

\subsection{Fitting resonance parameters}
\label{subsec:resonancefit}

To describe the resonant SM contribution, we introduce six magnitude factors $a_{P(S),R}$ and six phases $\delta_{P(S),R}$ in  $\mathcal{C}_9^{\mathcal{R},P(S)}$ (
see Eq.~\eqref{eq:BWwc}), along with a relative magnitude factor $a_{p^2,\omega}$ and phase $\delta_{p^2,\omega}$ for the P-wave lineshape (see Eq.~\eqref{eq:PwaveLineshape}). 
With available data~\cite{LHCb:2017uns,LHCb:2021yxk} there is no sensitivity to the  relative phase between S- and P-wave contributions, and we set this phase to zero.
 Observables sensitive to this phase have been discussed in Ref.~\cite{Fajfer:2023tkp}. 
In branching ratios, only relative phases between resonances matter. A relative phase between resonant SM and NP only appears via interference terms and cannot be determined through a fit without significant NP in the data. Therefore, in the SM fit, we set $\delta_{S,\phi}=\delta_{P,\phi}=\pi$ and vary the SM-NP strong phase for fits that include NP Wilson coefficients.\footnote{ \label{foot:null}The undetermined strong phase  affects upper limits, because a single observable $\langle S(A)_j \rangle_{[q^2_j,q^2_k]}$ can vanish even in the presence of large NPs by tuning  the overall resonance strong phase. Therefore, multiple measurements with different dependence on this  phase are required to improve the NP reach.}

We employ the  SM measurements  listed in Tab.~\ref{tab:4bodydata},  including branching ratios and the angular symmetries $\langle S_{2,3,4}\rangle$.
Available data  include the differential distributions $\mathrm{d}\Gamma/\mathrm{d}\sqrt{q^2}$ and $\mathrm{d}\Gamma/\mathrm{d}\sqrt{p^2}$ \cite{LHCb:2021yxk} which are provided, however, without systematic uncertainties, accounting for unfolded detector effects and bin-to-bin   correlations. Given the complexity of disentangling  resonances from short-distance NP effects we suggest making this type of   information  available in the future,
in addition to the double differential distribution.

As observed in Ref.~\cite{Fajfer:2023tkp}, using a similar ansatz as Eq.~\eqref{eq:Hamp},  the fit to the $D \to\pi^+\pi^-\mu^+\mu^-$  data~\cite{LHCb:2017uns, LHCb:2021yxk}  is not satisfactory. This complexity, already noted in~\cite{Fajfer:2023tkp}, may arise from missing theoretical descriptions that, if included, could potentially explain the kinematic regions of the data that are not well understood.
This concerns  in particular the high $\pi \pi$ and  the low $\mu \mu$ mass bins, which constitute  hence  rather ``bad''  bins.
Following a similar approach as Ref.~\cite{Fajfer:2023tkp}, we define ``good'' bins as those with $\sqrt{q^2} > 0.6\,\mathrm{GeV}$ for $\mathrm{d}\Gamma/\mathrm{d}\sqrt{q^2}$, $\sqrt{p^2} < 0.9\,\mathrm{GeV}$ for $\mathrm{d}\Gamma/\mathrm{d}\sqrt{p^2}$, and $\sqrt{q^2} \geq 0.565\,\mathrm{GeV}$ for $\mathcal{B}$ and $\langle S_{2,3,4} \rangle$. 
Kinematic cuts in both $q^2$ and $p^2$, without having access to the double differential and thus correlations, introduces potentially inconsistencies. We  estimate this effect to be small,  by comparing the outcome of  fits
with all SM measurements and just ``good" bins, such as scenario 1 and 2 defined below.

We perform fits in scenarios with different resonance models and data sets.
Scenario 1,2 and 5 exclude the $\omega$ in the P-wave as in~\cite{Fajfer:2023tkp}.
Furthermore, we consider  two exploratory scenarios,  9 and 10,  where we allow for effective contributions to the matrix element of  $\mathcal{O}_9$ and $\mathcal{O}_{10}$, denoted by $r_9$ and  $r_{10}$, respectively.  $r_9$ can be  interpreted as a constant SM contribution from the strong interaction   to  $\mathcal{C}_9$, 
beyond the phenomenological ansatz for $\mathcal{C}_9^R$. 
We also consider effective contributions  $r_{10}$ to the axial-vector current  $\mathcal{C}_{10}$, to check if this is an artifact of the fit.
We analyze the following scenarios:

\begin{enumerate}
    \item Including all SM measurements and excluding $\omega\to\ell^+\ell^-$ in P-wave with $a_{P,\omega}=\delta_{P,\omega}=0$.
    \item Including only the ``good'' bins of all SM measurements and excluding $\omega\to\ell^+\ell^-$ in P-wave with $a_{P,\omega}=\delta_{P,\omega}=0$ (same scenario as in Ref.~\cite{Fajfer:2023tkp}).
    \item Including all SM measurements and all parameters.
    \item Including only the ``good'' bins of all SM measurements and all parameters.
    \item Including only $\langle S_{2,3,4} \rangle$ \& $\mathcal{B}$ measurements and excluding $a_{P,\omega}$, $\delta_{P,\omega}$ and $a_{p^2,\omega}$, $\delta_{p^2,\omega}$. 
     \item Including only the ``good'' bins of $\langle S_{2,3,4} \rangle$ \& $\mathcal{B}$ measurements, excluding $a_{p^2,\omega}$, $\delta_{p^2,\omega}$ and imposing the isospin relations $a_{P(S),\omega}/a_{P(S),\rho} = 1/3$, $\delta_{P(S),\omega}=\delta_{P(S),\rho}+\pi$.
    \item Including only $\langle S_{2,3,4} \rangle$ \& $\mathcal{B}$ measurements and excluding $a_{p^2,\omega}$ and $\delta_{p^2,\omega}$.
    \item Including only the ``good'' bins of $\langle S_{2,3,4} \rangle$ \& $\mathcal{B}$ measurements and excluding $a_{p^2,\omega}$, $\delta_{p^2,\omega}$.
    \item Including all SM measurements  and all parameters together with  effective contributions $r_9, r_{10}$  to the matrix element of  $\mathcal{O}_9$, $\mathcal{O}_{10}$.
      \item Including the ``good'' bins  and all parameters together with  $r_9, r_{10}$.
      \end{enumerate}

\begin{figure}[h!]
    \centering
    \begin{subfigure}[b]{0.49\textwidth}
        \centering
        \includegraphics[width=\textwidth]{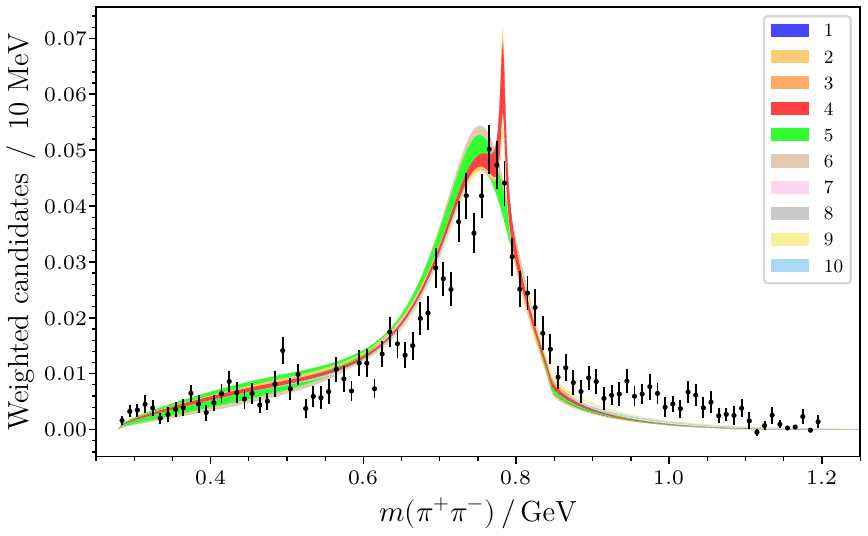}
    \end{subfigure}
    \begin{subfigure}[b]{0.49\textwidth}
        \centering
        \includegraphics[width=\textwidth]{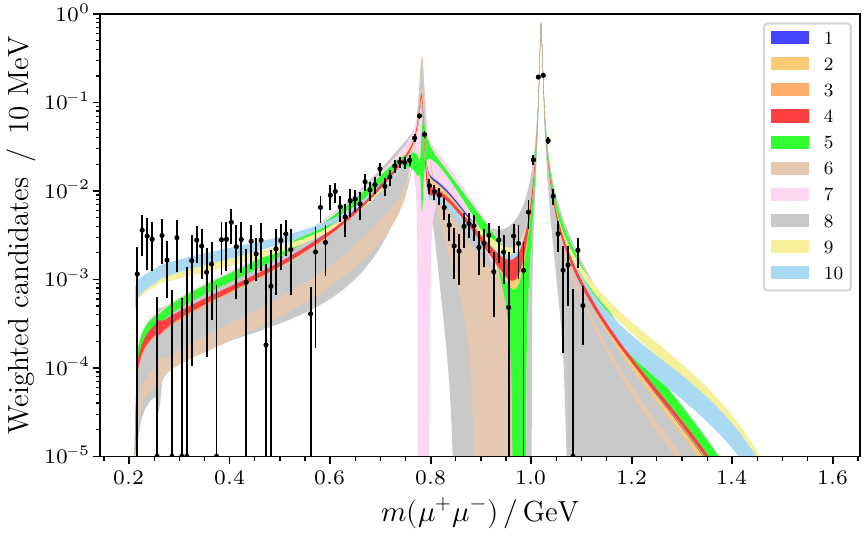}
    \end{subfigure}
    \caption{ 
    Differential branching fractions for $D \to \pi^+ \pi^- \mu^+ \mu^-$ decays in invariant masses $m(\pi^+\pi^-)=\sqrt{p^2}$ (left panel) and  $m(\mu^+\mu^-)=\sqrt{q^2}$ (right panel) for
      the  scenarios detailed in Sec.~\ref{subsec:resonancefit}.  Data points (black) are from  LHCb \cite{LHCb:2021yxk}.}
    \label{fig:SMfit_dGamma_dx_compare}
\end{figure}

We perform  fits of the resonance parameters for all  scenarios using the methodology described in Ref.~\cite{Bause:2022rrs}. Tab.~\ref{tab:SMfit} shows the resulting best-fit values and their  $1\sigma$ uncertainties, together with the reduced chi-square $\chi^2/\text{dof}$, the number of observables and the number of fit  parameters. 
The differential $m(\pi^+\pi^-)=\sqrt{p^2}$ and $m(\mu^+\mu^-)=\sqrt{q^2}$ spectra in the scenarios after the fit  together with data (black points)  are shown in 
Fig.~\ref{fig:SMfit_dGamma_dx_compare}. Our results are consistent with Refs.~\cite{DeBoer:2018pdx,Fajfer:2023tkp}.

\begin{table}[h!]
   \centering
   \begin{adjustbox}{max width=\textwidth}
       \begin{tabular}{c|c|c|c|c|c|c|c|c|c|c}
           \hline
           \hline
           \rowcolor{LightBlue} fit scenario &1&2&3&4&5&6&7&8&9&10\\
           \hline
           \hline
           $a_{P,\rho}/ \text{GeV}^2$ & \SI[parse-numbers=false]{0.270^{+0.019}_{-0.019}}{} & \SI[parse-numbers=false]{0.261^{+0.019}_{-0.019}}{} & \SI[parse-numbers=false]{0.322^{+0.028}_{-0.028}}{} & \SI[parse-numbers=false]{0.289^{+0.029}_{-0.030}}{} & \SI[parse-numbers=false]{0.29^{+0.04}_{-0.05}}{} & \SI[parse-numbers=false]{0.213^{+0.024}_{-0.028}}{} & \SI[parse-numbers=false]{0.32^{+0.05}_{-0.12}}{} & \SI[parse-numbers=false]{0.21^{+0.16}_{-0.26}}{} & \SI[parse-numbers=false]{0.285^{+0.024}_{-0.024}}{} & \SI[parse-numbers=false]{0.264^{+0.024}_{-0.024}}{}\\
           \hline
           $a_{P,\omega}/a_{P,\rho}$ & \SI[parse-numbers=false]{0}{} & \SI[parse-numbers=false]{0}{} & \SI[parse-numbers=false]{0.190^{+0.010}_{-0.011}}{} & \SI[parse-numbers=false]{0.199^{+0.014}_{-0.019}}{} & \SI[parse-numbers=false]{0}{} & \SI[parse-numbers=false]{1/3}{} & \SI[parse-numbers=false]{0.08^{+0.14}_{-0.10}}{} & \SI[parse-numbers=false]{0.3^{+1.7}_{-0.4}}{} & \SI[parse-numbers=false]{0.198^{+0.014}_{-0.016}}{} & \SI[parse-numbers=false]{0.208^{+0.014}_{-0.019}}{}\\
           \hline
           $a_{P,\phi}/a_{P,\rho}$ & \SI[parse-numbers=false]{0.291^{+0.012}_{-0.012}}{} & \SI[parse-numbers=false]{0.306^{+0.012}_{-0.013}}{} & \SI[parse-numbers=false]{0.217^{+0.020}_{-0.017}}{} & \SI[parse-numbers=false]{0.256^{+0.030}_{-0.024}}{} & \SI[parse-numbers=false]{0.26^{+0.04}_{-0.02}}{} & \SI[parse-numbers=false]{0.37^{+0.05}_{-0.05}}{} & \SI[parse-numbers=false]{0.24^{+0.14}_{-0.03}}{} & \SI[parse-numbers=false]{0.38^{+0.45}_{-0.19}}{} & \SI[parse-numbers=false]{0.252^{+0.018}_{-0.016}}{} & \SI[parse-numbers=false]{0.295^{+0.023}_{-0.019}}{}\\
           \hline
           $\delta_{P,\rho}$ & \SI[parse-numbers=false]{2.67^{+0.18}_{-0.16}}{} & \SI[parse-numbers=false]{2.69^{+0.22}_{-0.18}}{} & \SI[parse-numbers=false]{2.38^{+0.23}_{-0.18}}{} & \SI[parse-numbers=false]{2.34^{+0.32}_{-0.20}}{} & \SI[parse-numbers=false]{3.4^{+1.3}_{-1.5}}{} & \SI[parse-numbers=false]{3.3^{+1.7}_{-2.0}}{} & \SI[parse-numbers=false]{3.4^{+1.4}_{-1.7}}{} & \SI[parse-numbers=false]{2.9^{+4.9}_{-4.9}}{} & \SI[parse-numbers=false]{2.17^{+0.19}_{-0.19}}{} & \SI[parse-numbers=false]{2.12^{+0.22}_{-0.21}}{}\\
           \hline
           $\delta_{P,\omega}$ & \SI[parse-numbers=false]{0}{} & \SI[parse-numbers=false]{0}{} & \SI[parse-numbers=false]{5.09^{+0.26}_{-0.24}}{} & \SI[parse-numbers=false]{4.9^{+0.4}_{-0.4}}{} & \SI[parse-numbers=false]{0}{} & $\delta_{P,\rho} +\pi$ & \SI[parse-numbers=false]{5.9^{+2.0}_{-2.0}}{} & \SI[parse-numbers=false]{4.9^{+5.5}_{-5.5}}{} & \SI[parse-numbers=false]{4.81^{+0.28}_{-0.31}}{} & \SI[parse-numbers=false]{4.85^{+0.24}_{-0.24}}{}\\
           \hline
           $a_{p^2,\omega}$ & \SI[parse-numbers=false]{0.0033^{+0.0007}_{-0.0007}}{} & \SI[parse-numbers=false]{0.0036^{+0.0008}_{-0.0008}}{} & \SI[parse-numbers=false]{0.0025^{+0.0007}_{-0.0007}}{} & \SI[parse-numbers=false]{0.0030^{+0.0008}_{-0.0008}}{} & \SI[parse-numbers=false]{0}{} & \SI[parse-numbers=false]{0}{} & \SI[parse-numbers=false]{0}{} & \SI[parse-numbers=false]{0}{} & \SI[parse-numbers=false]{0.0025^{+0.0007}_{-0.0008}}{} & \SI[parse-numbers=false]{0.0030^{+0.0008}_{-0.0008}}{}\\
           \hline
           $\delta_{p^2,\omega}$ & \SI[parse-numbers=false]{4.62^{+0.27}_{-0.29}}{} & \SI[parse-numbers=false]{4.71^{+0.27}_{-0.30}}{} & \SI[parse-numbers=false]{4.6^{+0.3}_{-0.4}}{} & \SI[parse-numbers=false]{4.7^{+0.3}_{-0.3}}{} & \SI[parse-numbers=false]{0}{} & \SI[parse-numbers=false]{0}{} & \SI[parse-numbers=false]{0}{} & \SI[parse-numbers=false]{0}{} & \SI[parse-numbers=false]{4.8^{+0.3}_{-0.4}}{} & \SI[parse-numbers=false]{4.9^{+0.3}_{-0.3}}{}\\
           \hline
           $a_{S,\rho}/a_{P,\rho}$ & \SI[parse-numbers=false]{42.5^{+4.4}_{-4.3}}{} & \SI[parse-numbers=false]{43.9^{+4.9}_{-4.7}}{} & \SI[parse-numbers=false]{31.1^{+8.7}_{-9.4}}{} & \SI[parse-numbers=false]{43.2^{+10.7}_{-11.2}}{} & \SI[parse-numbers=false]{47.8^{+14.6}_{-18.3}}{} & \SI[parse-numbers=false]{43.1^{+13.8}_{-13.5}}{} & \SI[parse-numbers=false]{44.9^{+29.4}_{-17.1}}{} & \SI[parse-numbers=false]{63.1^{+293.2}_{-64.0}}{} & \SI[parse-numbers=false]{35.7^{+7.1}_{-6.8}}{} & \SI[parse-numbers=false]{51.3^{+6.5}_{-7.5}}{}\\
           \hline
           $a_{S,\omega}/a_{S,\rho}$ & \SI[parse-numbers=false]{0.271^{+0.023}_{-0.020}}{} & \SI[parse-numbers=false]{0.271^{+0.027}_{-0.022}}{} & \SI[parse-numbers=false]{0.07^{+0.07}_{-0.07}}{} & \SI[parse-numbers=false]{0.07^{+0.07}_{-0.07}}{} & \SI[parse-numbers=false]{0.09^{+0.08}_{-0.10}}{} & \SI[parse-numbers=false]{1/3}{} & \SI[parse-numbers=false]{0.07^{+0.12}_{-0.12}}{} & \SI[parse-numbers=false]{0}{} & \SI[parse-numbers=false]{0.10^{+0.07}_{-0.07}}{} & \SI[parse-numbers=false]{0.00^{+0.05}_{-0.00}}{}\\
           \hline
           $a_{S,\phi}/a_{S,\rho}$ & \SI[parse-numbers=false]{0.08^{+0.06}_{-0.08}}{} & \SI[parse-numbers=false]{0.04^{+0.08}_{-0.04}}{} & \SI[parse-numbers=false]{0.31^{+0.14}_{-0.08}}{} & \SI[parse-numbers=false]{0.20^{+0.08}_{-0.06}}{} & \SI[parse-numbers=false]{0.06^{+0.09}_{-0.06}}{} & \SI[parse-numbers=false]{0.00^{+0.26}_{-0.00}}{} & \SI[parse-numbers=false]{0.06^{+0.09}_{-0.06}}{} & \SI[parse-numbers=false]{0.09^{+0.08}_{-0.09}}{} & \SI[parse-numbers=false]{0.24^{+0.07}_{-0.06}}{} & \SI[parse-numbers=false]{0.00^{+0.14}_{-0.00}}{}\\
           \hline
           $\delta_{S,\rho}$ & \SI[parse-numbers=false]{4.14^{+0.18}_{-0.16}}{} & \SI[parse-numbers=false]{4.14^{+0.21}_{-0.19}}{} & \SI[parse-numbers=false]{3.7^{+0.9}_{-1.1}}{} & \SI[parse-numbers=false]{3.5^{+1.3}_{-1.4}}{} & \SI[parse-numbers=false]{1.6^{+3.3}_{-3.3}}{} & \SI[parse-numbers=false]{2.3^{+2.5}_{-2.5}}{} & \SI[parse-numbers=false]{1.6^{+3.3}_{-3.3}}{} & \SI[parse-numbers=false]{0.3^{+2.6}_{-2.6}}{} & \SI[parse-numbers=false]{5.8^{+0.6}_{-0.6}}{} & \SI[parse-numbers=false]{0.4^{+0.5}_{-0.5}}{}\\
           \hline
           $\delta_{S,\omega}$ & \SI[parse-numbers=false]{0}{} & \SI[parse-numbers=false]{0}{} & \SI[parse-numbers=false]{1.1^{+1.4}_{-1.4}}{} & \SI[parse-numbers=false]{1.0^{+1.7}_{-1.8}}{} & \SI[parse-numbers=false]{5.6^{+3.6}_{-3.6}}{} & $\delta_{S,\rho} +\pi$ & \SI[parse-numbers=false]{5.3^{+3.7}_{-3.7}}{} & \SI[parse-numbers=false]{0}{} & \SI[parse-numbers=false]{3.4^{+1.1}_{-1.0}}{} & \SI[parse-numbers=false]{0}{}\\
           \hline
           $\textrm{Re}\,r_{9}$ & - & - & - & - & - & - & - & - & \SI[parse-numbers=false]{2.9^{+0.4}_{-0.4}}{} & \SI[parse-numbers=false]{2.2^{+0.8}_{-1.0}}{}\\
           \hline
           $\textrm{Im}\,r_{9}$ & - & - & - & - & - & - & - & - & \SI[parse-numbers=false]{0.5^{+1.1}_{-1.0}}{} & \SI[parse-numbers=false]{2.2^{+0.9}_{-1.2}}{}\\
           \hline
           $\textrm{Re}\,r_{10}$ & - & - & - & - & - & - & - & - & \SI[parse-numbers=false]{0.0^{+0.8}_{-0.8}}{} & \SI[parse-numbers=false]{-0.0^{+0.9}_{-0.9}}{}\\
           \hline
           $\textrm{Im}\,r_{10}$ & - & - & - & - & - & - & - & - & \SI[parse-numbers=false]{0.0^{+0.8}_{-0.8}}{} & \SI[parse-numbers=false]{-0.0^{+0.9}_{-0.9}}{}\\
           \hline
           $\chi^2$ & 581.17 & 288.13 & 516.54 & 250.15 & 66.42 & 60.93 & 65.89 & 49.72 & 469.20 & 231.52\\
           \hline
           \# obs. & 193 & 124 & 193 & 124 & 18 & 17 & 18 & 14 & 193 & 124\\
           \hline
           \# ang. obs. & 3 & 3 & 3 & 3 & 3 & 3 & 3 & 3 & 3 & 3\\
           \hline
           \# ang. bins & 5 & 4 & 5 & 4 & 5 & 5 & 5 & 4 & 5 & 4\\
           \hline
           \# $\mathrm{d}\Gamma/\mathrm{d}\sqrt{q^2}$ bins & 85 & 49 & 85 & 49 & 0 & 0 & 0 & 0 & 85 & 49\\
           \hline
           \# $\mathrm{d}\Gamma/\mathrm{d}\sqrt{p^2}$ bins & 92 & 62 & 92 & 62 & 0 & 0 & 0 & 0 & 92 & 62\\
           \hline
           \# $\mathcal{B}$ bins & 1 & 1 & 1 & 1 & 3 & 2 & 3 & 2 & 1 & 1\\
           \hline
           \# fit parameters & 9 & 9 & 12 & 12 & 8 & 6 & 10 & 8 & 16 & 15\\
           \hline
           $\chi^2/\text{dof}$ & 3.16 & 2.51 & 2.85 & 2.23 & 6.64 & 5.54 & 8.24 & 8.29 & 2.65 & 2.12\\
           \hline
           \hline
       \end{tabular}
   \end{adjustbox}
   \caption{Best fit values and uncertainties of the resonance parameters in  different fit scenarios detailed in the main text.
   An exact zero  ($0$) means  a parameters is set to zero as an input to  the fit. 
   To highlight that the $r_{9,10}$ parameters are only included in scenario 9 and 10 we use a  `-'  for the other scenarios here.}
         \label{tab:SMfit}
\end{table}

We observe the following  general pattern 
\begin{align} \nonumber
    a_{P,\rho} &\simeq 0.3 \, \text{GeV}^2~,\quad a_{P,\phi}/a_{P,\rho}\simeq 0.3~,\quad\delta_{P,\rho}\simeq 3~,\quad a_{S,\rho}/a_{P,\rho}\simeq 40~,\\
   \delta_{P,\omega} & \sim 0 \, , \quad  \delta_{S,\omega} \sim 0 \,  (\text{except scenario 9}) \, ,\quad a_{p^2,\omega}\lesssim 0.003 \, ,
\end{align}
where we identify $0$ with $2 \pi$ for phases.
The remaining parameters vary between the scenarios. We identify five groups where the
best-fit values of resonance parameters are in agreement: scenarios 1 and 2, scenarios 3
and 4, scenarios 5 and 7, scenarios
6 and 8 (with disagreement in S-wave parameters), and scenarios 9 and 10 (with disagreement in $a_{S,\rho}/a_{P,\rho}$).

The ratio $a_{P,\omega}/a_{P,\rho}$
is either set to zero or consistent with zero in scenarios 1, 2, 5, and 7. All other scenarios have  $a_{P,\omega}/a_{P,\rho} \sim 0.2-0.3$.
The value of $a_{S,\omega}$ is only large if we set $a_{P,\omega}$ to zero with the exception of scenario 6 that uses isospin relations.  This could be disentangled using the observables proposed in Ref.~\cite{Fajfer:2023tkp}, which are more sensitive to S-wave contributions.

Scenarios  5,6,7,8, are  based on a reduced set of observables,  $\langle S_{2,3,4} \rangle$ and $\mathcal{B}$ only, and data are not sufficient to fully determine all resonance parameters. In this case  we exclude them from  the fit, see Tab.~\ref{tab:SMfit}. Technically, in the code we set them to zero. These scenarios all  have
$\chi^2/\text{dof} > 5$, while all other scenarios have $\chi^2/\text{dof} \sim 2-3$. This highlights the importance of differential decay distributions for the theoretical modelling.

The best fit for $r_{9,10}=0$  is obtained in scenario 4, with $\chi^2/\text{dof}=2.23$, closely followed by 2 and 3.  
Interestingly, we find that the  overall best $\chi^2/\text{dof}=2.12$ has scenario 10, favoring  a large $r_9 \sim 2$ with an order one phase, and vanishing $r_{10}\sim0$.
As  large values of NP in $\mathcal{C}_{9}\sim 2 $ are in tension with limits   \eqref{eq:c9bound} from other $c\to u\mu^+\mu^-$ modes, this points to  missing hadronic contributions  to  the operator $\mathcal{O}_9$ in $D \to \pi^+ \pi^- \mu^+ \mu^-$ .
The right panel of  Fig.~\ref{fig:SMfit_dGamma_dx_compare} supports also that the presence of $r_9 \sim 2$ in scenario 10  (light blue) and 9 (lemon) improves the agreement with data 
(black)  \cite{LHCb:2021yxk} at low $q^2$. Both scenarios 9 and 10 feature  also larger high dimuon mass tails, where, however, no data is available and rates decline.
 Further theoretical  investigation is desirable  but is beyond the scope of this work.

\begin{figure}[h!]
    \centering
    \includegraphics[width=0.79\textwidth]{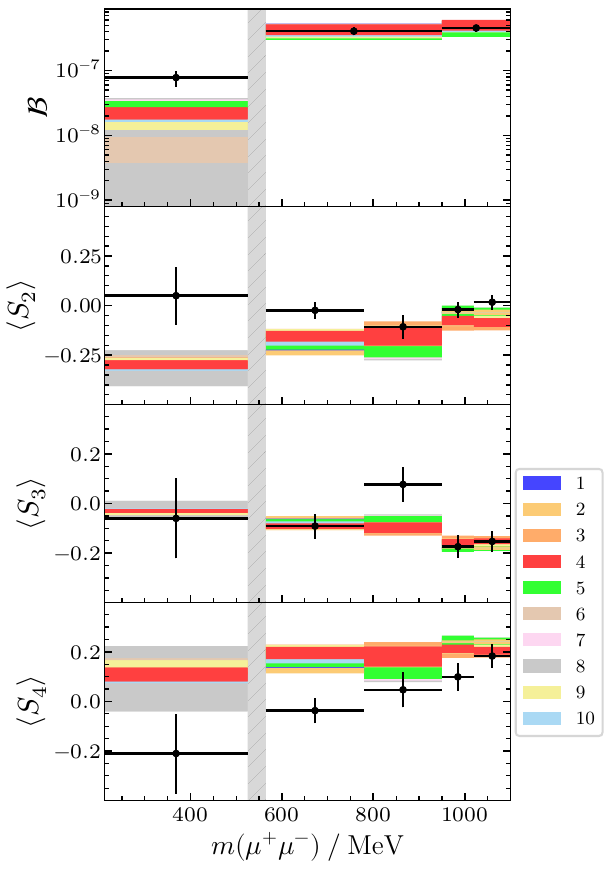}
    \caption{ 
      Data  (black) on the $D \to \pi^+ \pi^- \mu^+ \mu^-$ binned branching ratio \cite{LHCb:2017uns} and CP-averaged observables $\langle S_{2,3,4} \rangle$  \cite{LHCb:2021yxk} 
    and SM predictions using the best-fit values of the resonance parameters for the different scenarios.  Scenarios largely overlap. For the ones with lowest $\chi^2/\text{dof}$, scenario 10 (light  blue)
    is in part underneath scenario 4 (red), see text. Note, $\langle S_4 \rangle =- \langle S_4^\text{LHCb} \rangle$, see footnote \ref{foot:angles}.}
    \label{fig:angular_obs_SM}
\end{figure}

The CP-averaged angular observables $\langle S_{2,3,4} \rangle$ together with the binned  branching ratio using the fit values of the resonance parameters for the different  scenarios are displayed in Fig.~~\ref{fig:angular_obs_SM}. The predictions of the various scenarios are in part largely overlapping. For the scenarios with lowest $\chi^2/\text{dof}$,
see Tab.~\ref{tab:SMfit}, 
scenario 10 (light blue)  is underneath scenario 4 (red), if it is not visible.
We consider uncertainties in $D\to\rho$ and $D\to f_0(500)$ FFs, the Blatt-Wei{\ss}kopf factor $r_\text{BW}$ of the lineshapes, as well as the masses and widths; see Appendix~\ref{app:FFs} for details. We observe that the first two low-$q^2$ bins in $\langle S_2\rangle$ and $\langle S_4\rangle$
as well as the low-$q^2$ bin of $\mathcal{B}$
are in tension with the data. For this reason, we only consider the  other  bins with higher dilepton mass in  the global fits to $\mathcal{C}_{7,9,10}^{(\prime)}$
Moreover, our framework predicts  $\langle S_9^\text{SM} \rangle = 0$, \footnote{This can  be lifted with transversity-dependence, see (\ref{eq:IiSM}).}
while LHCb finds deviations from zero for some of the high-$q^2$ bins with the largest value being $\langle S_9 \rangle = (16.9\pm 4.4 )\%$ for $[0.950\text{-}1.02]\,\mathrm{GeV}$~\cite{LHCb:2021yxk}, a local  anomaly of $3.8  \sigma$. A NP explanation  would require  large  values of $\mathcal{C}_9^{(\prime)}$, which are already excluded \eqref{eq:c9bound}. Therefore we do not take  data on $\langle S_9 \rangle$ into account  as this would  just add a constant to the $\chi^2$-function.

\begin{figure}[h!]
        \centering
        \begin{subfigure}[b]{0.476\textwidth}
            \centering
            \includegraphics[width=\textwidth]{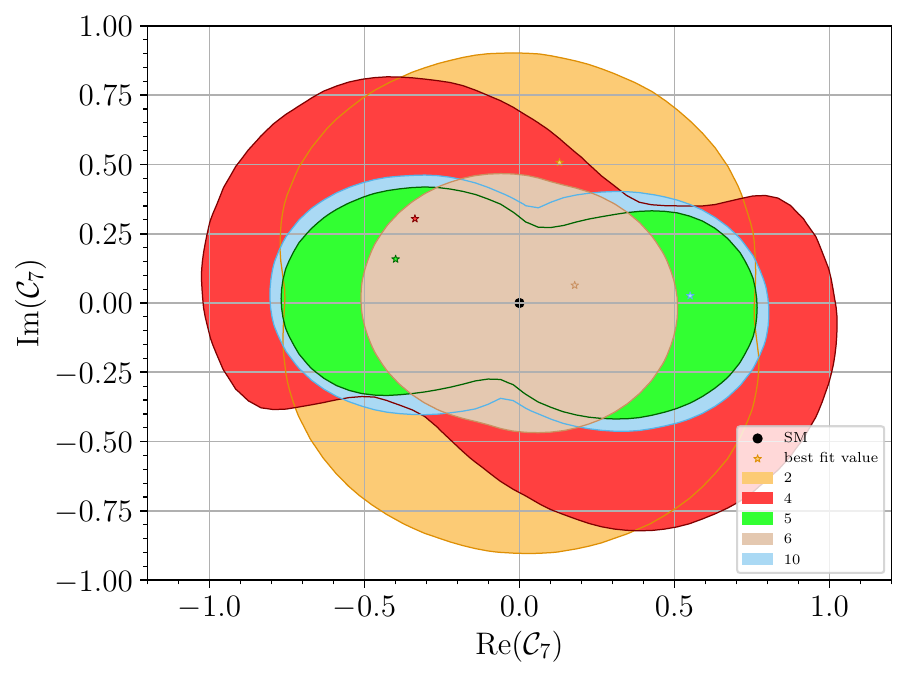}
            \caption{Fit of $\mathcal{C}_{7}$ to $\langle A_{2-4,8,9} \rangle$, $\langle S_8\rangle$ and $\mathcal{B}$.}
        \end{subfigure}
        \begin{subfigure}[b]{0.49\textwidth}
            \centering
            \includegraphics[width=\textwidth]{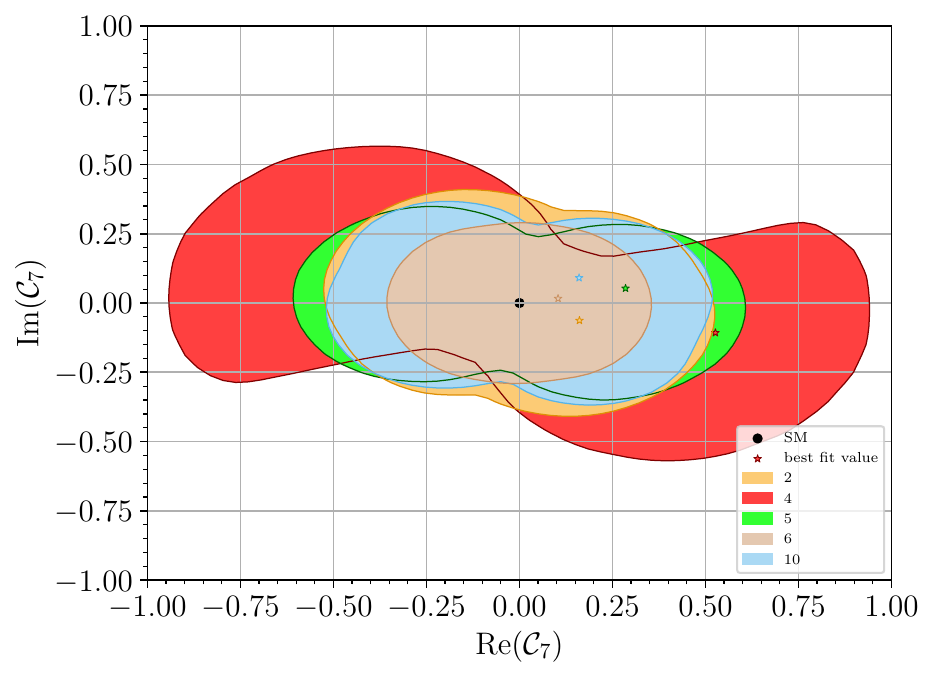}
            \caption{Fit of $\mathcal{C}_{7}$ to $\langle S(A)_{2-4,8}\rangle$, $\langle A_9\rangle$ and $\mathcal{B}$.}
        \end{subfigure}
        \begin{subfigure}[b]{0.49\textwidth}
            \centering
            \includegraphics[width=\textwidth]{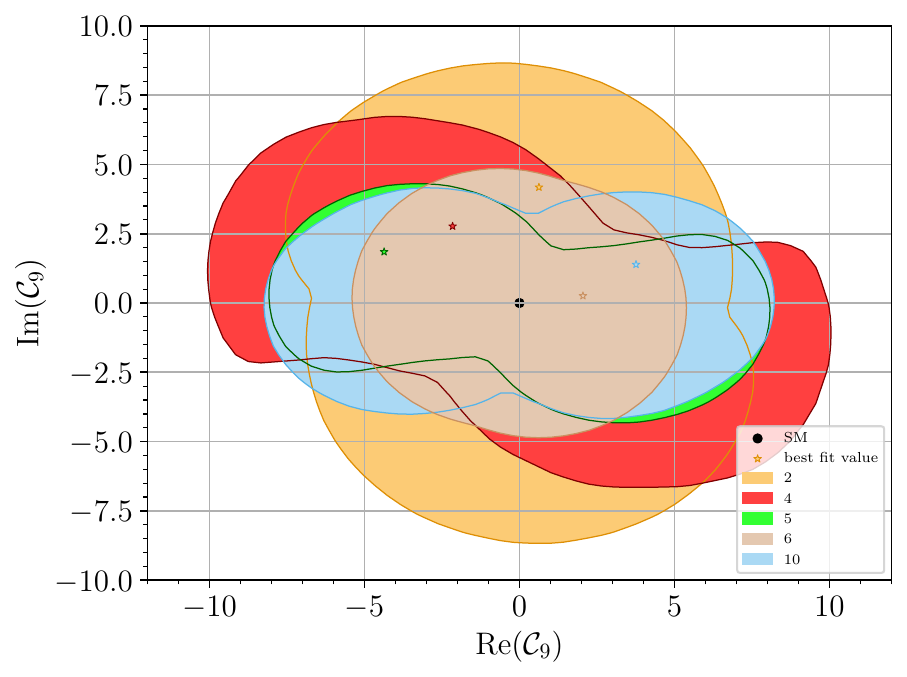}
            \caption{Fit of $\mathcal{C}_{9}$ to $\langle A_{2-4,8,9}\rangle$, $\langle S_8\rangle$ and $\mathcal{B}$.
            }
        \end{subfigure}
        \begin{subfigure}[b]{0.469\textwidth}
            \centering
            \includegraphics[width=\textwidth]{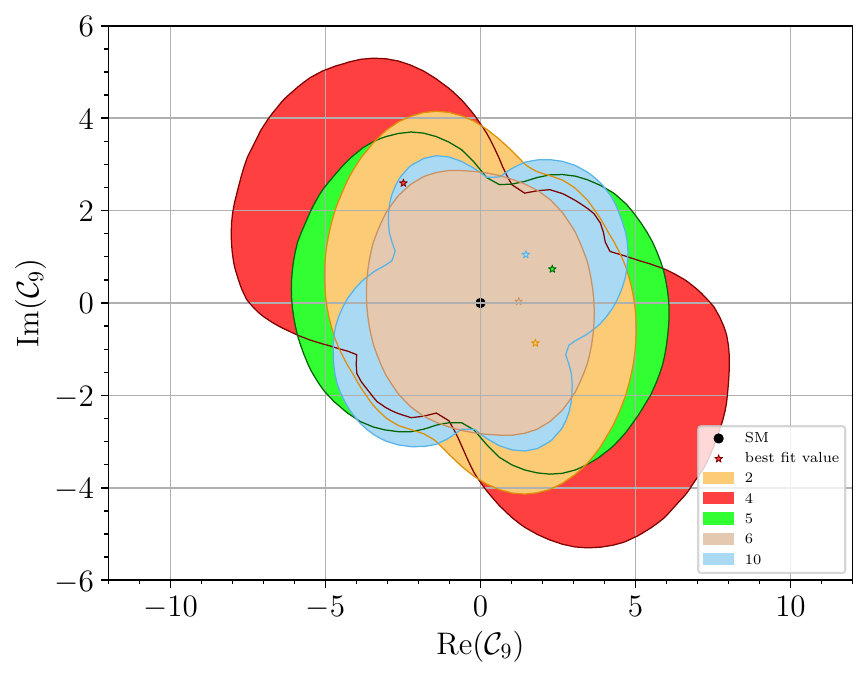}
            \caption{Fit of $\mathcal{C}_{9}$ to $\langle S(A)_{2-4,8}\rangle$, $\langle A_9\rangle$ and $\mathcal{B}$.}
        \end{subfigure}
        \begin{subfigure}[b]{0.49\textwidth}
            \centering
            \includegraphics[width=\textwidth]{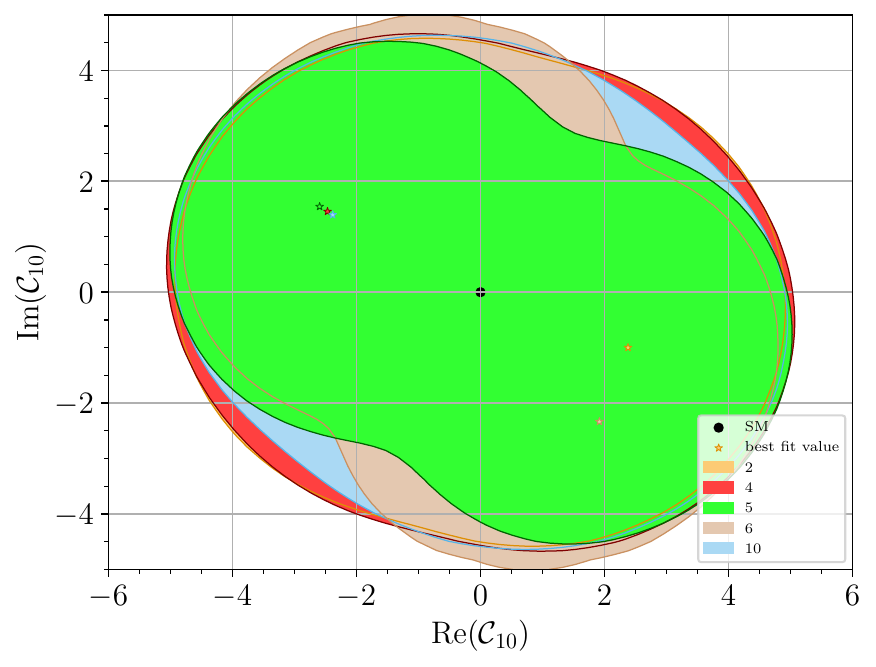}
            \caption{Fit of $\mathcal{C}_{10}$ to $\langle S(A)_{5,6,7}\rangle$ and $\mathcal{B}$.}
        \end{subfigure}
        \begin{subfigure}[b]{0.485\textwidth}
            \centering
            \includegraphics[width=\textwidth]{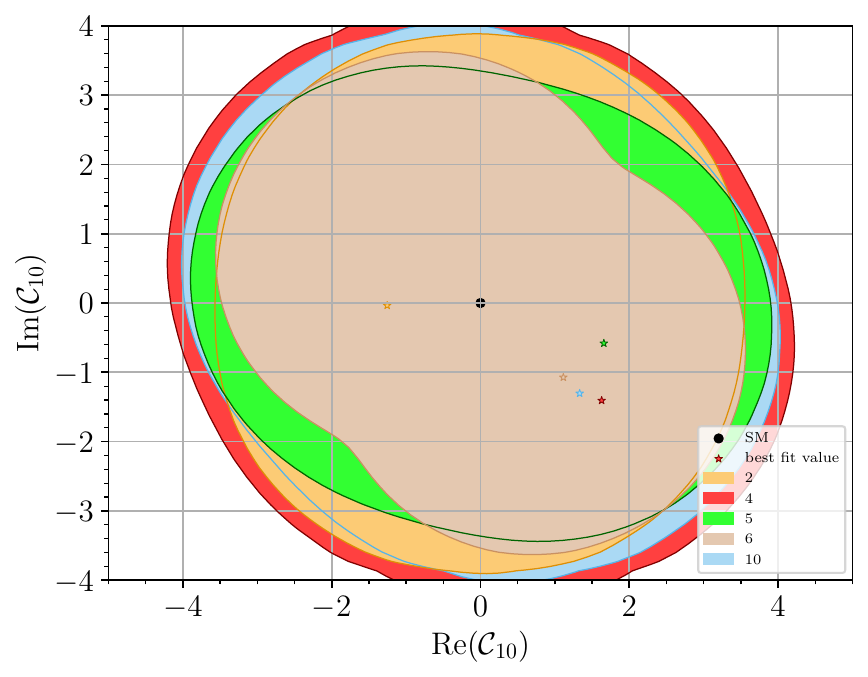}
            \caption{Fit of $\mathcal{C}_{10}$ to $\langle S_{2,3,4}\rangle$, $\langle S(A)_{5,6,7}\rangle$ and $\mathcal{B}$.}
        \end{subfigure}
        \caption{Fits of $\mathcal{C}_{7},\mathcal{C}_{9}$, and $\mathcal{C}_{10}$ in fit scenarios 2,4,5,6 and 10, representing the groups with similar fit parameters, see main text.}
    \label{fig:NPfit_compare_SMscenarios}
\end{figure}

 Since we   use  ``good''  bins only and here the exploratory scenario 10 is very much in agreement with scenario 4,  we use the latter
and   its best-fit values of hadronic parameters  for  $D \to \pi^+ \pi^- \mu^+ \mu^-$ decays in the global fit worked out in Sec.~\ref{sec:fit}.
All best-fit values of hadronic parameters in scenario 4 and 10 (except $r_9$) agree within $\sim 1 \sigma$.

\subsection{Fits to NP  in different scenarios}
\label{subsec:fitsNPDPPmumu}

We perform global fits to NP driven by $\mathcal{C}_{7,9,10}$ in the various model scenarios defined in Sec.~\ref{subsec:resonancefit}.  We set the resonance parameters to the best-fit values  given in Tab.~\ref{tab:SMfit}. The experimental data included in NP fits cover the remaining angular observables not considered in the resonance fit,
 that is, the NP measurements in Tab.~\ref{tab:4bodydata}.
In each  fit scenario, we  only include  those  observables  that depend on the fit parameters.
For $\mathcal{C}_{7,9}$, we include $\langle S_{8} \rangle$ and $\langle A_{2,3,4,8,9}\rangle$, excluding the two lowest $q^2$-bins, which we are unable to explain for the observables $\langle S_{2,4}\rangle$, see Fig.~\ref{fig:angular_obs_SM}.
Likewise, for $\mathcal{C}_{10}$, we include $\langle S_{5,6,7}\rangle$ and $\langle A_{5,6,7}\rangle$ in the same $q^2$-bins.
To  fit the ratio-type observables  we also take
 data on the differential branching ratios
into account.  We further include $\langle S_{2,3,4}\rangle$ in additional fits
for $\mathcal{C}_{7,9,10}$. Using
 branching ratios or $\langle S_{2,3,4}\rangle$
in the resonance {\it and}
the NP fit  may introduce inconsistencies due to potential double counting. A more
consistent approach would be to simultaneously fit all parameters, including both resonance and NP contributions. However, with present data this approach is not feasible.
As $\langle S_{2,3,4}\rangle$ only slightly improves the fit,
we do not use these observables for global fits worked out  in Sec.~\ref{sec:fit}.

The results, shown in Fig.~\ref{fig:NPfit_compare_SMscenarios}, indicate that the bounds on $\mathcal{C}_{7,9}$ are very sensitive to the assumptions regarding the extraction of resonance parameters, whereas the bounds on $\mathcal{C}_{10}$ are rather stable over the model scenarios.
 Comparing scenario 4 (red) with the exploratory scenario 10 (light blue) we learn  that  the latter is significantly more constraining in $\mathcal{C}_{7,9}$ fits.
 Overall, the limits from $D \to \pi^+ \pi^- \mu^+ \mu^-$  are much weaker than
from 2- and 3-body decays obtained in Secs.~\ref{sec:Dll} and  \ref{sec:Lc}. Therefore,  we expect that present $D \to \pi^+ \pi^- \mu^+ \mu^-$  data have only a minor impact
on the global $|\Delta c|=|\Delta u|=1$  fits performed in Sec.~\ref{sec:fit}.

The reason for  the comparatively weak bounds despite  existing advanced measurements  are twofold:
Firstly, the experimental sensitivity \cite{LHCb:2021yxk}  is not quite at the level where viable NP in $\mathcal{C}_{10}^{(\prime)}$ can cause a signal \cite{DeBoer:2018pdx}, also Fig.~\ref{fig:nulltests_binned}.
Secondly, and this holds of course in general for null tests, but is very critical for $D \to \pi^+ \pi^- \mu^+ \mu^-$   decays,  the sensitivity to NP, i.e.~the smallest value of a NP coefficient that causes a null test  observation, is lower than the limit derived from non-observation of a NP signal. The difference between these two is given by the SM uncertainties from strong phases, whose impact is pronounced  in the null tests.
For the case of 4-body decays and resonances in more than one channel, these are substantial, as discussed in Sec.~\ref{subsec:resonancefit}.

While  constraints from  $D \to \pi^+ \pi^- \mu^+ \mu^-$  observables are weaker compared to those  from other decays,
 the rich structure of the differential distribution of the 4-body decay complements other $c\to u\ell^+\ell^-$ observables. This aids to  
 resolve flat directions in the global fit, see Sec.~\ref{sec:comp} for further details.

\subsection{Beyond S- and P-wave \label{sec:beyondSP}}

Because of the issues  of the SM fit discussed in Sec.~\ref{subsec:resonancefit}, we consider  additional  contributions beyond the resonance ansatz \eqref{eq:BWwc}.

We begin analyzing  the impact of  D-wave resonances such  as the $f_2(1270)$, which decays mostly to $\pi \pi$.
Due to its high mass, it mostly contributes to low $q^2$ \cite{DeBoer:2018pdx}.
We find that $\langle S_{8,9}\rangle = 0$ even for D-waves assuming universality between different polarizations. The observables $\langle S_{2,3,4} \rangle$ receive D-wave contributions, specifically $\langle S_2 \rangle_{D} \sim (\left|\mathcal{F}_{\parallel D} \right|^2+\left|\mathcal{F}_{\perp D} \right|^2-3\left|\mathcal{F}_{0 D} \right|^2)$, $\langle S_3 \rangle_D \sim (\left|\mathcal{F}_{\parallel D} \right|^2-\left|\mathcal{F}_{\perp D} \right|^2)$ and $\langle S_4 \rangle_D \sim \mathrm{Re}(\mathcal{F}_{0\,D} \mathcal{F}_{\parallel\,D}^\ast)$, as well as a $D$-$S$-interference contribution only for $\langle S_4 \rangle_{\text{$D$-$S$}} \sim \mathrm{Re}(\mathcal{F}_{0\,S} \mathcal{F}_{\parallel\,D}^\ast)$ 
\cite{Das:2014sra}. However, we could not identify some parameter region for the D-wave that significantly affects $\langle S_4 \rangle$, as desirable, see Fig.~\ref{fig:angular_obs_SM}, while leaving $\langle S_{2,3} \rangle$  unchanged,  or not altering the differential branching fraction.  Likewise, an improvement of the SM prediction of the differential branching ratio does not appear to be possible with the $f_2$ resonance as its mass is above the
 region $\sqrt{p^2} \in [0.8,1.1]\,\mathrm{GeV}$ in which the decay distributions are difficult to explain, see  Fig.~\ref{fig:SMfit_dGamma_dx_compare}.

Bremsstrahlung contributions have been estimated in Ref.~\cite{Cappiello:2012vg} through Low's theorem and although they appear in all angular observables except the null tests they are restricted to low $q^2$ and are too small to address the discrepancies in the differentials. Although $\langle S_9 \rangle$ receives contributions, they are negligible for high-$q^2$ and the tension with the experimental data similar to $\langle S_{2,4} \rangle$ remains.

Cascade contributions  $D^0\to \pi^\mp a_1^\pm(\to \pi^\pm \rho^0(\to \mu^+\mu^-))$ could play a significant role as their branching fractions are sizeable.\footnote{ We believe that the first mention of axial-vector resonance contributions  is by Oscar Cata.} 
However they are more difficult to model than the S- and P-wave resonances and beyond the scope of this work. Nevertheless an experimental double differential including correlations would be useful to decipher the resonances and to perform studies on these cascade contributions in the future.

\section[Global \texorpdfstring{$\boldsymbol{c\to u}$}{c->u} fits]{Global $\boldsymbol{c\to u}$ fits}
\label{sec:fit}

We perform one-dimensional (1D) and two-dimensional (2D) fits for various combinations and subsets of NP Wilson coefficients $\mathcal{C}_7^{(\prime)}, \mathcal{C}_9^{(\prime)}$, and $\mathcal{C}_{10}^{(\prime)}$ using experimental information from $|\Delta c| = |\Delta u| = 1$ transitions,
branching ratio data  of $D \to \mu^+ \mu^-$~\cite{CMS-PAS-BPH-23-008}, $D^+ \to \pi^+ \mu^+ \mu^-$~\cite{LHCb:2020car},  $\Lambda_c \to p \mu^+ \mu^-$~\cite{LHCb:2024hju}, and branching ratios  \cite{LHCb:2017uns} and angular observables \cite{LHCb:2021yxk}
of $D \to \pi^+ \pi^- \mu^+ \mu^-$  decays.
 We use the same fit procedure as Ref.~\cite{Bause:2022rrs}. 
Resonances in $D \to \pi^+ \pi^- \mu^+ \mu^-$  decays are described with the best-fit values of scenario 4, see Sec.~\ref{subsec:resonancefit}.
1D and 2D fit results are given  in Secs~\ref{sec:1dfits} and \ref{sec:2dfits}, respectively.
 The main results of the fits are presented in Tabs.~\ref{tab:fitres_1d} and  \ref{tab:fitres_2d}. Constraints on the NP Wilson coefficients are given at the charm mass scale $\mu_c$.

\begin{figure*}[ht!]
    \centering
    \includegraphics[width=0.49\columnwidth]{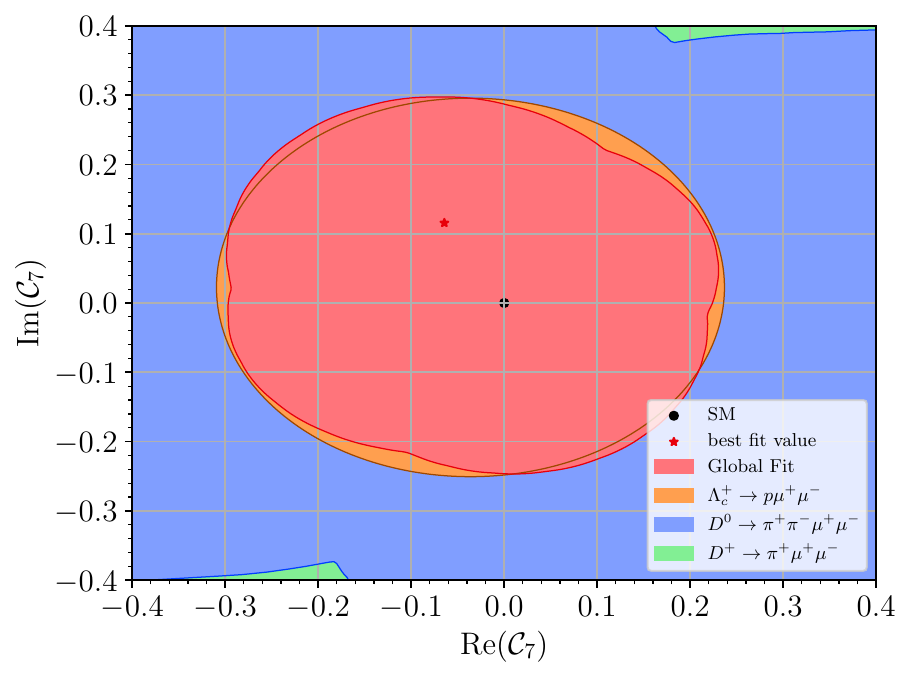}
    \includegraphics[width=0.49\columnwidth]{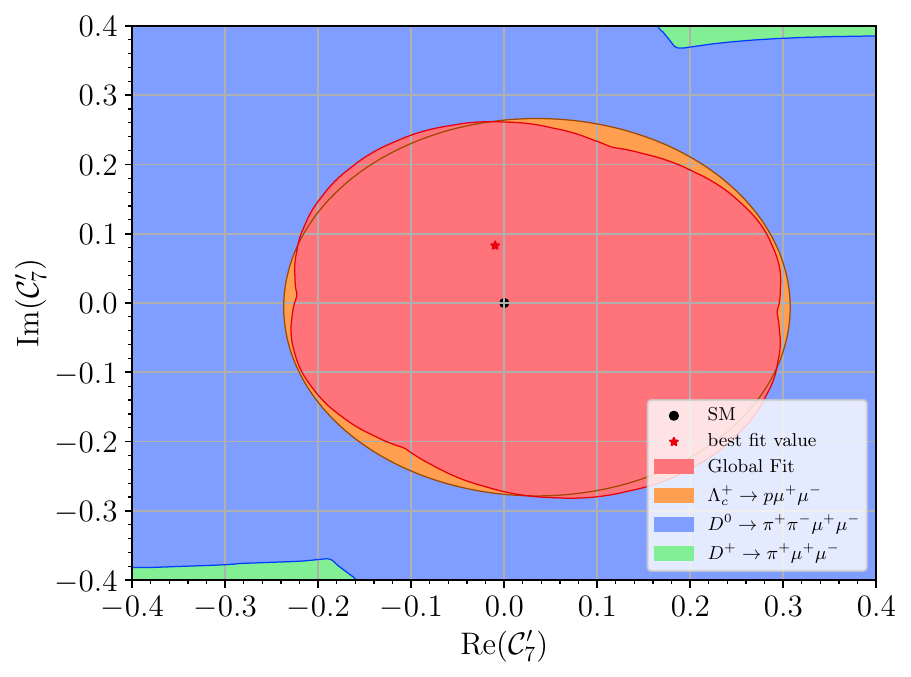}
    \caption{
   2D fits to $\mathcal{C}_7$ (left panel) and $\mathcal{C}_7^\prime$ (right panel), corresponding to scenarios $H_{13}$ and $H_{16}$ in Tab.~\ref{tab:fitres_2d}. Shown are  the 1$\sigma$ allowed regions of the individual observables and the combined 1 $\sigma$ region (red). The red  (black) star represents the best-fit (SM) value. 
    }
    \label{fig:dipoles}
\end{figure*}

\begin{figure*}[htbp!]
    \centering
    \includegraphics[trim={0cm 0.17cm 0cm 0.2cm},width=0.47\columnwidth]{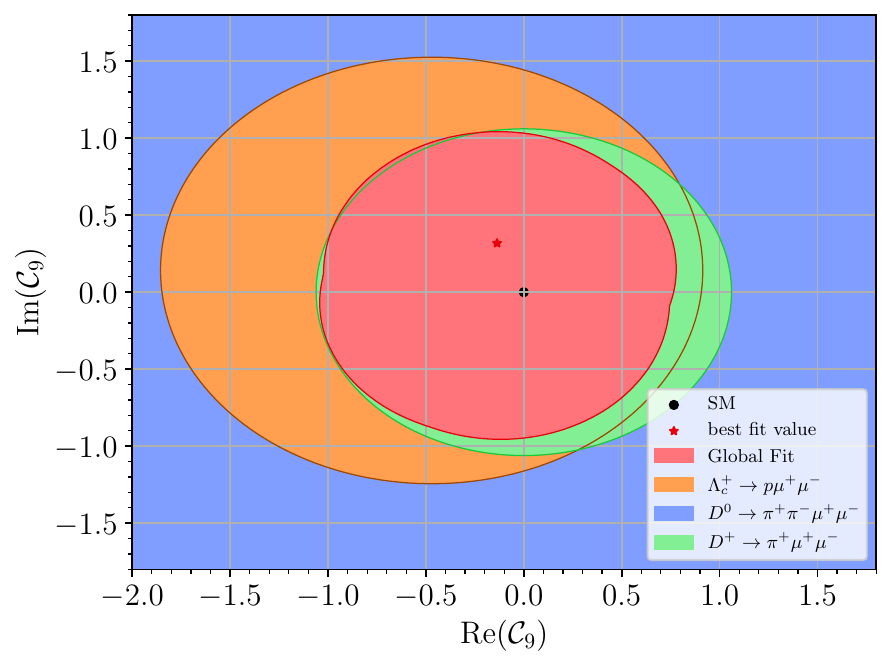}
    \includegraphics[trim={0cm 0.17cm 0cm 0.2cm},width=0.47\columnwidth]{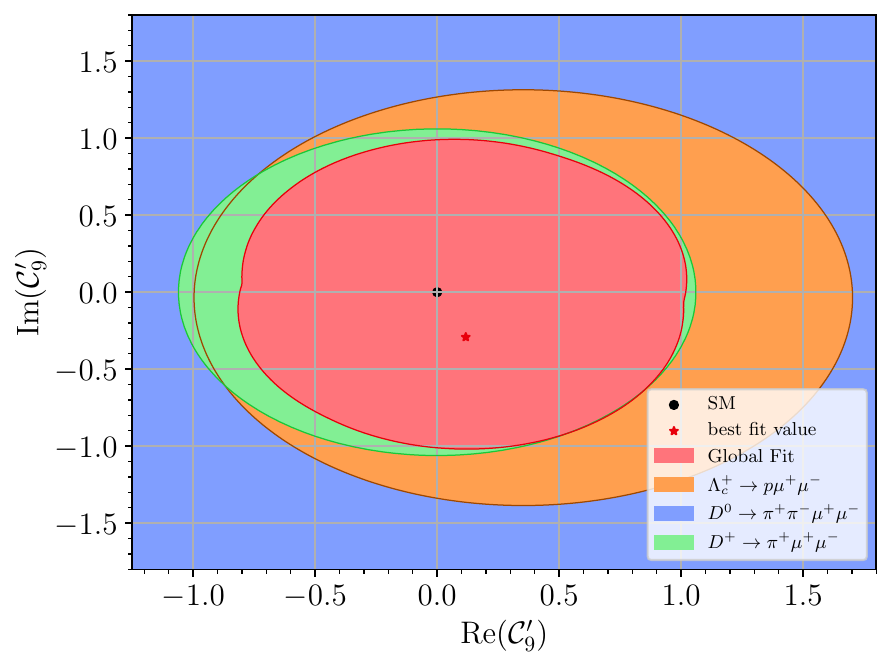}
    \includegraphics[trim={0cm 0.17cm 0cm 0.2cm},width=0.47\columnwidth]{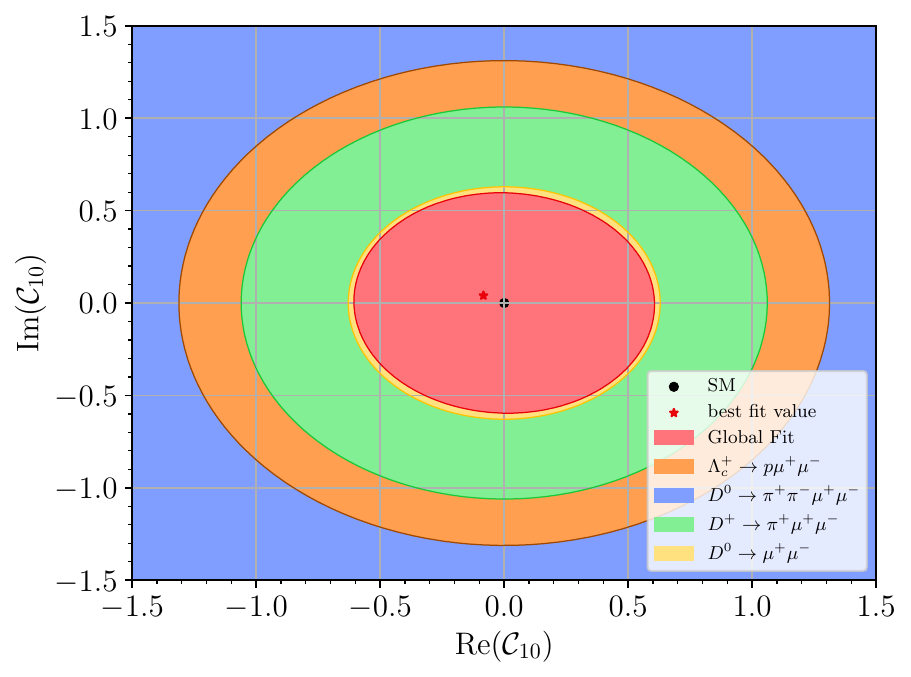}
    \includegraphics[trim={0cm 0.17cm 0cm 0.2cm},width=0.47\columnwidth]{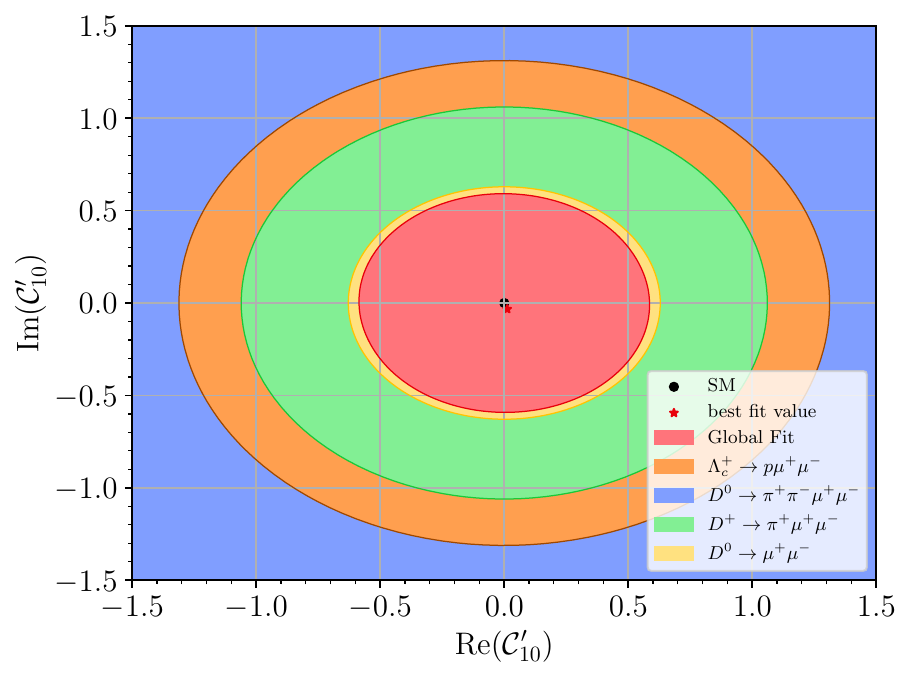}
    \includegraphics[trim={0cm 0.17cm 0cm 0.2cm},width=0.47\columnwidth]{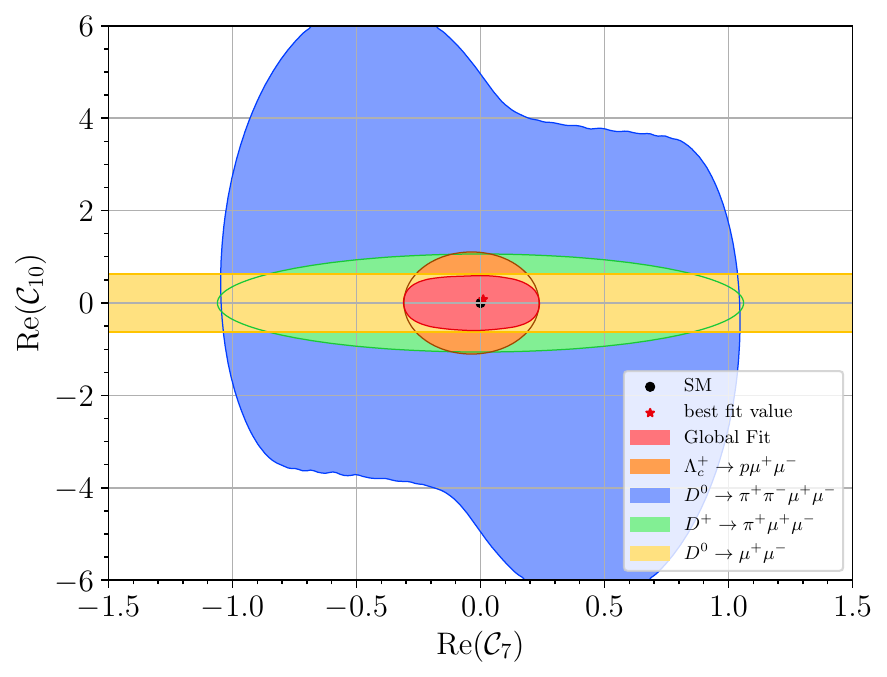}
    \includegraphics[trim={0cm 0.17cm 0cm 0.2cm},width=0.47\columnwidth]{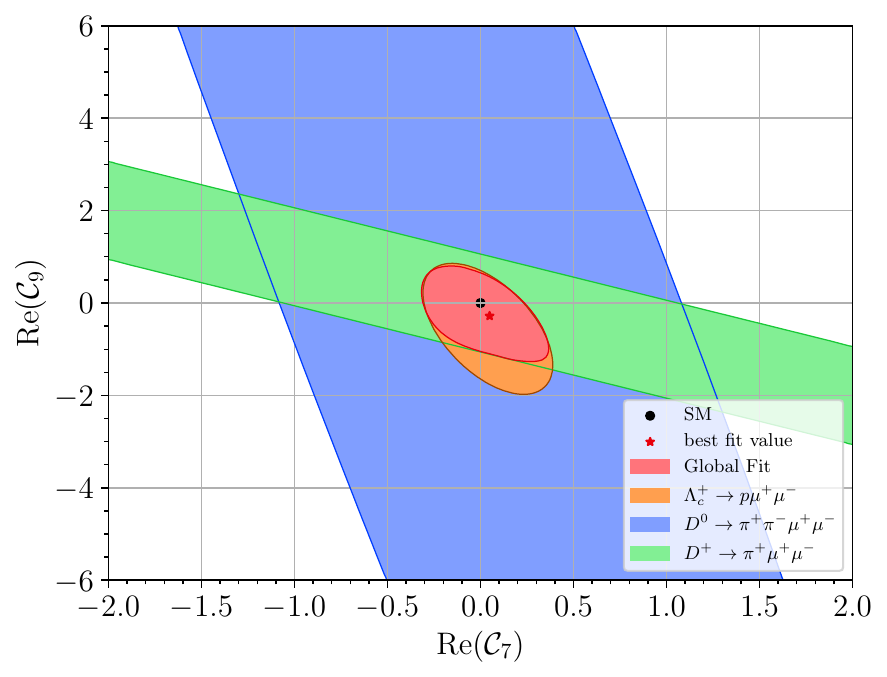}
    \includegraphics[trim={0cm 0.17cm 0cm 0.2cm},width=0.47\columnwidth]{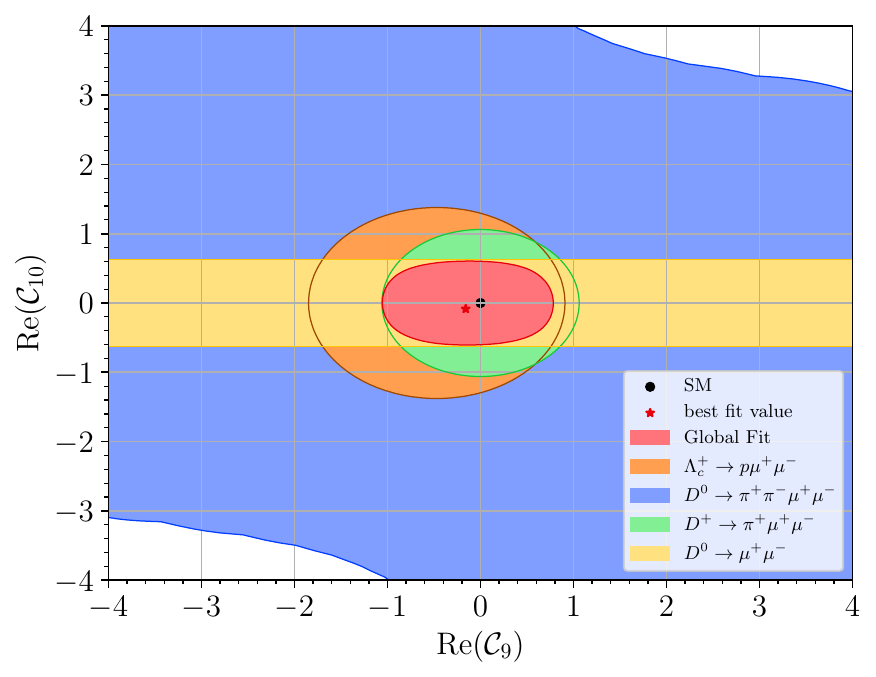}
    \includegraphics[trim={0cm 0.17cm 0cm 0.2cm},width=0.47\columnwidth]{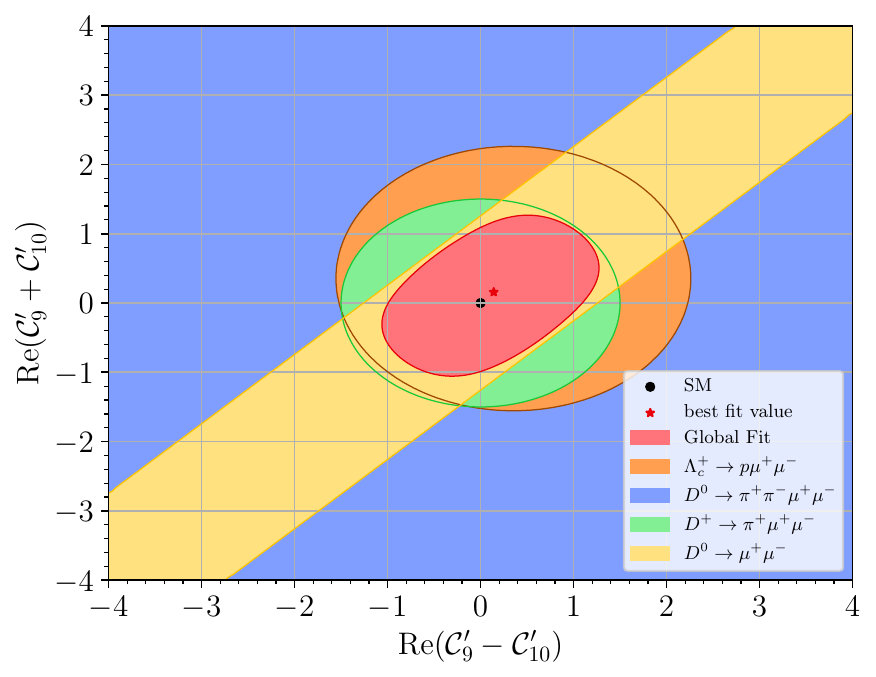}
    \caption{Contours of the 2D fits to semileptonic four-fermion operators,  and mixed ones on $(\text{Re}\,  \mathcal{C}_7, \text{Re} \, \mathcal{C}_{10})$, and 
    $(\text{Re} \,  \mathcal{C}_7, \text{Re} \, \mathcal{C}_9)$, see Tab.~\ref{tab:fitres_2d}.
Also shown in the bottom right plot  is  $\text{Re} (\mathcal{C}_9'+\mathcal{C}_{10}')$ vs $\text{Re} (\mathcal{C}_9'-\mathcal{C}_{10}')$; similar description as Fig.~\ref{fig:dipoles}.}
    \label{fig:2dfit_contours}
\end{figure*}

\subsection{One-dimensional fits}\label{sec:1dfits}

The 1D fit results  are reported in Tab.~\ref{tab:fitres_1d}. The first column lists the names ($H_{1,\ldots,12}$), followed by the second column with the Wilson coefficient
that is fitted. The best-fit values, along with their 1$\sigma$ and 2$\sigma$ confidence intervals, are provided in the third, fourth and fifth columns, respectively. The last two columns give the goodness-of-fit indicators, including the best-fit $\chi^2$ value and the $p$-value of the fit. 

We recall that in a specific fit we only include observables which depend on the fit parameters, that is, the degrees of freedom in general  vary between the different fits.
The $\chi^2$ value of the SM hypothesis  is $\chi^2_{\text{SM}} = 17.0$  for fits to  $\mathcal{C}_7^{(\prime)}$ or $\mathcal{C}_9^{(\prime)}$ and
$\chi^2_{\text{SM}} = 22.3$ for fits to  $\mathcal{C}_{10}^{(\prime)}$.
In all cases, the minimal  $\chi^2$ values in the fit scenarios $\chi^2_{H_i,\,\text{min}}$ are close to the corresponding SM one, hence, the fits are  good, supporting the SM and allowing one to put  robust constraints on NP.

\begingroup
\renewcommand*{\arraystretch}{1.6}
\begin{table*}[ht!]
 \centering
 \resizebox{0.95\textwidth}{!}{
 \begin{tabular}{c|c|c|c|c|c|c}
  \hline
  \hline
  \rowcolor{LightBlue}
  scenario  & fit parameter & best fit &$1\sigma$  &$2\sigma$  &$\chi^2_{H_i,\,\text{min}}$ & $p$-value (\%) \\ 
  \hline
  \hline
  $H_{1}$ &                                                   $\text{Re}\,\mathcal{C}_7$ &     \phantom{-}0.01 & [-0.26,+0.18] & [-0.35,+0.28] & 17.0 & 91.1 \\
  $H_{2}$ &                                                   $\text{Re}\,\mathcal{C}_9$ & -0.21 & [-0.85,+0.55] & [-1.2,+0.95] & 16.9 & 91.2  \\
  $H_{3}$ &                                                $\text{Re}\,\mathcal{C}_{10}$ &     -0.09 & [-0.50,+0.50] & [-0.71,+0.71] & 22.3 & 76.8  \\
  $H_{4}$ &                                            $\text{Re}\,\mathcal{C}_7^\prime$ &     \phantom{-}0.05 & [-0.18,+0.25] & [-0.28,+0.35] & 16.9 & 91.2 \\
  $H_{5}$ &                                            $\text{Re}\,\mathcal{C}_9^\prime$ &     \phantom{-}0.15 & [-0.61,+0.82] & [-1.0,+1.2] & 16.9 & 91.1 \\
  $H_{6}$ &                                         $\text{Re}\,\mathcal{C}_{10}^\prime$ &     \phantom{-}0.00 & [-0.45,+0.45] & [-0.69,+0.69] & 22.3 & 76.7  \\
  $H_{7}$ &                                                   $\text{Im}\,\mathcal{C}_7$ &     \phantom{-}0.09 & [-0.20,+0.24] & [-0.29,+0.33] & 16.6 & 92.1  \\
  $H_{8}$ &                                                   $\text{Im}\,\mathcal{C}_9$ &     \phantom{-}0.32 & [-0.76,+0.85] & [-1.1,+1.2] & 16.8 & 91.5  \\
  $H_{9}$ &                                                $\text{Im}\,\mathcal{C}_{10}$ &     \phantom{-}0.06 & [-0.47,+0.47] & [-0.69,+0.70] & 22.3 & 76.7  \\
  $H_{10}$ &                                            $\text{Im}\,\mathcal{C}_7^\prime$ &     -0.09 & [-0.22,+0.21] & [-0.31,+0.30] & 16.6 & 92.0  \\
  $H_{11}$ &                                            $\text{Im}\,\mathcal{C}_9^\prime$ &     -0.28 & [-0.82,+0.80] & [-1.2,+1.1] & 16.8 & 91.5  \\
  $H_{12}$ &                                         $\text{Im}\,\mathcal{C}_{10}^\prime$ &     -0.03 & [-0.46,+0.46] & [-0.69,+0.69] & 22.3 & 76.7  \\
  \hline  
  \hline
  \end{tabular}
  }
  \caption{Fit results for Wilson coefficients in the 1D scenarios ($H_{1,...,12}$) at the scale $\mu_c$. Best-fit values and $1\sigma$ ($2\sigma$) uncertainties are displayed in the third and fourth (fifth) columns. The indicators of the goodness-of-fit are provided in the last columns, the $\chi^2$ function evaluated at the best-fit point and the $p-$value.
   }
  \label{tab:fitres_1d}
\end{table*}
\endgroup

\subsection{Two-dimensional fits}\label{sec:2dfits}

We explore scenarios consisting of pairs of coefficients from $\mathcal{C}_7^{(\prime)}$, $\mathcal{C}_9^{(\prime)}$, $\mathcal{C}_{10}^{(\prime)}$  in $H_{13,...,22}$. 
The results of the 2D fits are presented in Tab.~\ref{tab:fitres_2d}.  The
$\chi^2$ value of the SM hypothesis  for fits involving  $\mathcal{C}_7^{(\prime)}$  and  $\mathcal{C}_{10}^{(\prime)}$, 
or   $\mathcal{C}_9^{(\prime)}$  and  $\mathcal{C}_{10}^{(\prime)}$   is   $\chi^2_{\text{SM}} = 37.4$. 
As for the 1D case we obtain good fits, and consistency with the SM.

 In Figs.~\ref{fig:dipoles} and~\ref{fig:2dfit_contours} we show the allowed regions for 2D fits, for coefficients of  dipole  $\mathcal{C}_7^{(\prime)}$ and of  semi-leptonic  operators $\mathcal{C}_{9,10}^{(\prime)}$, respectively.
 Also shown in Fig.~\ref{fig:2dfit_contours} are  constraints on  $\text{Re} (\mathcal{C}_9'+\mathcal{C}_{10}')$ vs $\text{Re} (\mathcal{C}_9'-\mathcal{C}_{10}')$,
     illustrating constraints for right-handed up-type quarks and left- and right-handed leptons.  
   This corresponds to a simplified $SU(2)_L$-invariant setting where renormalization group effects from the NP scale $\Lambda_\text{NP}$ down  to the electroweak scale
   have not been included. The latter are, however,  small, less than 5\% for $\Lambda_\text{NP}=10\,$TeV~\cite{Bause:2021cna}.
 High-$p_T$ Drell-Yan tails also provide contraints on semileptonic 4-fermion  $|\Delta c|=|\Delta u|=1$ operators in this framework, however, they are insensitive to interference effects~\cite{Fuentes-Martin:2020lea}.  Assuming $SU(2)_L$,
operators with left-handed charm quarks are subject to strong constraints from kaon decays~\cite{Bause:2020auq}.

As already anticipated in Sec.~\ref{subsec:fitsNPDPPmumu} the global  fit is dominated by 2- und 3-body decays, and $D \to \pi^+ \pi^- \mu^+ \mu^-$ is not competitive.
If  dipole operators are present,  the most important constraint is from the $\Lambda_c \to p \mu^+ \mu^-$ branching ratio.
In fits with $\mathcal{C}_{10}^{(\prime)}$, the strongest constraint is from $D \to \mu^+ \mu^-$; for $\mathcal{C}_9, \mathcal{C}_{10}$ the $D^+ \to \pi^+ \mu^+ \mu^-$ data is the most constraining, followed by data on the baryonic decays.

\begingroup
\renewcommand*{\arraystretch}{1.6}
\begin{table*}[ht!]
\centering
\resizebox{0.95\textwidth}{!}{
\begin{tabular}{c|c|c|c|c|c|c}
 \hline
 \hline
 \rowcolor{LightBlue}
 scen.  & fit parameters  & best fit & $1\sigma$ & $2\sigma$ & $\chi^2_{H_i,\,\text{min}}$  & $p$-v. (\%) \\
 \hline
 \hline
$H_{13}$ &                                ($\text{Re}\,\mathcal{C}_7$,\,$\text{Im}\,\mathcal{C}_7$) &   (-0.06,+0.12) & ([-0.24,+0.17],[-0.19,+0.25]) & ([-0.34,+0.27],[-0.29,+0.33]) & 16.5 & 90.0 \\
$H_{14}$ &                              ($\text{Re}\,\mathcal{C}_9$,\,$\text{Im}\,\mathcal{C}_9$) &   (-0.14,+0.32) & ([-0.82,+0.55],[-0.76,+0.86]) & ([-1.2,+0.94],[-1.1,+1.2]) & 16.8 & 89.0  \\
 $H_{15}$ &

($\text{Re}\,\mathcal{C}_{10}$,\,$\text{Im}\,\mathcal{C}_{10}$) &   (-0.09,+0.04) & ([-0.48,+0.48],[-0.46,+0.46]) & ([-0.70,+0.70],[-0.69,+0.69]) & 22.3 & 72.3  \\
 $H_{16}$ &                                ($\text{Re}\,\mathcal{C}_7^\prime$,\,$\text{Im}\,\mathcal{C}_7^\prime$) &   (-0.01,0.08) & ([-0.17,+0.24],[-0.24,+0.22]) & ([-0.27,+0.34],[-0.32,+0.30]) & 16.6 & 89.4  \\
$H_{17}$ &                              ($\text{Re}\,\mathcal{C}_9^\prime$,\,$\text{Im}\,\mathcal{C}_9^\prime$) &   (+0.12,-0.29) & ([-0.58,+0.80],[-0.83,+0.80]) & ([-0.99,+1.2],[-1.2,+1.1]) & 16.8 & 88.9  \\
 $H_{18}$ &

($\text{Re}\,\mathcal{C}_{10}^\prime$,\,$\text{Im}\,\mathcal{C}_{10}^\prime$) &   (+0.01,-0.032) & ([-0.45,+0.45],[-0.46,+0.46]) & ([-0.69,+0.68],[-0.69,+0.69],) & 22.3 & 72.1  \\
$H_{19}$ &

($\text{Re}\,\mathcal{C}_{7}$,\,$\text{Re}\,\mathcal{C}_{9}$) &   (+0.049,-0.28) & ([-0.26,+0.29],[-1.0,+0.56]) & ([-0.36,+0.42],[-1.5,+0.98]) & 16.9 & 88.7
\\
$H_{20}$ &

($\text{Re}\,\mathcal{C}_{7}$,\,$\text{Re}\,\mathcal{C}_{10}$) &   (+0.01,+0.08) & ([-0.26,+0.18],[-0.48,+0.47]) & ([-0.35,+0.28],[-0.69,+0.69]) & 37.4 & 90.6  \\
$H_{21}$ &

($\text{Re}\,\mathcal{C}_{9}$,\,$\text{Re}\,\mathcal{C}_{10}$) &   (-0.16,-0.09) & ([-0.85,+0.55],[-0.48,+0.48]) & ([-1.2,+1.0],[-0.70,+0.70]) & 37.3 & 90.7  \\
$H_{22}$ & ($\text{Re}\left\{\mathcal{C}_{9}^\prime - \mathcal{C}_{10}^\prime\right\}$,\,$\text{Re}\left\{\mathcal{C}_{9}^\prime + \mathcal{C}_{10}^\prime\right\}$) & (0.14,0.15) & ([-0.75,1.0],[-0.75,1.0]) & ([-1.3,1.5],[-1.3,1.5]) & 37.4 & 90.6 \\

 \hline
 \hline
 \end{tabular}
 }
 \caption{2D fit results with similar description as Tab.~\ref{tab:fitres_1d}.
 }
 \label{tab:fitres_2d}
\end{table*}
\endgroup

\section{Future prospects \label{sec:future}}

We study future perspectives of the $|\Delta c|=|\Delta u|=1$ global fit. We analyze the  physics potential of $\Lambda_c \to p\,\ell^+\ell^-$ decays in Sec.~\ref{sec:Lcfuture}.
In Sec.~\ref{sec:comp} we work out  complementarities between  2- , 3- and 4-body $c \to u \mu^+ \mu^-$ decays.

\subsection[\texorpdfstring{$\Lambda_c\to p\,\ell^+\ell^-$ physics potential }{Lambda c -> p l+ l- future prospects}]{$\boldsymbol{\Lambda_c\to p\,\ell^+\ell^-}$ physics potential
\label{sec:Lcfuture}}

To study the future sensitivity of  $\Lambda_c \to p\,\ell^+\ell^-$  we define CP-symmetries and CP-asymmetries of the binned forward-backward asymmetry \eqref{eq:AFB},
 \eqref{eq:AFBbar} 
as
\begin{align}
    \Delta \langle A_{\text{FB}} \rangle_{\QRanges} &=  \frac{1}{2}\left(\langle A_{\text{FB}} \rangle_{\QRanges} + \langle \bar{A}_{\text{FB}} \rangle_{\QRanges}\right) \:,\\
    \Sigma \langle A_{\text{FB}} \rangle_{\QRanges} &=  \frac{1}{2}\left(\langle A_{\text{FB}} \rangle_{\QRanges} - \langle \bar{A}_{\text{FB}} \rangle_{\QRanges}\right) \:.
\end{align}
Since  $A_{\text{FB}}$ is  CP-odd,  $ \Delta \langle A_{\text{FB}} \rangle$ is a CP-asymmetry and $ \Sigma \langle A_{\text{FB}} \rangle$ the CP-average.
Looking only at the $\mathcal{C}_{10}$ contribution and using $K_{1c}  \propto  \mathrm{Re}\left( \mathcal{C}_9^R \mathcal{C}_{10}^\ast \right)$~\cite{Golz:2021imq},  one obtains $\Delta \langle A_{\text{FB}} \rangle \propto \mathrm{Im}\left(\mathcal{C}_{10}\right)$ and 
$\Sigma \langle A_{\text{FB}} \rangle \propto \mathrm{Re}\left(\mathcal{C}_{10}\right)$. 

The values  of $\Delta \langle A_{\text{FB}} \rangle$ and 
$\Sigma \langle A_{\text{FB}} \rangle$ depend strongly  on the  strong phases $\delta_\rho$, $\delta_\omega$ and $\delta_\phi$. The isospin relation fixes the 
relative phase $\delta_\rho - \delta_\omega = \pi$ 
and a more precise measurement of the branching ratio 
constrains the relative phase $\delta_\phi - \delta_\rho$, see  Fig.~\ref{fig:Lc-br}. The remaining overall 
phase of the resonances  $\mathcal{C}_9^R$  which one could choose to be $\delta_\phi$, 
is actually a relative strong phase with respect to the NP contribution and affects  $A_{\text{FB}}$.
However, as it enters  in the  branching ratio only in  subdominant interference terms it cannot be efficiently constrained with  branching ratio data,
and remains as one of the dominant  uncertainties in $A_{\text{FB}}$.

\begingroup
\renewcommand*{\arraystretch}{1.6}
\begin{table}[h!]
    \centering
     \resizebox{\textwidth}{!}{
    \begin{tabular}{c|c|c|c|c}
        \hline\hline
        \rowcolor{LightBlue} bin & $q^2\,/\,\mathrm{GeV}^2$ region & $\mathcal{B}_{\text{SM},\text{bin}}$ & $\langle A_{\text{FB}} \rangle_{\text{bin}}$, $\mathcal{C}_{10}=0.30$ & $\langle A_{\text{FB}} \rangle_{\text{bin}}$, $\mathcal{C}_{10}=0.05$\\
        \hline\hline
        $\omega/\rho$ left optimized & $[0.045,0.478]$ & $[0.15,0.41]\times 10^{-7}$ & $[-0.06,0.06]$ & $[-0.010,0.010]$\\
        \hline
        $\omega/\rho$ left & $[0.346,0.613]$ & $[0.58,1.07]\times 10^{-7}$ & $[-0.029,0.029]$ & $[-0.005,0.005]$\\
        \hline
        $\omega/\rho$ right & $[0.613,0.932]$ & $[0.41,1.05]\times 10^{-7}$ & $[-0.05,0.05]$ & $[-0.009,0.009]$\\
                \hline
        $\rho$ part $2$ \cite{LHCb:2024hju}& $[0.677,0.932]$ & $[0.15,0.67]\times 10^{-7}$ & $[-0.10,0.10]$ & $[-0.017,0.017]$\\
        \hline
        $\phi$ \cite{LHCb:2024hju} & $[0.959,1.122]$ & $[2.3,3.7]\times 10^{-7}$ & $[-0.011,0.011]$ & $[-0.0018,0.0018]$\\
\hline
        $\phi$ left & $[0.959,1.039]$ & $[1.0,2.0]\times 10^{-7}$ & $[-0.018,0.018]$ & $[-0.0030,0.0030]$\\
        \hline
        $\phi$ right & $[1.039,1.122]$ & $[1.1,2.0]\times 10^{-7}$ & $[-0.018,0.018]$ & $[-0.0031,0.0031]$\\
        \hline
        $\phi$ right optimized & $[1.052,1.818]$ & $[0.15,0.70]\times 10^{-7}$ & $[-0.10,0.10]$ & $[-0.017,0.017]$\\
     \hline
             $\sqrt{q^2}>1.25\,\mathrm{GeV}$\cite{Golz:2021imq} & $[1.562,1.818]$ & $[0.11,1.84]\times 10^{-9}$ & 
             $[-0.29,0.29]$ & $[-0.14,0.14]$\\
        \hline
        \hline
    \end{tabular}
}
\caption{$\mathcal{B}$ and $\langle A_{\text{FB}} \rangle$ 
in $\Lambda_c \to p\,\mu ^+\mu^-$  decays for different $q^2$ bins and  NP benchmarks. We explore bins to the left  (below $q^2=m_R^2$ ) and to the right  (above $q^2=m_R^2$)
of a resonance to  prevent large cancellations from  $A_{\text{FB}}$ changing sign, and further optimized ones to increase the NP signal (compare $\phi$ right optimized and $\phi$ right). NP effects in the branching ratio are negligible except for the $\sqrt{q^2} > 1.25\,\mathrm{GeV}$ bin and $\mathcal{C}_{10}=0.30$ for which it is within 
$[0.5,2.2 ]\times 10^{-9}$.}
    \label{tab:BR_and_AFB_binned}
\end{table}
\endgroup

In Tab.~\ref{tab:BR_and_AFB_binned} we provide  branching ratios and $\Sigma \langle A_{\text{FB}} \rangle$ 
for various bins and  benchmark values of  $\mathcal{C}_{10}$,  with  varied  strong phases.
Besides the $q^2$-bins considered in~\cite{LHCb:2024hju,Golz:2021imq}, we 
introduce additional  ones  to increase the NP reach. Specifically, we  separate 
the resonance contributions  into a bin   above `right' and one below `left' to avoid  cancellations from  sign changes close to the resonances
$m_{\omega,\rho}^2\sim 0.613\,\mathrm{GeV}^2$ or $m_{\phi}^2\sim 1.039\,\mathrm{GeV}^2$. We further explore  the bins ``$\omega/\rho$ left optimized" and
``$\phi$ right optimized",  which extend to the kinematic endpoints and have cuts  around the resonances to increase the sensitivity of $\langle A_{\text{FB}} \rangle$. As a trade-off,  the branching fraction decreases.
 As the impact of NP on the branching ratio  is negligible, $\langle A_{\text{FB}} \rangle$
 scales linearly  with  $\mathcal{C}_{10}$.  The sole  exception  is  the $\sqrt{q^2}>1.25\,\mathrm{GeV}$ bin  for largish NP  coefficient $ \mathcal{C}_{10} = 0.30$ near  maximum
  (\ref{eq: C10max}).
 However, even in this case $\mathcal{B}=[0.5,2.2 ]\times 10^{-9}$,  which is  close to the SM value.

As $\langle A_{{\rm FB}} \rangle$  is normalized to the branching ratio, generically a bin with a small branching ratio has a large asymmetry, and vice versa.
An extreme case is the  $\sqrt{q^2}>1.25$ GeV bin, with branching ratio of up to  $2 \times 10^{-9}$, and an asymmetry $\langle A_{{\rm FB}} \rangle \lesssim (1-3)  \cdot    \mathcal{C}_{10}$.
On the other hand, the NP sensitivity depends also on the resonance dynamics.
In the ``$\phi$ right optimized"-bin  or the '$\rho$ part 2'-bin with larger branching ratios,  reaching $10^{-7}$, we find   
$\langle A_{{\rm FB}} \rangle \lesssim 0.3 \cdot    \mathcal{C}_{10}$.
Other bins have  similar branching  ratios, but  less efficient  transfer of NP, for instance,   the other $\phi$ bins have   $
\langle A_{{\rm FB}} \rangle \lesssim 0.06  \cdot    \mathcal{C}_{10}$ or less.

To study the impact of a future measurement of $\langle A_{{\rm FB}} \rangle$ , we assume that, in each of the bins used,  the 
 $\Lambda_c \to p\,\mu ^+\mu^-$ branching ratio is measured with a precision of $\Delta \mathcal{B} /\mathcal{B}= 7\%$ $(2 \%)$.
Using Gaussian-error propagation, we 
approximate the corresponding experimental sensitivity  in  $\langle A_{\text{FB}} \rangle$   as  $\Delta \mathcal{B}/ \mathcal{B} = 7 \%$ $(2 \%)$.
 This implies sensitivity to  NP contributions in $\mathcal{C}_{10} \sim 0.2$ $(0.06)$  from one of the  best bins, i.e.,
the ``$\phi$ right optimized"-bin  or the '$\rho$ part 2', see Tab.~\ref{tab:BR_and_AFB_binned}.
The sensitivity is set by the highest sensitivity of a single null test observable. This reach can be matched from $D \to \mu^+ \mu^-$ if the present limit (\ref{eq:CMS})  
is improved by a factor 7 (75). We recall complementarity of the decay modes in the presence of both $\mathcal{C}_{10}$ and $\mathcal{C}_{10}^\prime$. We study this also further in Sec.~\ref{sec:comp}.
Also note that $\mathcal{B}(D \to \mu^+ \mu^-)$ is quadratic in the NP coefficient, while the null test $\langle A_{\text{FB}} \rangle$ is  linear, hence favorable for smaller NP contributions.

Working out upper limits is more involved due to correlations between bins  and observables and accidental suppressions by phase tunings.
Recall that  a single observable,  $\Sigma\langle A_{\text{FB}} \rangle$ or $\Delta\langle A_{\text{FB}} \rangle$  in a single bin, can give a null result even in the presence of large NP
with  tuning between the overall resonance phase and NP, see also footnote \ref{foot:null}.
On the other hand, measurements of both $\Sigma\langle A_{\text{FB}} \rangle$ and $\Delta\langle A_{\text{FB}} \rangle$ in more than a single bin fixes this  and prevents flat directions. Improvements of NP constraints are  therefore  not only  driven by the most precise measurement but rather  by multiple measurements and their correlations with the strong phases.  We emphasize that this assumes negligible  kinematic dependence of the strong phases, as in the ansatz (\ref{eq:BWwc}).

\begin{figure*}[ht!]
    \centering
    \begin{subfigure}[b]{0.44\textwidth}
    \centering
    \includegraphics[width=\textwidth]{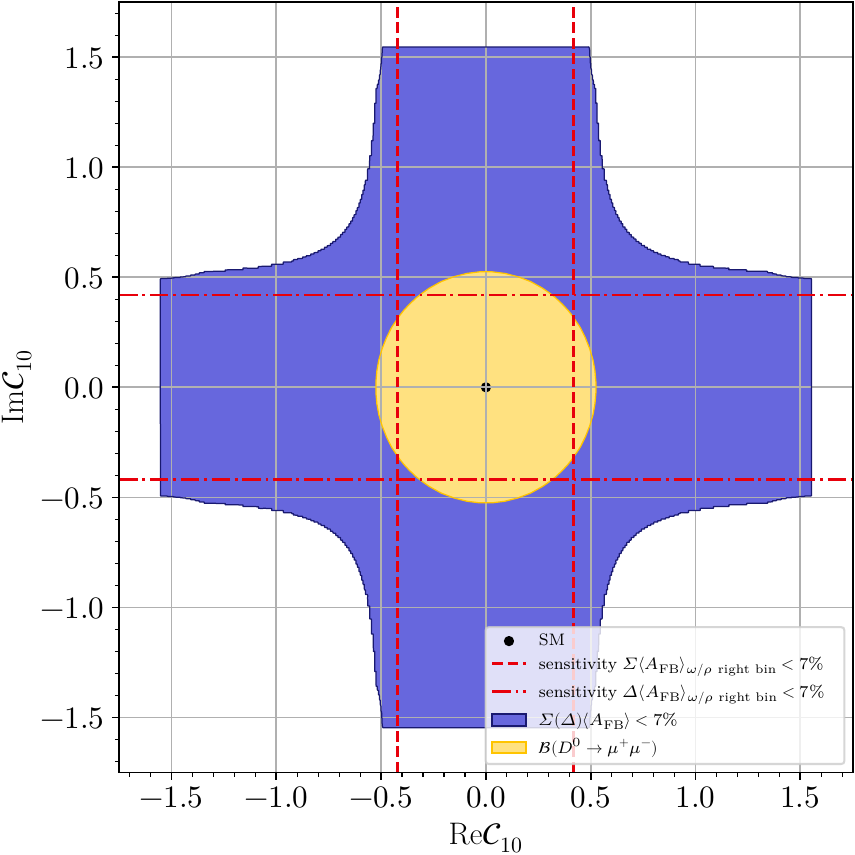}
    \caption{Bins $\omega/\rho$ left, $\omega/\rho$ right, $\phi$ left and $\phi$ right}
    \end{subfigure}
    \begin{subfigure}[b]{0.44\textwidth}
    \centering
    \includegraphics[width=\textwidth]{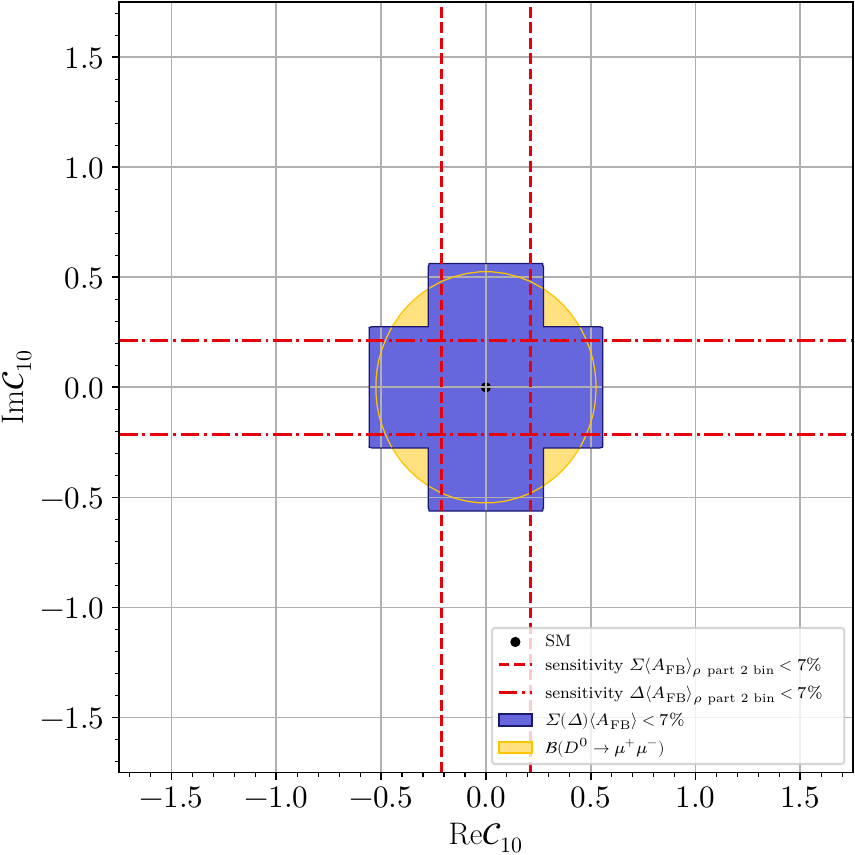}
    \caption{Bins $\omega/\rho$ left optimized, $\rho$ part 2 and $\phi$ right optimized}
    \end{subfigure}
      \caption{
   Allowed region of  (blue area) and sensitivity to  (red lines)    $\mathcal{C}_{10}$ with future measurements of $\Sigma\langle A_{\text{FB}} \rangle$ and  $\Delta\langle A_{\text{FB}} \rangle$  at the level of $7 \%$ in each bin used, as specified in the captions, and 
    with  bins defined in Tab.~\ref{tab:BR_and_AFB_binned}. In the right panel (b) optimized  bins are used which improve the NP reach compared to bins in panel (a).
    The yellow circle denotes the allowed region by current data on $\mathcal{B}(D \to \mu^+ \mu^-)$.}
        \label{fig:AFB_future}
\end{figure*}

In Fig.~\ref{fig:AFB_future} we show the allowed region of  $\mathcal{C}_{10}$ from  current data on $D^0\to \mu^+\mu^-$  (\ref{eq:Dllcons}) (yellow), and compare it to the one from a hypothetical limit on  $\Sigma \langle A_{\text{FB}} \rangle$
and $\Delta \langle A_{\text{FB}} \rangle$  at the level of $7 \%$  (blue) using the bins $\omega/\rho$ left, $\omega/\rho$ right, $\phi$ left and $\phi$ right (left panel (a)) or 
    $\omega/\rho$ left optimized, $\rho$ part 2, and $\phi$ right optimized (right panel (b)) from Tab.~\ref{tab:BR_and_AFB_binned}. 
The symmetry of the blue cross results from  invariance of  $\Sigma \langle A_{\text{FB}} \rangle$
and $\Delta \langle A_{\text{FB}} \rangle$ under a simultaneous sign change of $\mathcal{C}_{10}$ and $\delta_{\rho,\omega,\phi} \to \delta_{\rho,\omega,\phi} + \pi$ \cite{Golz:2021imq}.
Comparison of Fig.~\ref{fig:AFB_future} (a) to (b) confirms  that the optimized bins used in (b) have better  NP reach, with a factor three improvement on the upper limits.
The sensitivity to NP from the hypothetical measurements  of $\Sigma\langle A_{\text{FB}} \rangle$ or $\Delta\langle A_{\text{FB}} \rangle$  is shown by red vertical and horizontal  lines.
As already argued in Sec.~\ref{subsec:fitsNPDPPmumu} for null tests in $D \to \pi^+ \pi^- \mu^+ \mu^-$, 
the upper limits on the Wilson coefficients  are larger than the sensitivity.

\subsection{Complementarity  \label{sec:comp}}

We analyze  the complementarity   between the decay modes for the NP coefficients
$\mathrm{Re}\left\{\mathcal{C}_{10}\pm \mathcal{C}_{10}^{\prime}\right\}$ of the  GIM-protected operators $\mathcal{O}_{10}, \mathcal{O}_{10}^\prime$.
They enter the observables as follows
 \begin{align} \nonumber
\mathcal{B}(D \to \mu^+ \mu^-) &\propto |\mathcal{C}_{10} -\mathcal{C}_{10}^{\prime}|^2 \, ,  \\  \nonumber
\mathcal{B}(D \to \pi \mu^+ \mu^-)_{\text{NP}}  &\propto |\mathcal{C}_{10} +\mathcal{C}_{10}^{\prime}|^2 \, ,  \\ \nonumber
\mathcal{B}(\Lambda_c \to p \mu^+ \mu^-)_{\text{NP}}  &\propto |\mathcal{C}_{10} \pm \mathcal{C}_{10}^{\prime}|^2  \, ,  \\  \label{eq:obs}
 \Sigma \langle A_{\text{FB}}  (\Lambda_c \to p \mu^+ \mu^-)\rangle & \propto \mathrm{Re} \,  \mathcal{C}_{10} \, , \\ \nonumber
 \Delta \langle A_{\text{FB}}  (\Lambda_c \to p \mu^+ \mu^-)\rangle & \propto \mathrm{Im} \,  \mathcal{C}_{10} \, , \\ \nonumber
 I_{7} (D \to \pi \pi \mu^+ \mu^-) &\propto  \mathrm{Re,Im}\left\{\mathcal{C}_{10} - \mathcal{C}_{10}^{\prime}\right\}  \, , \\
  I_{5,6}(D \to \pi \pi \mu^+ \mu^-)   &\propto   \mathrm{Re,Im}\left\{\mathcal{C}_{10} \pm  \mathcal{C}_{10}^{\prime}\right\}    \, . \nonumber
\end{align}
For the angular observables we only give those terms  that interfere with $\mathcal{C}_9^R$ to indicate  resonance enhancement  and  omit  here possible subleading NP effects  in the denominators, that is, the branching ratios. We include them however in the numerics to prohibit  artifacts in the fit.
Note that then angular observables (\ref{eq:obs}) are linear in NP, whereas branching ratios are quadratic, making the ratio-type observables more sensitive to smaller NP coefficients.
If both $\pm$ signs are present their relative strength depends on the hierarchy and interplay of transversity amplitudes, which depend on form factors and the $q^2$-region.

\begin{figure*}[ht!]
    \centering
\includegraphics[width=0.70\textwidth]{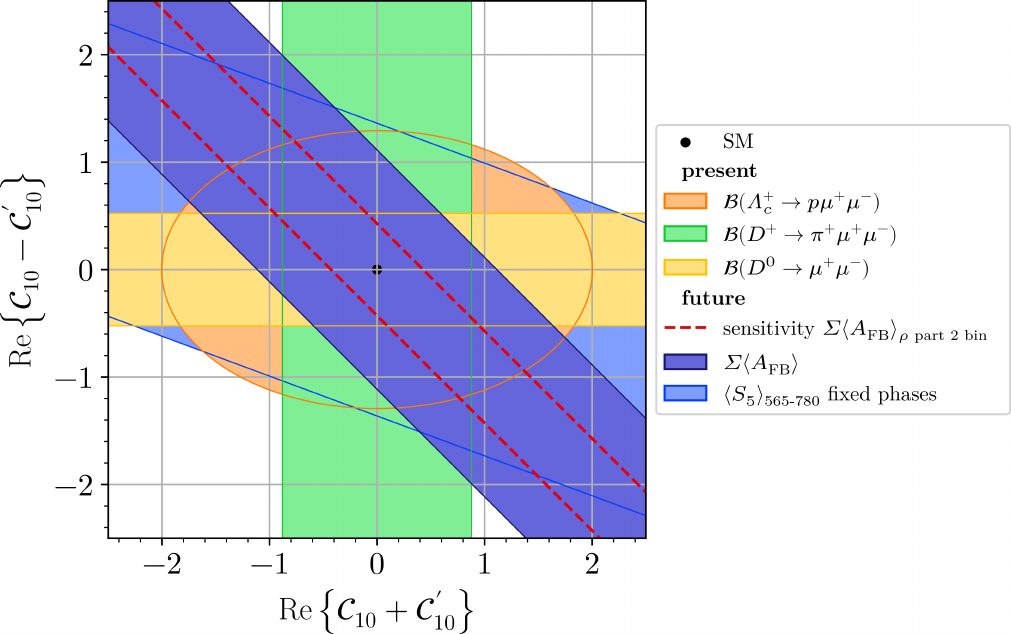}
    \caption{Complementarity of constraints in  $\mathrm{Re}\left\{\mathcal{C}_{10}\pm \mathcal{C}_{10}^{\prime}\right\}$ 
    from current branching ratio data  on
$D \to \mu^+ \mu^-$  (yellow), $D^+ \to \pi^+ \mu^+ \mu^-$ (green) and $\Lambda_c \to p \mu^+ \mu^-$  (orange), together with  hypothetical future  measurements of
    $ \Sigma \langle A_{\text{FB}} \rangle$ at 7\% level (dark blue) and  $\langle S_5 \rangle $ at  $0. 7\%$   (light blue), see text.}\label{fig:AFB_future_complimentary}
    \label{fig:complement}
\end{figure*}

Current constraints on the branching ratios of
$D \to \mu^+ \mu^-$ (yellow), $D^+ \to \pi^+ \mu^+ \mu^-$ (green) and $\Lambda_c \to p \mu^+ \mu^-$  (orange) are shown  in Fig.~\ref{fig:complement}.
Also shown is the impact of future null test measurements $ \Sigma \langle A_{\text{FB}} \rangle$ at 7\% level (dark blue), as in Fig.~\ref{fig:AFB_future},
and the $D \to \pi^+ \pi^- \mu^+ \mu^-$-observable $\langle S_5 \rangle $ at   $0. 7\%$  for illustration  in  the $\sqrt{q^2}$-bin $[0.565\text{-}0.780]\,\mathrm{GeV}$~(light blue)
 for fixed  strong phases 
$\delta_{P,\phi} = \delta_{S,\phi} = 2/5 \pi$ and $\delta_i = \bar{\delta}_{i, \text{best-fit 4}} + \delta_{P,\phi}$ for all others, 
assuming they are known from other observables of the 4-body decay. Therefore, SM uncertainites are significantly  reduced and the upper limits are close to the sensitivity to NP.
Note, other $q^2$ bins would exhibit  different slopes in the $\mathrm{Re}\left\{\mathcal{C}_{10}\pm \mathcal{C}_{10}^{\prime}\right\}$-plane.
We illustrate the potential of $D \to \pi^+ \pi^- \mu^+ \mu^-$ with the angular coefficient $I_5$, because
$I_7$ is sensitive to the difference of  $\mathcal{C}_{10}$ and $\mathcal{C}_{10}^{\prime}$, just  like the $D \to \mu^+ \mu^-$ branching ratio,  and  $I_{6}$ probes also  both linear combinatons, although with $I_6 \sim H_\parallel H^*_\perp$ instead of  $I_5 \sim H_0 H^*_\perp$, see (\ref{eq:IH}), implying different form factors.
As apparent from Fig.~\ref{fig:complement}, each observable  shown gives different type of constraints, demonstrating  synergies between different decays.

In a $SU(2)_L$-invariant setting, strong constraints from kaon decays on $\mathcal{C}_9- \mathcal{C}_{10}$ from dineutrino decays and  $\mathcal{C}_9+ \mathcal{C}_{10}$  
for decays to 
dimuons exist~\cite{Bause:2020auq}. 
Sizable NP in charm is then  limited to the right-handed quark sector, here in  $\mathcal{C}_{10}^{\prime}$.
In this case $I_{5,6,7}$  in $D \to \pi^+ \pi^-  \mu^+ \mu^-$  are still resonance-enhanced, whereas $A_{\text{FB}}$  in $\Lambda_c \to p \mu^+ \mu^-$  decays is not.

\section{Conclusions}\label{sec:conclusions}

We perform the first  global analysis of rare charm decays using multiple decays and observables to exploit  complementarities to strengthen  the  bounds on
$|\Delta c|=|\Delta u|=1$ NP coefficients. The outcome of  the 1D and 2D fits are given in Tab.~\ref{tab:fitres_1d} and Tab.~\ref{tab:fitres_2d}, respectively.
They are  based on branching ratio data
of $D \to \mu^+ \mu^-$, $D \to \pi \mu^+ \mu^-$, $D \to \pi^+ \pi^-  \mu^+ \mu^-$  and  $\Lambda_c \to p \mu^+ \mu^-$ decays, and  for the first time, angular observables,
including the null test  coefficients
$I_{5,6,7}$ of  $D \to \pi^+ \pi^-  \mu^+ \mu^-$    decays. Let us summarize status and findings:

\begin{enumerate}[leftmargin=*]
    \item Null test observables 
are induced by NP, and amplified by  interference with  SM resonance amplitudes, see Fig.~\ref{fig:nulltests_binned}. Hence, the null tests
exhibit resonance structure, and binning is key, as demonstrated in  Fig.~\ref{fig:AFB_future} for a future measurement of the  forward-backward asymmetry in $\Lambda_c \to p \mu^+ \mu^-$  decays.
Choosing optimized  $q^2$-bins improves the limits on NP  here by a factor of three.

\item While a null test can cleanly signal NP, an interpretation in terms of Wilson coefficients relies on the  understanding of
decay amplitudes.
Requisite hadronic parameters of semileptonic $D \to \pi \mu^+ \mu^-$, $D \to \pi^+ \pi^-  \mu^+ \mu^-$  and  $\Lambda_c \to p \mu^+ \mu^-$  decays, which cannot be obtained with sufficient precision from other means, are extracted from  observables which are largely
insensitive to NP. These are generically differential branching ratios, with their characteristic resonance peaks as in Fig.~\ref{fig:Lc-br}.

\item  We perform fits for $D \to \pi^+ \pi^-  \mu^+ \mu^-$ decays in ten model scenarios to  advance the theoretical description.
We find, in agreement with \cite{Fajfer:2023tkp},  but in more general settings,  that  the measured differential decay spectra  of  $D \to \pi^+ \pi^-  \mu^+ \mu^-$  in the dimuon and dipion mass \cite{LHCb:2021yxk}   are difficult to match theoretically using scalar and vector resonances alone, see Fig.~\ref{fig:SMfit_dGamma_dx_compare}.
An interesting outcome is that the fit with the best $\chi^2/\text{dof}$, scenario 10, favors an effective contribution to the matrix element of $\mathcal{O}_9$ of about $r_9 \sim 2$,
 with an order one phase.
Further study is desirable but beyond the scope of this work.
The constraining-NP power of  the 4-body decays, despite its many GIM-protected null tests, is limited due to the complicated hadronic structure involving a large number of  parameters, inducing significant hadronic uncertainties. In this way, $D \to \pi^+ \pi^-  \mu^+ \mu^-$  is also  a laboratory for hadronic dynamics, and probes different QCD frameworks, similar to the more simple radiative 
$D \to \pi \pi \gamma$ decays \cite{Adolph:2020ema}.
Improved data on the double differential decay rate $d^2 \Gamma/dp^2 dq^2$, including correlations provided by the experiment, would help here, as well as further modelling efforts.

\item Current constraints on NP  are hence dominated by 2-body $D \to \mu^+ \mu^-$ (if applicable, i.e. probing $\mathcal{C}_{10}-\mathcal{C}_{10}^\prime$), and the 3-body decays, see  Figs.~\ref{fig:dipoles} and~\ref{fig:2dfit_contours},
which displays synergies and added value between decay modes.

\item The decay  $\Lambda_c \to p \mu^+ \mu^-$ is the  rising star with the simplicity of a 3-body decays  and fewer hadronic parameters than multi-body decays 
however richer angular structure than $D \to \pi \mu^+ \mu^-$. In particular, the baryonic  decay features null test  distributions,
and has  high sensitivity to radiative  dipole operators  at  low $q^2$ via $\Lambda_c \to p \gamma^*$  \cite{Golz:2021imq}.

\item Combination of various decay modes and observables allows to exploit complementarity, see Fig.~\ref{fig:complement}.
 Due to its unique sensitivity to NP couplings in the angular distributions, $D \to \pi^+ \pi^-  \mu^+ \mu^-$  decays have potential to become
very  important for searches  in the high luminosity era.
We  hope to come back to a future version of the global  fit with improved data, also on dielectron modes.
Rare charm decays  are suitable for study at 
LHCb~\cite{Cerri:2018ypt}, Belle II~\cite{Belle-II:2018jsg},  and  future colliders ~\cite{Lyu:2021tlb,FCC:2018byv}, as well as high-$p_T$ experiments such as CMS \cite{CMS-PAS-BPH-23-008}.
\end{enumerate}

\acknowledgments

We thank Rigo Bause, Stefan de Boer, Marcel Golz, Tom Magorsch, Dominik Mitzel, Lara Nollen, Luiz Vale Silva and Eleftheria Solomonidi for useful discussions. The work of HG is supported by the European Union -- NextGeneration EU and by the University of Padua under the 2021 STARS Grants@Unipd programme (Acronym and title of the project: CPV-Axion -- Discovering the CP-violating axion)
as well as by the European Union -- Next Generation EU and by the Italian Ministry of University and Research (MUR) via the PRIN 2022 project n.~2022K4B58X -- AxionOrigins.
GH would like to thank the CERN Theory Department  for kind  hospitality and support during the finalization of this work.

\newpage
\appendix

\section{Form factors}\label{app:FFs}

Here we present the non-resonant and resonant form factors entering in Eqs.~\eqref{eq:I1}-\eqref{eq:I9}, Sections~\ref{app:noresFFs} and \ref{app:resFFs}, respectively.
In general the transversity form factors $\mathcal{F}_{0,\parallel,\perp}$ can be expanded in terms of the 
associated Legendre polynomials $P_\ell^m(\cos \theta_{P_1})$ and coefficients $F_{(0,\parallel,\perp),\,\ell}(q^2,p^2)$:
\begin{equation}\label{eq:FFsLeg}
\begin{aligned}
    \mathcal{F}_0 &= \sum_{\ell=0}F_{0,\,\ell}(q^2,p^2)\, P_\ell^{m=0}(\cos \theta_{P_1})\,, \\
    \mathcal{F}_\parallel &= \sum_{\ell=1}F_{\parallel,\,\ell}(q^2,p^2)\, \frac{P_\ell^{m=1}(\cos \theta_{P_1})}{\sin\theta_{P_1}} \,, \\
    \mathcal{F}_\perp &= \sum_{\ell=1}F_{\perp,\,\ell}(q^2,p^2)\, \frac{P_\ell^{m=1}(\cos \theta_{P_1})}{\sin\theta_{P_1}}\,,
\end{aligned}
\end{equation}
where the coefficients $F_{(0,\parallel,\perp)\,\ell}(q^2,p^2)$ can be calculated via the completeness relation \begin{equation}\label{eq:compFFs}
    \begin{aligned}
        F_{0,\,\ell} (q^2,p^2) &= \frac{2\ell+1}{2} \int_{-1}^1 \mathcal{F}_0 (q^2,p^2,\cos\theta_{P_1})\, P_\ell^0 (\cos \theta_{P_1}) \,\mathrm{d}\!\cos\theta_{P_1} \:,\\
        F_{(\parallel,\perp),\,\ell} (q^2,p^2) &= \frac{2\ell+1}{2} \frac{(\ell-1)!}{(\ell+1)!} \int_{-1}^1
         \mathcal{F}_{\parallel,\perp} (q^2,p^2,\cos\theta_{P_1})\,P_\ell^1 (\cos \theta_{P_1}) \sin \theta_{P_1} \,\mathrm{d}\!\cos\theta_{P_1} \:.
    \end{aligned}
\end{equation}

\subsection{Non-resonant form factors}\label{app:noresFFs}

The non-resonant transversity form factors are given by\footnote{To speed up the numerical integration, 
we expand the non-resonant transversity form factors $\mathcal{F}_{0,\parallel,\perp}$ in terms of $P_\ell^m(\cos \theta_{P_1})$ and $F_{(0,\parallel,\perp)\ell}(q^2,p^2)$ using Eqs.~\eqref{eq:FFsLeg}. This allows for analytical integration over $\cos\theta_{P_1}$.
The coefficients $F_{(0,\parallel,\perp)\ell}(q^2,p^2)$ are computed using the completeness relation~\eqref{eq:compFFs}. Depending on the integration range of the observable, we expand to different orders in $\ell$. This approximation is particularly effective for integrals of the form $\int_{-1}^{+1} f(\cos\theta_{P_1})\,\mathrm{d}\cos\theta_{P_1}$, where we consider $\ell \leq 3$. For $\left[\int_{0}^{+1} - \int_{-1}^{0}\right] f(\cos\theta_{P_1})\,\mathrm{d}\cos\theta_{P_1}$, we require higher orders with $\ell \leq 6$.}
\begin{equation}
\label{eq:transversity}
\begin{aligned}
    \mathcal{F}_0         &= \frac{\mathcal{N}_{\text{nr}}}{2} \left[
        \sqrt{\lambda_D}\, w_+ + \frac{1}{p^2} \left((m_{P_1}^2-m_{P_2}^2)\sqrt{\lambda_D} - (m_D^2-q^2-p^2)\,\sqrt{\lambda_h}\,\cos\theta_{P_1}\right) w_-
    \right] \:,\\
    \mathcal{F}_\perp     &= \frac{\mathcal{N}_{\text{nr}}}{2} \sqrt{\lambda_D \,\lambda_h \,\frac{q^2}{p^2}}\, h \:,\quad\quad
    \mathcal{F}_\parallel = \mathcal{N}_{\text{nr}} \sqrt{\lambda_h\,\frac{q^2}{p^2}}\, w_-\:, \\
    \mathcal{N}_{\text{nr}} &= \frac{G_F\,\alpha_e}{2^7 \,\pi^4\, m_D} \sqrt{\pi\, \frac{\sqrt{\lambda_D\,\lambda_h}}{m_D\, p^2}} \:,
\end{aligned}
\end{equation}
with $\lambda_D \equiv \lambda(m_D^2,q^2,p^2)$ and $\lambda_h \equiv \lambda(p^2,m_{P_1}^2,m_{P_2}^2)$ where  $\lambda(a,b,c)\equiv a^2+b^2+c^2-2\,(ab+ac+bc)$ is the Källén function. For the form factors $w_\pm$ and $h$ we refer to \cite{DeBoer:2018pdx}.

\subsection{Resonant form factors}\label{app:resFFs}

In the following subsections, we present the P- and S-wave components $F_{(0,\parallel,\perp),\,P(S)}(q^2,p^2)$ of the resonant transversity form factors $\mathcal{F}_{0,\,P(S)}^{\,\text{res}}$, related via Eq. \eqref{eq:FFsLeg}~.

\subsubsection{P-wave}

For the P-wave, we use the transversity form factors from Refs.~\cite{DeBoer:2018pdx,Fajfer:2023tkp}:
\begin{align}
 &F_{0,\,P}^{\text{res}}=-3 \sqrt{\frac{3-\beta_\ell^2}{2}} N_V(p^2,q^2)\frac{(m_D^2-p^2-q^2)(m_D+m_{\rho^0})^2A_1(q^2)-\lambda_DA_2(q^2)}{2(m_D+m_{\rho^0})\sqrt{p^2q^2} P_{\rho^0/\omega}(p^2)}\,,\nonumber\\
 &F_{\parallel,\,P}^{\text{res}}=-\frac3{\sqrt2}N_V(p^2,q^2) \sqrt{\frac{3-\beta_\ell^2}{2}} \sqrt2(m_D+m_{\rho^0})\frac{A_1(q^2)} {P_{\rho^0/\omega}(p^2)}\,,\\
 &F_{\perp,\,P}^{\text{res}}=\frac3{\sqrt2}N_V(p^2,q^2)\beta_\ell \frac{\sqrt{2\lambda_D }}{m_D+m_{\rho^0}}\frac{ V(q^2) }{P_{\rho^0/\omega}(p^2)}\,, \nonumber
\end{align}
with 
\begin{align}
 &N_V=G_F\alpha_e\sqrt{\frac{ \beta_\ell\, q^2\,\sqrt{\lambda_h\,\lambda_D }}{3\,(4\pi)^5\,m_D^3\,p^2}}\,, \quad \beta_\ell=\sqrt{1- \frac{4\, m_\ell^2}{q^2}} \,.
\end{align}
Compared to the procedure employed in Ref.~\cite{DeBoer:2018pdx}, the masses $m_V^2$ are replaced by $p^2$, which is suitable for off-resonance effects following Ref.~\cite{Fajfer:2023tkp}. We take the $D\to \rho$ form factors $A_{1,2}, V$ from Ref.~\cite{Melikhov:2000yu} and neglect
differences to the $D\to \omega$ form factors.

For the $p^2$ line shape $P_{\rho^0/\omega}(p^2)$ we use the Gounaris-Sakurai line shape employed by Refs.~\cite{BESIII:2018qmf,Fajfer:2023tkp}:
\begin{equation} \label{eq:PwaveLineshape}
    \frac{1}{P_{\rho^0/\omega}(p^2)} = \frac{1}{\sqrt{\pi} } \frac{p^\ast(p^2)}{p^{\ast}_0} \frac{B(p^*) }{ B(p_0^*) } \frac{1}{P_{\rho^0}(p^2)} \left( 1 + a_{p^2,\omega} \, e^{i \delta_{p^2,\omega}} \, \text{RBW}_\omega (p^2) \right) \,,
\end{equation}
with $p^\ast (p^2) = \sqrt{\lambda (p^2,m_\pi^2,m_\pi^2)} / ( 2 \sqrt{p^2} )$, and $ p^\ast_0 = p^\ast (m^2_{\rho^0}) $.
For the definitions of the functions  in Eq.~\eqref{eq:PwaveLineshape} see  Ref.~\cite{Fajfer:2023tkp}.

\subsubsection{S-wave}

For the S-wave expressions, we use Refs.~\cite{BESIII:2024lnh,Fajfer:2023tkp}
\begin{align}
        F_{0,\,S}^{\text{res}} &= - N \frac{\sqrt{\beta_\ell\,(3-\beta_\ell^2)}\lambda_h^{1/4}\lambda_D^{3/4} }{2\sqrt{2}\sqrt{p^2}} \frac{\tilde{f}_+(q^2)}{P_\sigma(p^2)} \,,\quad
        N = \frac{\alpha_e\,G_F}{128\pi^{7/2}m_D^{3/2}} \,, \quad
        \tilde{f}_+(q^2) &= \frac{1}{1-\frac{q^2}{m_A^2}} 
\end{align}
with $m_A=2.42\,\text{GeV}$. We use the same line shape $P_\sigma(p^2) = 1/\mathcal{A}_S(p^2)$ for the $\sigma$-resonance as in Ref.~\cite{Fajfer:2023tkp}.

\section[\texorpdfstring{$\boldsymbol{D \to  P_1 P_2 \,\ell^+ \ell^-}$  angular observables}{D -> P_1 P_2  l+ l- angular observables}]{$\boldsymbol{D \to  P_1 P_2 \, \ell^+ \ell^-}$  angular observables}\label{app:full}

We expand the angular observables  of $D\,\to\, P_1\, P_2\, \ell^+\,\ell^-$ in terms of the resonant S-, P-wave and non-resonant transversity 
form factors.
Using Eq.~\eqref{eq:Hamp} for the transversity amplitudes the angular coefficients given by Eq.~\eqref{eq:IH} become\footnote{For simplicity, we do not give the explicit contribution from dipole operators $\mathcal{C}_7^{(\prime)}$, but they can be included via the following shift $\mathcal{C}_\pm^{L,R}(q^2) \to \mathcal{C}_\pm^{L,R}(q^2)\,+\,\kappa\, \frac{2m_c m_D}{q^2}\,(\mathcal{C}_7\pm\mathcal{C}_7^{\prime})$.}
\begin{align}\label{eq:I1}
    I_1\,&=\,
+\frac{1}{8} \left|\mathcal{C}_9^{\,\mathcal{R},\,S}\right|^2 \left|\mathcal{F}_{0,\,S}^{\,\text{res}}\right|^2
+\left| \mathcal{C}_9^{\,\mathcal{R},\,P}\right|^2 \left(\frac{1}{8}\left|\mathcal{F}_{0,\,P}^{\,\text{res}}\right|^2
+\frac{3}{16} \left(\left|\mathcal{F}_{\parallel,\,P}^{\,\text{res}}\right|^2+\left|\mathcal{F}_{\perp,\,P}^{\,\text{res}}\right|^2\right)
\sin^2\theta_{P_1}\right)\nonumber
\\ &\quad  \nonumber
+\frac{1}{4}{\rm Re} \left\{ \mathcal{C}_9^{\mathcal{R},P} \left(\mathcal{C}_9^{\mathcal{R},S}\right)^\ast \right\}\, {\rm Re}[\mathcal{F}_{0,\,S}^{\,\text{res}} \mathcal{F}_{0,\,P}^{\,\text{res},\ast}]
+\frac{1}{4}{\rm Im} \left\{ \mathcal{C}_9^{\mathcal{R},P} \left(\mathcal{C}_9^{\mathcal{R},S}\right)^\ast \right\}\, {\rm Im}[\mathcal{F}_{0,\,S}^{\,\text{res}} \mathcal{F}_{0,\,P}^{\,\text{res},\ast}]
\\ &\quad
+\frac{3}{16} \rho_1^+ \left|\mathcal{F}_{\perp}\right|^2 \sin^2\theta_{P_1}
+\rho_1^-\, \left(\frac{1}{8}\left|\mathcal{F}_{0}\right|^2
+\frac{3}{16} \left|\mathcal{F}_{\parallel}\right|^2
\sin^2\theta_{P_1}\right)
\\ &\quad \nonumber
-\frac{1}{4} {\rm Im} \rho_{3,\,S}^-\, {\rm Im}[\mathcal{F}_{0,\,S}^{\,\text{res}} \mathcal{F}_{0}^{\,\ast}]
+\frac{1}{4} {\rm Re}\rho_{3,\,S}^-\, {\rm Re}[\mathcal{F}_{0,\,S}^{\,\text{res}} \mathcal{F}_{0}^{\,\ast}] 
\\  &\quad  \nonumber
+\frac{3}{8} \left( {\rm Re} \rho_{3,\,P}^+ {\rm Re}[\mathcal{F}_{\perp,\,P}^{\,\text{res}} 
\mathcal{F}_{\perp}^{\,\ast}] 
-{\rm Im} \rho_{3,\,P}^+ {\rm Im}[\mathcal{F}_{\perp,\,P}^{\,\text{res}} \mathcal{F}_{\perp}^{\,\ast}] 
\right) \sin^2\theta_{P_1} 
\\ &\quad \nonumber
+{\rm Im} \rho_{3,\,P}^-\, \left(-\frac{1}{4} {\rm Im}[\mathcal{F}_{0,\,P}^{\,\text{res}} \mathcal{F}_{0}^{\,\ast}]
-\frac{3}{8} {\rm Im}[\mathcal{F}_{\parallel,\,P}^{\,\text{res}}\mathcal{F}_{\parallel}^{\,\ast}] \sin^2\theta_{P_1}\right)
\\ &\quad \nonumber
+{\rm Re} \rho_{3,\,P}^-\, \left(\frac{1}{4} {\rm Re}[\mathcal{F}_{0,\,P}^{\,\text{res}} \mathcal{F}_{0}^{\,\ast}]
+\frac{3}{8}{\rm Re}[\mathcal{F}_{\parallel,\,P}^{\,\text{res}} \mathcal{F}_{\parallel}^{\,\ast}] \sin^2\theta_{P_1}\right) 
    \,, \\
%
I_2\,&=\, 
-\frac{1}{8} \left|\mathcal{C}_9^{\,\mathcal{R},\,S}\right|^2 \left|\mathcal{F}_{0,\,S}^{\,\text{res}}\right|^2
+\left| \mathcal{C}_9^{\,\mathcal{R},\,P}\right|^2 \left(
    -\frac{1}{8} \left|\mathcal{F}_{0,\,P}^{\,\text{res}}\right|^2
+\frac{1}{16}
\left(\left|\mathcal{F}_{\parallel,\,P}^{\,\text{res}}\right|^2+\left|\mathcal{F}_{\perp,\,P}^{\,\text{res}}\right|^2\right) \sin^2\theta_{P_1}\right)
\nonumber\\ &\quad \nonumber
-\frac{1}{4} {\rm Re} \left\{ \mathcal{C}_9^{\mathcal{R},P} \left(\mathcal{C}_9^{\mathcal{R},S}\right)^\ast \right\}\, {\rm Re}[\mathcal{F}_{0,\,S}^{\,\text{res}} \mathcal{F}_{0,\,P}^{\,\text{res},\ast}]
-\frac{1}{4} {\rm Im} \left\{ \mathcal{C}_9^{\mathcal{R},P} \left(\mathcal{C}_9^{\mathcal{R},S}\right)^\ast \right\}\, {\rm Im}[\mathcal{F}_{0,\,S}^{\,\text{res}} \mathcal{F}_{0,\,P}^{\,\text{res},\ast}]
\\ &\quad \nonumber
+\frac{1}{16}
\rho_1^+ \left|\mathcal{F}_{\perp}\right|^2 \sin^2\theta_{P_1}
+\rho_1^-\, \left(
    -\frac{1}{8}
\left|\mathcal{F}_{0}\right|^2
+\frac{1}{16} \left|\mathcal{F}_{\parallel}\right|^2 \sin^2\theta_{P_1}\right)
\\  &\quad 
-\frac{1}{4}
{\rm Re}\rho_{3,\,S}^-\, {\rm Re}[\mathcal{F}_{0,\,S}^{\,\text{res}} \mathcal{F}_{0}^{\,\ast}]
+\frac{1}{4}
{\rm Im} \rho_{3,\,S}^-\, {\rm Im}[\mathcal{F}_{0,\,S}^{\,\text{res}} \mathcal{F}_{0}^{\,\ast}]
\\ &\quad \nonumber
+\frac{1}{8}\left(
 {\rm Re} \rho_{3,\,P}^+ {\rm Re}[\mathcal{F}_{\perp,\,P}^{\,\text{res}} \mathcal{F}_{\perp}^{\,\ast}] 
- {\rm Im} \rho_{3,\,P}^+ {\rm Im}[\mathcal{F}_{\perp,\,P}^{\,\text{res}} \mathcal{F}_{\perp}^{\,\ast}]
\right) \sin^2\theta_{P_1}
\\ &\quad \nonumber
+{\rm Im} \rho_{3,\,P}^-\, \left(\frac{1}{4} {\rm Im}[\mathcal{F}_{0,\,P}^{\,\text{res}}
\mathcal{F}_{0}^{\,\ast}]
-\frac{1}{8} {\rm Im}[\mathcal{F}_{\parallel,\,P}^{\,\text{res}} \mathcal{F}_{\parallel}^{\,\ast}] \sin^2\theta_{P_1}\right)
\\ &\quad \nonumber
+{\rm Re} \rho_{3,\,P}^-\, \left(-\frac{1}{4}{\rm Re}[\mathcal{F}_{0,\,P}^{\,\text{res}} \mathcal{F}_{0}^{\,\ast}]
+\frac{1}{8} {\rm Re}[\mathcal{F}_{\parallel,\,P}^{\,\text{res}} \mathcal{F}_{\parallel}^{\,\ast}] \sin^2\theta_{P_1}\right)
    \,,\nonumber \\
%
    I_3\,&=\, \biggl[
+\frac{1}{8} \rho_1^+ \left|\mathcal{F}_{\perp}\right|^2
-\frac{1}{8}
\rho_1^-\, \left|\mathcal{F}_{\parallel}\right|^2 
-\frac{1}{8} \left| \mathcal{C}_9^{\,\mathcal{R},\,P}\right|^2 (\left|\mathcal{F}_{\parallel,\,P}^{\,\text{res}}\right|^2-\left|\mathcal{F}_{\perp,\,P}^{\,\text{res}}\right|^2) \nonumber
\\ &\quad
+\frac{1}{4} {\rm Im} \rho_{3,\,P}^-\, {\rm Im}[\mathcal{F}_{\parallel,\,P}^{\,\text{res}} \mathcal{F}_{\parallel}^{\,\ast}] 
-\frac{1}{4} {\rm Im} \rho_{3,\,P}^+ {\rm Im}[\mathcal{F}_{\perp,\,P}^{\,\text{res}}\mathcal{F}_{\perp}^{\,\ast}] 
\\ &\quad \nonumber
-\frac{1}{4} {\rm Re} \rho_{3,\,P}^-\, {\rm Re}[\mathcal{F}_{\parallel,\,P}^{\,\text{res}} \mathcal{F}_{\parallel}^{\,\ast}] 
+\frac{1}{4} {\rm Re} \rho_{3,\,P}^+ {\rm Re}[\mathcal{F}_{\perp,\,P}^{\,\text{res}} \mathcal{F}_{\perp}^{\,\ast}] 
\biggr] \sin^2\theta_{P_1} 
    \,, \nonumber
\end{align}

\begin{align}
        I_4\,&=\,\biggl[
-\frac{1}{4} \left| \mathcal{C}_9^{\,\mathcal{R},\,P}\right|^2 {\rm Re}[\mathcal{F}_{0,\,P}^{\,\text{res}} \mathcal{F}_{\parallel,\,P}^{\,\text{res},\ast}]
-\frac{1}{4} \rho_1^-\, {\rm Re}[\mathcal{F}_{0} \mathcal{F}_{\parallel}^{\,\ast}] \nonumber
\\ &\quad \nonumber
+\frac{1}{4} {\rm Im} \left\{ \mathcal{C}_9^{\mathcal{R},P} \left(\mathcal{C}_9^{\mathcal{R},S}\right)^\ast \right\} \,
{\rm Im}[\mathcal{F}_{\parallel,\,P}^{\,\text{res}} \mathcal{F}_{0,\,S}^{\,\text{res},\ast}] 
-\frac{1}{4} {\rm Re} \left\{ \mathcal{C}_9^{\mathcal{R},P} \left(\mathcal{C}_9^{\mathcal{R},S}\right)^\ast \right\} \, {\rm Re}[\mathcal{F}_{0,\,S}^{\,\text{res}} \mathcal{F}_{\parallel,\,P}^{\,\text{res},\ast}] 
\nonumber\\ &\quad
+\frac{1}{4} {\rm Im} \rho_{3,\,P}^-\, ({\rm Im}[\mathcal{F}_{\parallel,\,P}^{\,\text{res}} \mathcal{F}_{0}^{\,\ast}]+{\rm Im}[\mathcal{F}_{0,\,P}^{\,\text{res}}
\mathcal{F}_{\parallel}^{\,\ast}]) 
\\ &\quad \nonumber
+\frac{1}{4} {\rm Im} \rho_{3,\,S}^-\, {\rm Im}[\mathcal{F}_{0,\,S}^{\,\text{res}} \mathcal{F}_{\parallel}^{\,\ast}] 
-\frac{1}{4} {\rm Re}\rho_{3,\,S}^-\, {\rm Re}[\mathcal{F}_{0,\,S}^{\,\text{res}}
\mathcal{F}_{\parallel}^{\,\ast}] 
\\ &\quad \nonumber
+\frac{1}{4} {\rm Re} \rho_{3,\,P}^-\, (-{\rm Re}[\mathcal{F}_{\parallel,\,P}^{\,\text{res}}
\mathcal{F}_{0}^{\,\ast}]-{\rm Re}[\mathcal{F}_{0,\,P}^{\,\text{res}} \mathcal{F}_{\parallel}^{\,\ast}]) 
\biggr]\sin\theta_{P_1} \nonumber \\
%
        I_5\,&=\, \biggl[
+{\rm Re} \rho_2^+ {\rm Re}[\mathcal{F}_{0} \mathcal{F}_{\perp}^{\,\ast}] 
+{\rm Im} \rho_2^-\, {\rm Im}[\mathcal{F}_{0} \mathcal{F}_{\perp}^{\,\ast}]
\nonumber\\ &\quad \nonumber
+\frac{1}{2} {\rm Re} \rho_{4,\,P}^+ {\rm Re}[\mathcal{F}_{0,\,P}^{\,\text{res}}
\mathcal{F}_{\perp}^{\,\ast}] 
-\frac{1}{2} {\rm Im} \rho_{4,\,P}^+ {\rm Im}[\mathcal{F}_{0,\,P}^{\,\text{res}} \mathcal{F}_{\perp}^{\,\ast}] 
\nonumber\\ &\quad
+\frac{1}{2} {\rm Re} \rho_{4,\,S}^+ {\rm Re}[\mathcal{F}_{0,\,S}^{\,\text{res}} \mathcal{F}_{\perp}^{\,\ast}] 
-\frac{1}{2} {\rm Im} \rho_{4,\,S}^+{\rm Im}[\mathcal{F}_{0,\,S}^{\,\text{res}} \mathcal{F}_{\perp}^{\,\ast}] 
\\ &\quad \nonumber
+\frac{1}{2} {\rm Re} \rho_{4,\,P}^-\, {\rm Re}[\mathcal{F}_{\perp,\,P}^{\,\text{res}} \mathcal{F}_{0}^{\,\ast}]
-\frac{1}{2}
{\rm Im} \rho_{4,\,P}^-\, {\rm Im}[\mathcal{F}_{\perp,\,P}^{\,\text{res}} \mathcal{F}_{0}^{\,\ast}] 
\biggr] \sin\theta_{P_1}
    \,, \\
%
        I_6\,&=\,\biggl[
{\rm Im} \rho_2^-\, {\rm Im}[\mathcal{F}_{\perp} \mathcal{F}_{\parallel}^{\,\ast}] 
-{\rm Re} \rho_2^+ {\rm Re}[\mathcal{F}_{\parallel}
\mathcal{F}_{\perp}^{\,\ast}] 
\nonumber\\ &\quad
+\frac{1}{2} {\rm Im} \rho_{4,\,P}^-\, {\rm Im}[\mathcal{F}_{\perp,\,P}^{\,\text{res}}
\mathcal{F}_{\parallel}^{\,\ast}] 
-\frac{1}{2} {\rm Re} \rho_{4,\,P}^-\, {\rm Re}[\mathcal{F}_{\perp,\,P}^{\,\text{res}} \mathcal{F}_{\parallel}^{\,\ast}] 
\\ &\quad \nonumber
+\frac{1}{2} {\rm Im} \rho_{4,\,P}^+ {\rm Im}[\mathcal{F}_{\parallel,\,P}^{\,\text{res}} \mathcal{F}_{\perp}^{\,\ast}] 
-\frac{1}{2} {\rm Re} \rho_{4,\,P}^+ {\rm Re}[\mathcal{F}_{\parallel,\,P}^{\,\text{res}} \mathcal{F}_{\perp}^{\,\ast}] 
\biggr] \sin^2\theta_{P_1}
    \,,\nonumber \\
%
    I_7\,&=\, \biggl[
+\delta\rho\, {\rm Im}[\mathcal{F}_{0} \mathcal{F}_{\parallel}^{\,\ast}] +\frac{1}{2} {\rm Re} \rho_{4,\,P}^-\, (-{\rm Im}[\mathcal{F}_{\parallel,\,P}^{\,\text{res}}
\mathcal{F}_{0}^{\,\ast}]+{\rm Im}[\mathcal{F}_{0,\,P}^{\,\text{res}} \mathcal{F}_{\parallel}^{\,\ast}]) 
\nonumber\\ &\quad
+\frac{1}{2} {\rm Im} \rho_{4,\,P}^-\, (-{\rm Re}[\mathcal{F}_{\parallel,\,P}^{\,\text{res}} \mathcal{F}_{0}^{\,\ast}]+{\rm Re}[\mathcal{F}_{0,\,P}^{\,\text{res}}
\mathcal{F}_{\parallel}^{\,\ast}]) 
\\ &\quad \nonumber
+\frac{1}{2} {\rm Re} \rho_{4,\,S}^-\, {\rm Im}[\mathcal{F}_{0,\,S}^{\,\text{res}}
\mathcal{F}_{\parallel}^{\,\ast}] 
+\frac{1}{2} {\rm Im} \rho_{4,\,S}^-\, {\rm Re}[\mathcal{F}_{0,\,S}^{\,\text{res}} \mathcal{F}_{\parallel}^{\,\ast}] \biggr] \sin\theta_{P_1} 
    \,,\nonumber \\
%
        I_8\,&=\,
\biggl[+\frac{1}{4} \left| \mathcal{C}_9^{\,\mathcal{R},\,P}\right|^2 {\rm Im}[\mathcal{F}_{\perp,\,P}^{\,\text{res}} \mathcal{F}_{0,\,P}^{\,\text{res},\ast}]
+\frac{1}{2} {\rm Re} \rho_2^-\, {\rm Im}[\mathcal{F}_{\perp} \mathcal{F}_{0}^{\,\ast}] 
+\frac{1}{2} {\rm Im} \rho_2^+ {\rm Re}[\mathcal{F}_{0}
\mathcal{F}_{\perp}^{\,\ast}] 
\nonumber\\ &\quad \nonumber
+\frac{1}{4} {\rm Re} \left\{ \mathcal{C}_9^{\mathcal{R},P} \left(\mathcal{C}_9^{\mathcal{R},S}\right)^\ast \right\} {\rm Im}[\mathcal{F}_{\perp,\,P}^{\,\text{res}} \mathcal{F}_{0,\,S}^{\,\text{res},\ast}]
+\frac{1}{4} {\rm Im} \left\{ \mathcal{C}_9^{\mathcal{R},P} \left(\mathcal{C}_9^{\mathcal{R},S}\right)^\ast \right\} {\rm Re}[\mathcal{F}_{0,\,S}^{\,\text{res}}
\mathcal{F}_{\perp,\,P}^{\,\text{res},\ast}]
\\ &\quad
-\frac{1}{4} {\rm Im} \rho_{3,\,S}^+ {\rm Re}[\mathcal{F}_{0,\,S}^{\,\text{res}} \mathcal{F}_{\perp}^{\,\ast}] 
-\frac{1}{4} {\rm Re} \rho_{3,\,S}^+ {\rm Im}[\mathcal{F}_{0,\,S}^{\,\text{res}} \mathcal{F}_{\perp}^{\,\ast}]
\\ &\quad \nonumber
+\frac{1}{4} {\rm Im} \rho_{3,\,P}^-\, {\rm Re}[\mathcal{F}_{\perp,\,P}^{\,\text{res}} \mathcal{F}_{0}^{\,\ast}] 
+\frac{1}{4} {\rm Re} \rho_{3,\,P}^-\, {\rm Im}[\mathcal{F}_{\perp,\,P}^{\,\text{res}}
\mathcal{F}_{0}^{\,\ast}] 
\\ &\quad \nonumber
-\frac{1}{4} {\rm Im} \rho_{3,\,P}^+ {\rm Re}[\mathcal{F}_{0,\,P}^{\,\text{res}} \mathcal{F}_{\perp}^{\,\ast}] 
-\frac{1}{4} {\rm Re} \rho_{3,\,P}^+ {\rm Im}[\mathcal{F}_{0,\,P}^{\,\text{res}}
\mathcal{F}_{\perp}^{\,\ast}] 
\biggr] \sin\theta_{P_1} 
    \,,\nonumber \\
%
    I_9\,&=\,- \biggl[
+\frac{1}{4} \left| \mathcal{C}_9^{\,\mathcal{R},\,P}\right|^2 {\rm Im}[\mathcal{F}_{\parallel,\,P}^{\,\text{res}} \mathcal{F}_{\perp,\,P}^{\,\text{res},\ast}]
-\frac{1}{2} {\rm Im} \rho_2^+
{\rm Re}[\mathcal{F}_{\parallel} \mathcal{F}_{\perp}^{\,\ast}] 
+\frac{1}{2} {\rm Re} \rho_2^-\, {\rm Im}[\mathcal{F}_{\parallel} \mathcal{F}_{\perp}^{\,\ast}] 
\nonumber\\ &\quad 
+\frac{1}{4} {\rm Re} \rho_{3,\,P}^+
{\rm Im}[\mathcal{F}_{\parallel,\,P}^{\,\text{res}} \mathcal{F}_{\perp}^{\,\ast}] 
+\frac{1}{4} {\rm Im} \rho_{3,\,P}^+ {\rm Re}[\mathcal{F}_{\parallel,\,P}^{\,\text{res}} \mathcal{F}_{\perp}^{\,\ast}]
\label{eq:I9}\\ &\quad \nonumber
-\frac{1}{4} {\rm Im} \rho_{3,\,P}^-\, {\rm Re}[\mathcal{F}_{\perp,\,P}^{\,\text{res}} \mathcal{F}_{\parallel}^{\,\ast}] 
-\frac{1}{4} {\rm Re} \rho_{3,\,P}^-\, {\rm Im}[\mathcal{F}_{\perp,\,P}^{\,\text{res}} \mathcal{F}_{\parallel}^{\,\ast}]
 \biggr]\sin^2\theta_{P_1} 
    \,,
\end{align}
where the short-distance effects are encoded in 
\begin{align} \nonumber
    \rho_1^\pm & =\frac{1}{2}\,\left[ |\mathcal{C}_\pm^L|^2 + |\mathcal{C}_\pm^R|^2  \right]= \left| \mathcal{C}_9^{\,\text{eff}} + \mathcal{C}_9 \pm  \mathcal{C}_9^{\,\prime} \right|^2 + |\mathcal{C}_{10}  \pm  \mathcal{C}_{10}^{\,\prime}|^2 \, , \\ \nonumber
    \delta \rho & =\frac{1}{4}\,\left[ |\mathcal{C}_-^R|^2 - |\mathcal{C}_-^L|^2  \right] = {\rm Re}\left[ \left(\mathcal{C}_9^{\,\text{eff}} + \mathcal{C}_9 -  \mathcal{C}_9^{\,\prime}  \right)\left(\mathcal{C}_{10}  -  \mathcal{C}_{10}^{\,\prime}\right)^* \right]  \, , \\ \nonumber
    {\rm Re} \rho_2^+ & =\frac{1}{4}\,{\rm Re}\left[   \mathcal{C}_+^R (\mathcal{C}_-^{R})^* - \mathcal{C}_-^L (\mathcal{C}_+^{L})^*\right]={\rm Re} \left[  (\mathcal{C}_9^{\,\text{eff}} + \mathcal{C}_9)\,   \mathcal{C}_{10}^{*}  - \mathcal{C}_9^{\,\prime}\,  \mathcal{C}_{10}^{\,\prime\,*} \right]  \, , \\
    {\rm Im} \rho_2^+ & = \frac{1}{4}\,{\rm Im}\left[\mathcal{C}_+^R (\mathcal{C}_-^{R})^* - \mathcal{C}_-^L (\mathcal{C}_+^{L})^* \right]= {\rm Im}\left[     \mathcal{C}_{10}^{\,\prime}\, \mathcal{C}_{10}^{*} 
    + \mathcal{C}_{9}^{\,\prime}\,       (\mathcal{C}_9 + \mathcal{C}_9^{\,\text{eff}} )^*
      \right]  \, ,       \label{eq:BSM-dep} \\ \nonumber
    {\rm Re} \rho_2^- & = \frac{1}{4}\,{\rm Re}\left[\mathcal{C}_+^R (\mathcal{C}_-^{R})^* + \mathcal{C}_-^L (\mathcal{C}_+^{L})^* \right] = \frac{1}{2} \left[  |\mathcal{C}_{10}|^2  -| \mathcal{C}_{10}^{\,\prime}|^2+ \left|\mathcal{C}_9^{\,\text{eff}} + \mathcal{C}_9 \right|^2  
        -\left| \mathcal{C}_{9}^{\,\prime} \right|^2 
    \right]  
    \, , \\ \nonumber
    {\rm Im} \rho_2^- & = \frac{1}{4}{\rm Im}\left[   \mathcal{C}_+^R (\mathcal{C}_-^{R})^* + \mathcal{C}_-^L (\mathcal{C}_+^{L})^*\right] = {\rm Im} \left[\mathcal{C}_{10}^{\,\prime}  \,(\mathcal{C}_9^{\,\text{eff}} + \mathcal{C}_9)^* - \mathcal{C}_{10} \,\mathcal{C}_9^{\,\prime\,*} \right]  \, , \\ \nonumber
    \rho_{3,\,S/P}^\pm &=\frac{1}{2}\,\mathcal{C}_9^{\mathcal{R},\,S/P} \,\left((\mathcal{C}_\pm^{R})^* + (\mathcal{C}_\pm^{L})^*\right) = \mathcal{C}_9^{\mathcal{R},\,S/P} \,\left(\mathcal{C}_9^{*}\pm \mathcal{C}_9^{\prime \ast}\right) \, ,\\ \nonumber
    \rho_{4,\,S/P}^\pm &=\frac{1}{2}\,\mathcal{C}_9^{\mathcal{R},\,S/P} \,\left((\mathcal{C}_\pm^{R})^* - (\mathcal{C}_\pm^{L})^*\right)= \mathcal{C}_9^{\mathcal{R},\,S/P} \,\left(\mathcal{C}_{10}^{*}\pm \mathcal{C}_{10}^{\prime \ast}\right) \, .
\end{align}
In the SM besides the resonant contributions $\mathcal{C}_9^{\mathcal{R},S(P)}$ only one combination of couplings
survives  $\rho_1^{\pm}\,=\,2\,\re\rho_2^-=| \mathcal{C}_9^{\,\text{eff}}|^2$, and 
$\delta\rho=\re\rho_2^+=\im \rho_2^\pm=\rho_{3,\,S/P}^\pm=\rho_{4,\,S/P}^\pm=0$. This is in stark contrast with 
corresponding $b\to d,s\,\ell^+\ell^-$ processes where two combinations are finite in the SM,  $\rho_1^{\pm}\,=2\re\rho_2^- =| \mathcal{C}_9^{\,\text{eff}}|^2 +|\mathcal{C}_{10}|^2$
and $\delta\rho= \re \rho_2^+= {\rm Re}  \mathcal{C}_9^{\,\text{eff}} \mathcal{C}_{10}^*$~\cite{Das:2014sra}. 
This difference originates from  
the GIM-condition  in charm $\mathcal{C}_{10}^{\text{SM}} = 0$  and 
leads to the null tests with $I_{5,6,7}^{\text{SM}}=0$~.

\section{Input for SM and NP fits}
\begin{table}
    \centering
    \begin{adjustbox}{max width=\textwidth}
    \renewcommand{\arraystretch}{2}
    \begin{tabular}{c | c | c | c}
        \hline 
        \hline 
        \rowcolor{LightBlue} ``good" bins &  SM measurements & NP measurements & excluded measurements \\
        \hline 
        \hline 
        $d\Gamma/dq^2|_{\sqrt{q^2} > 0.6\,\mathrm{GeV}}$ \cite{LHCb:2021yxk} & $d\Gamma/dq^2$ \cite{LHCb:2021yxk} & & \\ 
        \hline 
        $d\Gamma/dp^2|_{\sqrt{p^2} < 0.9\,\mathrm{GeV}}$ \cite{LHCb:2021yxk} & $d\Gamma/dp^2$ \cite{LHCb:2021yxk} & & \\ 
        \hline 
        $\mathcal{B}$ with $\sqrt{q^2}\in \{[0.565\text{-}0.95],[0.95\text{-}1.1]\}\,\mathrm{GeV}$ \cite{LHCb:2017uns} & binned $\mathcal{B}$ \cite{LHCb:2017uns} & & full-$q^2$ $\mathcal{B}$ \cite{LHCb:2017uns}\\ 
        \hline 
        $\langle S_{2,3,4} \rangle$ 
        \cite{LHCb:2021yxk} & binned $\langle S_{2,3,4} \rangle$  \cite{LHCb:2021yxk} & & $\langle S_9 \rangle$ \& full-$q^2$ $\langle S_{2,3,4} \rangle$ \cite{LHCb:2021yxk} \\ 
        with $\sqrt{q^2}\in \{[0.565\text{-}0.78],[0.78\text{-}0.95],[0.95\text{-}1.02],[1.02\text{-}1.1]\}\,\mathrm{GeV}$ & & & \\
        \hline 
        & & $\langle S_{5,6,7,8} \rangle$ \& $\langle A_{2-9} \rangle$ \cite{LHCb:2021yxk} &  full-$q^2$ $\langle S_{5,6,7,8}\rangle \& \langle A_{2-9}\rangle$ \cite{LHCb:2021yxk} \\
        & & with $\sqrt{q^2} \in \{[0.78\text{-}0.95],[0.95\text{-}1.02],[1.02\text{-}1.1]\}\,\mathrm{GeV}$ & and $\sqrt{q^2} \in \{[<0.565],[0.565\text{-}0.78]\}\mathrm{GeV}$\\
        \hline 
        \hline 
    \end{tabular}
    \end{adjustbox}
    \caption{Data on $D \to \pi^+ \pi^- \mu^+ \mu^-$ decays. We denote by SM measurements those dominated by SM contributions, whereas NP measurements are those that are negligible in the SM. The ``good" bins are a subset of the SM measurements. Excluded measurements
    are those not used in any of our fits.}
    \label{tab:4bodydata}
\end{table}
Here we present in Tab.~\ref{tab:4bodydata} the measurements used for the SM fits in Sec.~\ref{subsec:resonancefit} and NP fits in Sec.~\ref{subsec:fitsNPDPPmumu} taken from \cite{LHCb:2017uns,LHCb:2021yxk}.
We do not take angular observables $S_i,A_i$  and branching ratio in the full $q^2$ region into account as these data are less sensitive and to avoid double-counting when using the $q^2$-binned ones.

\bibliographystyle{JHEP}
\bibliography{biblio.bib}

\end{document}